\documentclass[preprint,12pt,review,numbers,sort&compress]{elsarticle}

\usepackage{amssymb}
\usepackage{subfigure}
\usepackage{caption}
\usepackage{multirow}
\usepackage{tgtermes}
\usepackage{amsmath}
\usepackage{placeins}
\usepackage{natbib}
\usepackage{array}
\usepackage{graphicx}
\usepackage{gensymb}
\usepackage{textcomp}
\usepackage{bm}
\usepackage{enumerate}
\usepackage[margin=2.5cm]{geometry}
\usepackage{color, soul}
\usepackage[justification=centering]{caption}
\usepackage{booktabs}

\newcommand{\tabincell}[2]{\begin{tabular}{@{}#1@{}}#2\end{tabular}}  

\journal{ }

\begin{document}

\begin{frontmatter}
\sethlcolor{yellow}

\title{A New Meshless ``Fragile Points Method'' and A Local Variational Iteration Method for General Transient Heat Conduction in Anisotropic Nonhomogeneous Media}

\author[add1]{Yue Guan\corref{cor1}}
\ead{yuguan@ttu.edu}
\cortext[cor1]{Corresponding author.}

\author[add2]{Rade Grujicic}
\author[add3]{Xuechuan Wang}
\author[add4]{Leiting Dong}
\author[add1]{Satya N. Atluri}

\address[add1]{Department of Mechanical Engineering, Texas Tech University, Lubbock, TX 79415, United States}
\address[add2]{Faculty of Mechanical Engineering, University of Montenegro, 81000 Podgorica, Montenegro}
\address[add3]{School of Astronautics, Northwestern Polytchnical University, Xi'an 710072, China}
\address[add4]{School of Aeronautic Science and Engineering, Beihang University, Beijing 100191, China}

\begin{abstract}

A new and effective computational approach is presented for analyzing transient heat conduction problems. The approach consists of a meshless Fragile Points Method (FPM) being utilized for spatial discretization, and a Local Variational Iteration (LVI) scheme for time discretization. Anisotropy and nonhomogeneity do not give rise to any difficulties in the present implementation. The meshless FPM is based on a Galerkin weak-form formulation and thus leads to symmetric matrices. Local, very simple, polynomial and discontinuous trial and test functions are employed. In the meshless FPM, Interior Penalty Numerical Fluxes are introduced to ensure the consistency of the method. The LVIM in the time domain is generated as a combination of the Variational Iteration Method (VIM) applied over a large time interval and numerical algorithms. A set of collocation nodes are employed in each finitely large time interval. The FPM + LVIM approach is capable of solving transient heat transfer problems in complex geometries with mixed boundary conditions, including pre-existing cracks. Numerical examples are presented in 2D and 3D domains. Both functionally graded materials and composite materials are considered. It is shown that, with suitable computational parameters, the FPM + LVIM approach is not only accurate, but also efficient, and has reliable stability under relatively large time intervals. The present methodology represents a considerable improvement to the current state of science in computational transient heat conduction in anisotropic nonhomogeneous media.

\end{abstract}

\end{frontmatter}

\section{Introduction}

Transient heat conduction analysis in nonhomogeneous media is of great interest in research and engineering applications \cite{Sladek2005, Quint2011, Zhang2019}. Typical nonhomogeneous media include conventional composite materials and functionally graded materials (FGMs), etc. In FGMs, the material properties vary gradually in space \cite{Sladek2008, Miyamoto2013}. These materials have been found widespread prospective applications in aerospace industry, computer circuit industry, etc\cite{Simsek2009}. Therefore, there is an increasing demand for a reliable, accurate and efficient numerical approach for heat conduction problems in nonhomogeneous materials. Due to various processing techniques, the FGM or composite material may exhibit isotropic, orthotropic or anisotropic properties \cite{Chen2000, Sladek2008}. Hence the anisotropy should also be taken into account.

Numerical methods for solving transient heat conduction problems usually have two discretization stages \cite{LeVeque1992, Afrasiabi2019}. First, spatial discretization is employed. This step reduces the original partial differential equation (PDE) to a set of ordinary differential equations (ODEs) in time. These ODEs, known as ``semi-discrete equations'', are then discretized in time using some standard ODE solvers. The most commonly used discretization methods are the Finite Element Method (FEM) \cite{Zienkiewicz2005} in the spatial domain and the finite difference schemes \cite{Shen2011} in the time domain. However, in commercial FEM codes, the material properties are usually considered to be uniform in each element. It also has difficulties in analyzing systems with fragmentation, e.g., crack propagation in thermally shocked brittle materials. The finite difference methods in the time domain, on the other hand, may have stability and accuracy problems while using large time steps. In general, the current commercial codes are far from perfect.

Apart from the FEM, other mesh-based numerical methods such as the Finite Volume Method (FVM) \cite{Chai1994} and Boundary Element Method (BEM) \cite{Wrobel2003, Wen2009} can also be utilized in spatial discretization in heat transfer problems. In BEM, the non-availability of fundamental solutions in nonhomogeneous anisotropic media is a serious limitation which is often insurmountable. The same as the FEM, the FVM and BEM approaches also have drawbacks in solving fragmentation problems such as thermal-shock induced crack propagation in brittle solids. Their accuracy is also threatened when mesh distortion occurs. Another category of methods, known as ``meshless methods'', are partly or completely free of mesh discretization. As a result, the human and computer cost in generating a high-quality contiguous mesh can be eliminated or reduced. This is a great improvement especially in 3D problems involving complex geometries. The Smoothed Particle Hydrodynamics (SPH) proposed by \citet{Randles1996} is one of the earliest meshless methods. Though its original formulation has a problem of stability and particle deficiency on and near the boundaries, a number of improved methods based on the SPH have been carried out, including the modified SPH method by \citet{Randles1996}, the Reproducing Kernel Particle Method (RKPM) by \citet{Liu1995} and the Corrective Smoothed Particle Method (CSPM) by \citet{Chen1999}, etc. These methods are extensively used in thermal analysis and fluid and solid mechanics. The SPH and its improved methods are based on a strong form, making it difficult to study their stability. Yet numerical tests have implied that the SPH method may turn unstable under random point distributions \cite{Graham2008}.

Another category of meshless methods is weak-form-based, namely, employing the variational principle to minimize the weighted residual of the governing differential equations. The Diffuse Element Method (DEM) \cite{Nayroles1992} was initially introduced as a generalization of the FEM by removing some limitations related to the trial functions and mesh generations. The original formulation fails in passing the patch test. Nevertheless, \citet{Krongauz1997} established an improved DEM based on Petrov-Galerkin formulation (PG DEM) which satisfies the patch test but resulting in asymmetric matrices. Based on the DEM, the Element-Free Galerkin (EFG) method was carried out by \citet{Belytschko1994} in which the shape functions are developed by Moving Least Squares (MLS) or Radial Basis Function (RBF) approximations. The MLS approximation does not have delta-function properties and hence imposition of essential boundary conditions is a serious limitation \cite{Zhu1998}. In addition, the integration of the Galerkin functional in EFG requires back-ground meshes and is tedious. Furthermore, the EFG is not necessarily objective when the back-ground meshes are rotated. After that, a Local Boundary Integral Equation (LBIE) Method is introduced by \citet{Zhu1999}. While analogous to the BEM, the LBIE method circumvents the problem of global fundamental solutions in nonhomogeneous anisotropic materials in BEM. It uses local fundamental solutions (assuming locally homogeneous material properties) or Heaviside functions as test functions. However, the method still has drawbacks in solving fragmentation problems and has asymmetric matrices. Finally, \citet{Atluri1998} proposed the Meshless Local Petrov-Galerkin (MLPG) approach in 1998. Compared with the EFG method, the MLPG approach employs the local Petrov-Galerkin weak-formulation instead of the global Galerkin weak-formulation. The MLPG method is a truly meshless method and has shown its capability and accuracy in 2D and 3D transient heat conduction analysis involving anisotropy, nonhomogeneity and temperature-dependent material properties \cite{Sladek2008, Shibahara2011}. Yet the MLPG method still has its limitations: the matrices are asymmetric; the computation of integration in the Petrov-Galerkin weak-form over local subdomains is complicated, as a result of the complex shape functions; and (the same as the EFG method) the essential boundary conditions cannot be imposed directly, since the MLS approximation usually does not pass its corresponding data points. A modified collocation method has to be applied to enforce the essential boundary conditions \cite{Zhu1998}.

In contrast to the complex, global continuous MLS approximated shape function used in the EFG and MLPG, local, simple, polynomial, piecewise continuous trial functions are applied in generating the Fragile Points Method (FPM) in \cite{LeitingDongTianYangKaileiWang2019}. Nevertheless, the method would become inconsistent if the Galerkin weak-form is employed directly with these discontinuous trial and test functions. The Numerical Flux Correction, which is widely used in Discontinuous Galerkin (DG) methods \cite{Arnold2001, Mozolevski2007}, is introduced to remedy the problem. Whereas on the other hand, the inherent discontinuity can also be a benefit. Since it is convenient to relax the continuity requirement between neighboring points, the FPM has a great potential in analyzing systems involving cracks, ruptures and fragmentations, such as in problems of thermal shock in brittle solids. The method has already shown its stability, accuracy and efficiency in solving 1D and 2D Poisson equations \cite{LeitingDongTianYangKaileiWang2019} and elasticity problems \cite{Yang2019}. In the current work, it is further extended to 2D and 3D heat transfer analysis for discretization in space. The Galerkin functional in the FPM can be integrated quite simply, and finally leads to symmetric matrices. Thus, the FPM based on a Galerkin weak-form is far more efficient than either the EFG or the MLPG method.

After the spatial discretization is carried out, a semi-discrete system is achieved. The final computing accuracy and efficiency are also significantly affected by the ODE solver employed in the time domain. Though nonlinearity is not emphasized in this paper, here we focus on generalized time discretization methods that can deal with linear as well as nonlinear ODEs. These methods can roughly be divided into two categories: 1). The finite difference method, which is simple and the most widely used, and 2). weak-form method based on weighted residual approximations \cite{Wang2018}. The most well-known finite difference schemes include: the central, forward and backward difference schemes \cite{Smith1985}, Runge-Kutta method \cite{Fehlberg1969}, Newmark-$\beta$ method \cite{Newmark1959} and Hilber-Hughes-Taylor (HHT)-$\alpha$ method \cite{Hilber1977}. The Houbolt's method \cite{Houbolt1950} based on a third-order interpolation is also famous in dynamic analysis. \textit{Ode45} , the most popular, highly optimized built-in ODE solver in MATLAB, is based on an explicit Runge-Kutta formula \cite{Vie1986, Dormand1986} and has excellent performance in solving low-dimensional nonlinear ODEs. However, we aim here at a high-dimensional semi-discrete system generated with the 2D or 3D FPM. The efficiency of the classic finite difference methods may not be sufficient then. Moreover, many finite difference methods may encounter stability problems when considering nonlinearity or under large time steps \cite{Newmark1959, Hilber1977, Wang2019b}.

On the other hand, the weak-form methods, though somewhat hard to implement, have a potential in analyzing high dimensional and nonlinear systems more efficiently. For periodic systems, the Harmonic Balance (HB) method \cite{Thomas2013} and the Spectral Time Domain Collocation (TDC) method \cite{Elgohary2014} were developed. However, periodic responses are rarely seen in heat conduction problems. Considering more generalized transient solutions, a series of analytical or semi-analytical asymptotic methods are introduced. \citet{He1999} proposed the Variational Iteration Method (VIM) which can be seen as an extension of the Newton-Raphson method to nonlinear algebraic equations (NAEs). The Adomian Decomposition Method (ADM) is then developed by \citet{Adomian1988}, in which the initial guess is corrected step by step by adding components of an Adomian polynomial. Following that, the Picard Iteration Method (PIM) is carried out by \citet{Fukushima1997} and modified by \citet{Woollands2015}. The method is also based on an initial guess and correctional iterative formula. The formula is relatively concise, yet computing the integral of nonlinear terms in each time step is a challenge. Though developed independently, the previous VIM, ADM and PIM approaches can be unified using a generalized Lagrange multiplier \cite{Wang2019a}. Based on that, \citet{Wang2019} employed the VIM over a finitely large time interval, along with a collocation method and numerical discretization, leading to the Local Variational Iteration Method (LVIM). Unlike the VIM, the LVIM is a numerical method, applicable to digital computation and has the potential in being implemented with parallel processing. The initial guess can be constructed in a relatively simple form. The method possesses excellent efficiency in solving nonlinear ODEs in fluid mechanics, structural mechanics, and astrophysics, etc \cite{Wang2019}. Several approximated algorithms can be generated to further improve the computing efficiency. In the current work, in order to maintain the best stability, only the classic LVIM based on the first kind of Chebyshev polynomials is considered. This method is also named as Chebyshev Local Iterative Collocation - 1 (CLIC-1) algorithm in \cite{Wang2019a}.

In this paper, we focus on transient heat conduction problems in anisotropic nonhomogeneous media. The system is discretized by using the FPM in space, and the LVIM in the time domain. Section~\ref{sec:FPM} presents the governing equations for heat conduction in anisotropic nonhomogeneous media, and the formulation of the FPM. The LVIM and its numerical implementations are introduced in section~\ref{sec:LVIM}. Numerous 2D and 3D examples, solved with the proposed FPM + LVIM approach, are carried out in section~\ref{sec:NR}, followed by a discussion on the computational parameters, and a brief concluding section.

\section{Fragile Points Method (FPM) based on a Galerkin Weak-Form and Point ``Stiffness'' Matrices} \label{sec:FPM}
\subsection{The heat conduction problem and governing equation}

Consider a transient heat conduction problem in a continuously anisotropic nonhomogeneous medium, which is governed by the following partial differential equation \cite{Sladek2008, Mackowski2011}:
\begin{align}
\begin{split}
\rho(\mathbf{x}) c(\mathbf{x})  \frac{\partial u}{\partial t} (\mathbf{x}, t) = \nabla \cdot \left[ \mathbf{k} \nabla u (\mathbf{x}, t) \right] + Q(\mathbf{x}, t), \quad \mathbf{x} \in \Omega, t \in \left[ 0, T \right],
\label{eqn:governing}
\end{split}
\end{align}
where $\Omega$ is the entire 2D or 3D domain under study, the coordinate vector $\mathbf{x} = \left[ x \; y \right]$ in 2D or $\mathbf{x} = \left[ x \; y \; z \right]$ in 3D, $u(\mathbf{x}, t)$ is the temperature field, and $Q(\mathbf{x}, t)$ is the density of heat sources. $\nabla$ is the gradient operator. The thermal conductivity tensor $\mathbf{k} (\mathbf{x})$, mass density $\rho (\mathbf{x})$ and specific heat capacity $c(\mathbf{x})$ are dependent on the spatial coordinates in nonhomogeneous media. The thermal conductivity tensor components $k_{ij}$ can also be directionally dependent for anisotropic materials.

Three main kinds of boundary conditions are considered:
\begin{align}
 \text{1).} \;  & \text{Dirichlet bc}:  & u(\mathbf{x}, t) = \widetilde{u}_D (\mathbf{x}, t), \quad \text{on} \; \Gamma_D &,  \notag \\
 \text{2).} \;  & \text{Neumann bc}: & q(\mathbf{x}, t) = \mathbf{k} \nabla u(\mathbf{x}, t)  \cdot \mathbf{n} =  \widetilde{q}_N (\mathbf{x}, t), \quad \text{on} \; \Gamma_N &, \notag \\
 \text{3).} \; & \text{Robin (convective) bc}: & q(\mathbf{x}, t) = h(\mathbf{x}) \left[ \widetilde{u}_R (\mathbf{x}) - u(\mathbf{x}, t) \right], \quad \text{on} \; \Gamma_R &,
 \label{eqn:BC0s}
\end{align}
where the global boundary $\partial \Omega = \Gamma_D \cup \Gamma_N \cup \Gamma_R$, $\mathbf{n}$ is the unit outward normal of $\partial \Omega$, $h(\mathbf{x})$ is the heat transfer coefficient, and $\widetilde{u}_R (\mathbf{x})$ is the temperature of the medium outside the convective boundary.

The initial condition is assumed as:
\begin{align}
\begin{split}
\left. u(\mathbf{x}, t) \right|_{t=0} = u(\mathbf{x}, 0) , \quad \text{in} \; \Omega \cup \partial \Omega.
\end{split}
\end{align}

\subsection{Local, polynomial, point-based discontinues trial and test functions}

\begin{figure}[htbp] 
  \centering 
    \subfigure[]{ 
    \label{fig:Schem_2D_01} 
    \includegraphics[width=0.48\textwidth]{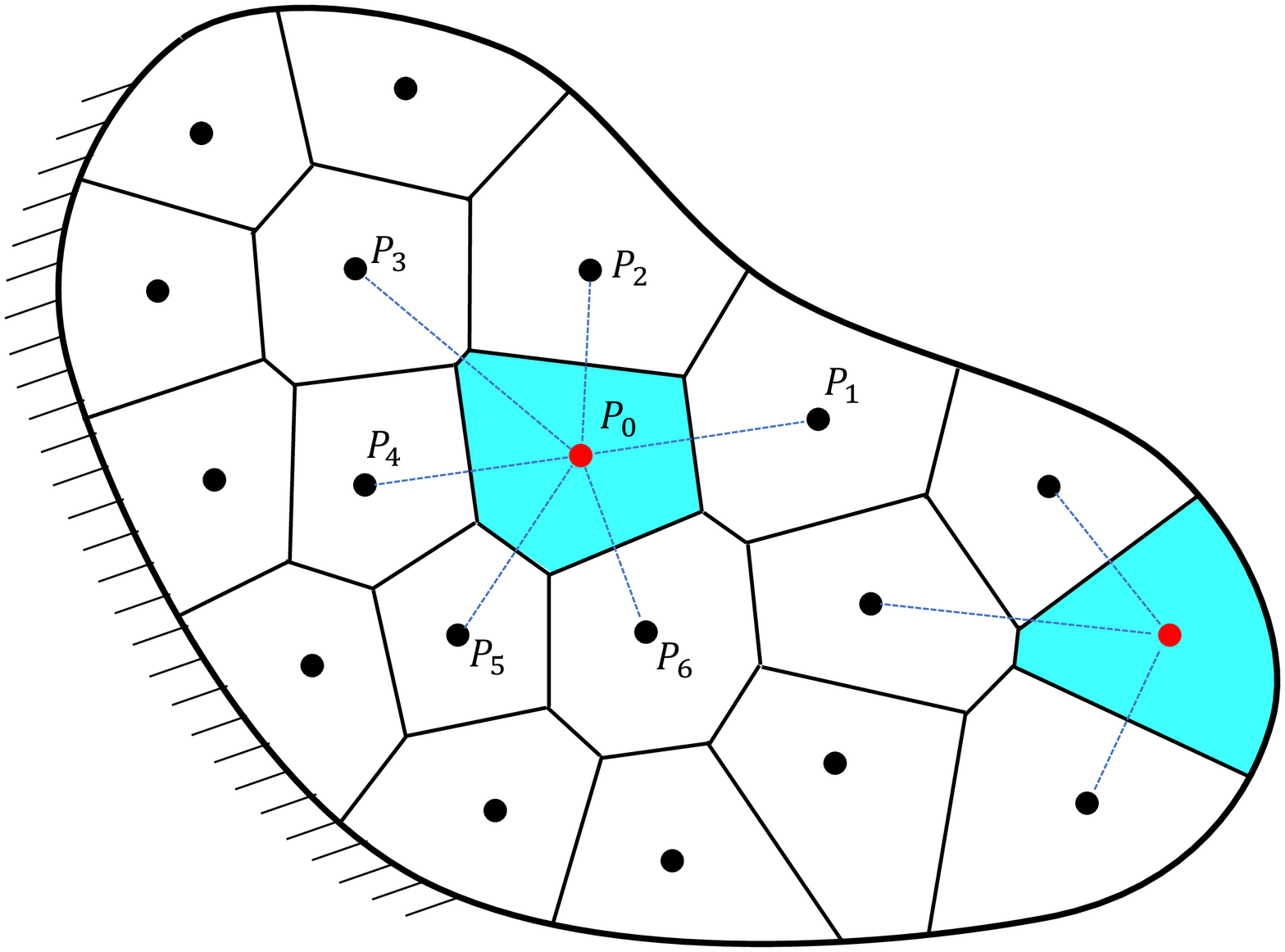}}  
    \subfigure[]{ 
    \label{fig:Schem_2D_02} 
    \includegraphics[width=0.48\textwidth]{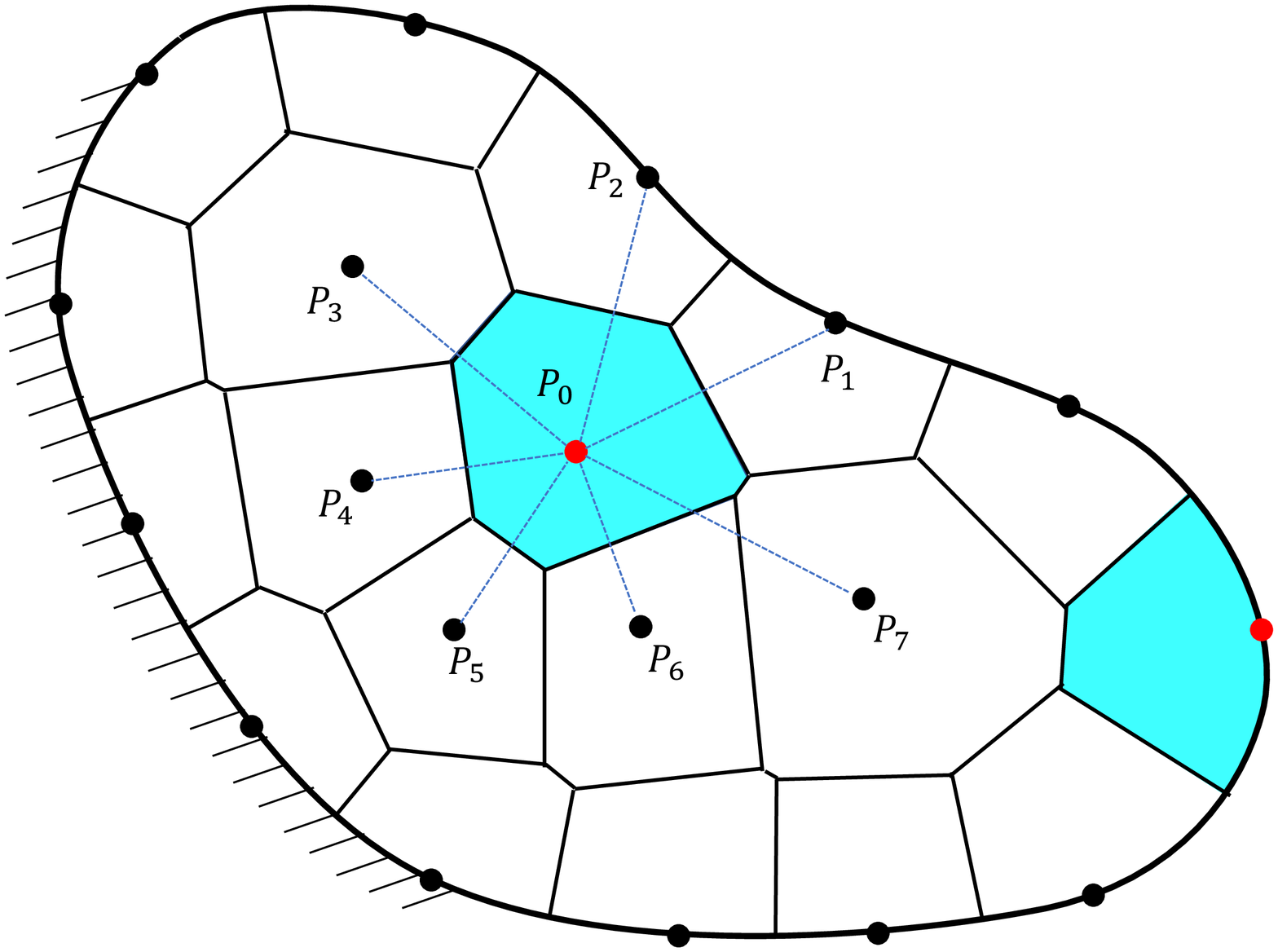}}  
  \subfigure[]{ 
    \label{fig:Schem_3D} 
    \includegraphics[width=0.75\textwidth]{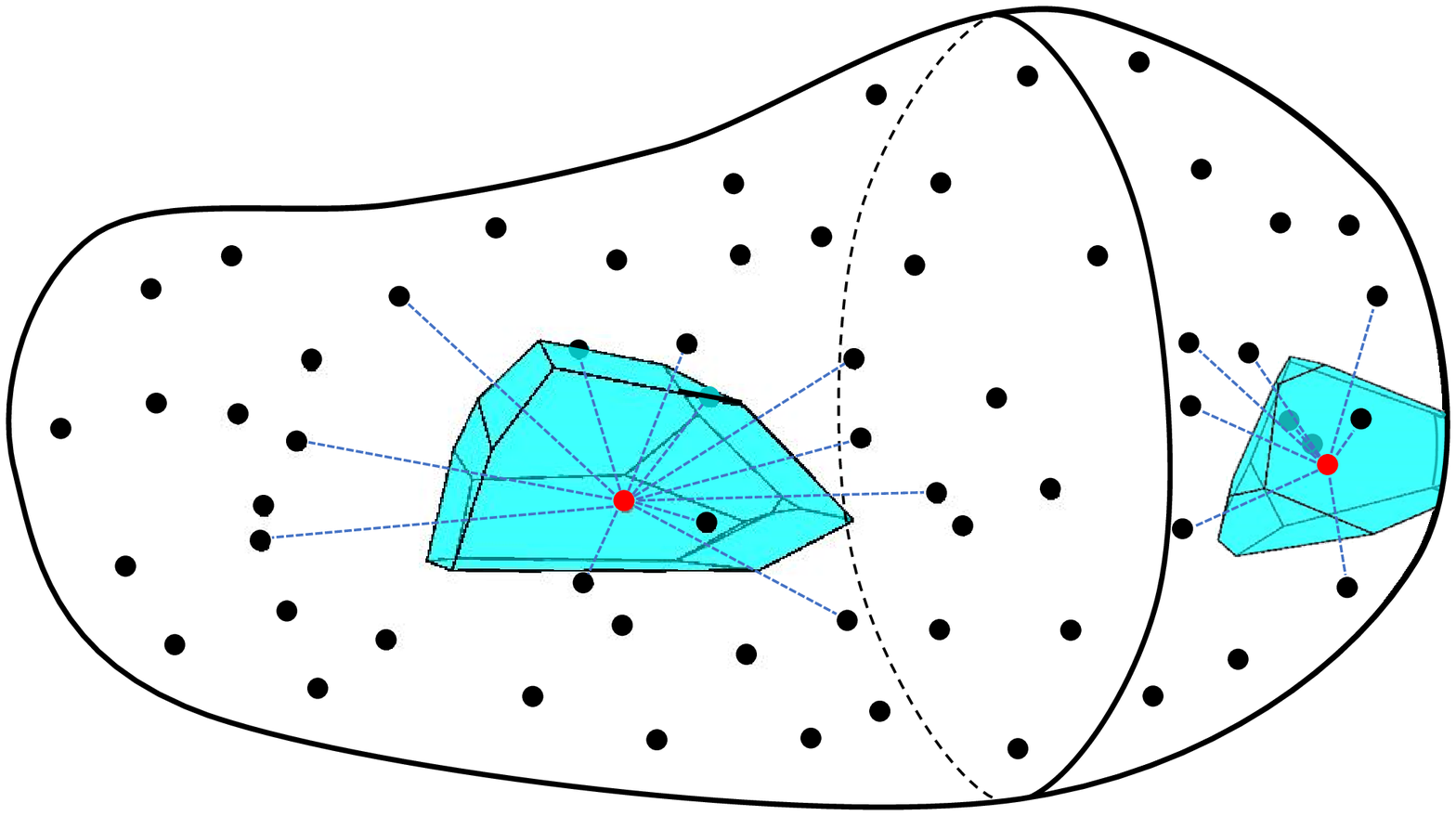}}  
  \caption{The domain $\Omega$ and its partitions. (a) 2D domain with points distributed inside it ($ P \in \Omega$). (b) 2D domain with points distributed inside it and on its boundary ($ P \in \Omega \cup \partial \Omega$). (c) 3D domain with points distributed inside it ($ P \in \Omega$).} 
  \label{fig:Schem} 
\end{figure}

In the domain $\Omega$, a set of random points are introduced. The global domain can then be partitioned into several confirming and nonoverlapping subdomains, with only one point in each subdomain (as shown in Fig.~\ref{fig:Schem}). The subdomains could be of arbitrary geometric shapes. And the partition is not unique. For simplicity, in this paper, the Voronoi Diagram partition \cite{Voronoi1908} is applied. Unlike the FEM and other element-based methods, the shape and trial functions in the present Fragile Points Method (FPM) are totally established based on the random points and are \textit{independent of the domain partition}. As a result, the temperature field could be discontinuous between subdomains, as well as the shape and trial functions. In FEM on the other hand, the trial and test functions are element-based, and are continuous at the interelement boundaries.

In each subdomain, we define the simple, local, polynomial trial function in terms of temperature $u$ and its gradient at the internal point. For instance, the trial function $u_h$ in subdomain $E_0$ which contains an internal point $P_0$ can be written as:
\begin{align}
\begin{split}
u_h (\mathbf{x}) = u_0 + (\mathbf{x} - \mathbf{x}_0) \cdot \nabla u \Big| _{P_0} , \quad \mathbf{x} \in E_0
\end{split}
\end{align}
where $u_0$ is the value of $u_h$ at $P_0$, and $\mathbf{x}_0$ is the coordinate vector of $P_0$. 

\begin{figure}[htbp] 
  \centering 
    \subfigure[]{ 
    \label{fig:Support_3D_01} 
    \includegraphics[width=0.48\textwidth]{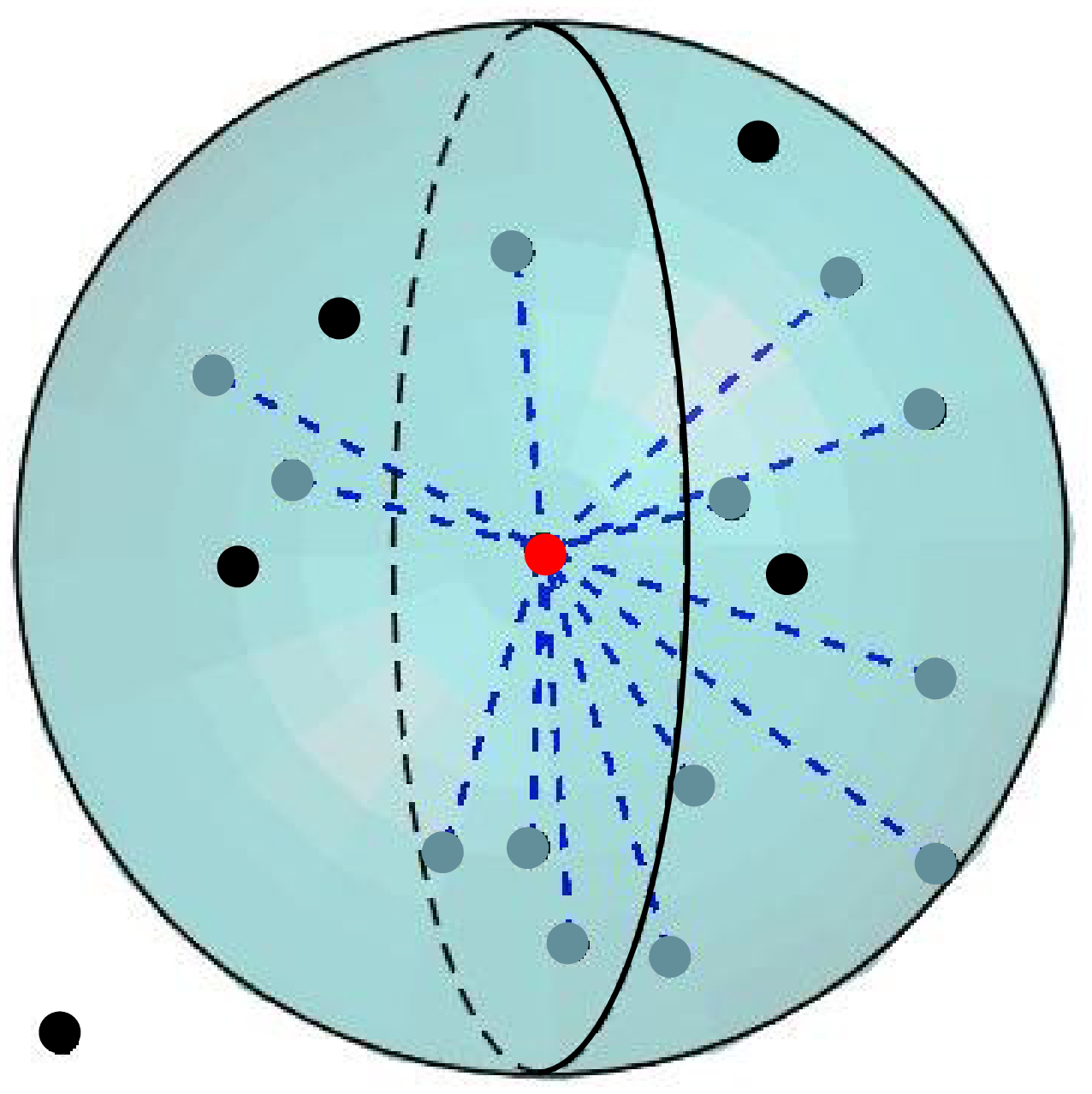}}  
    \subfigure[]{ 
    \label{fig:Support_3D_02} 
    \includegraphics[width=0.48\textwidth]{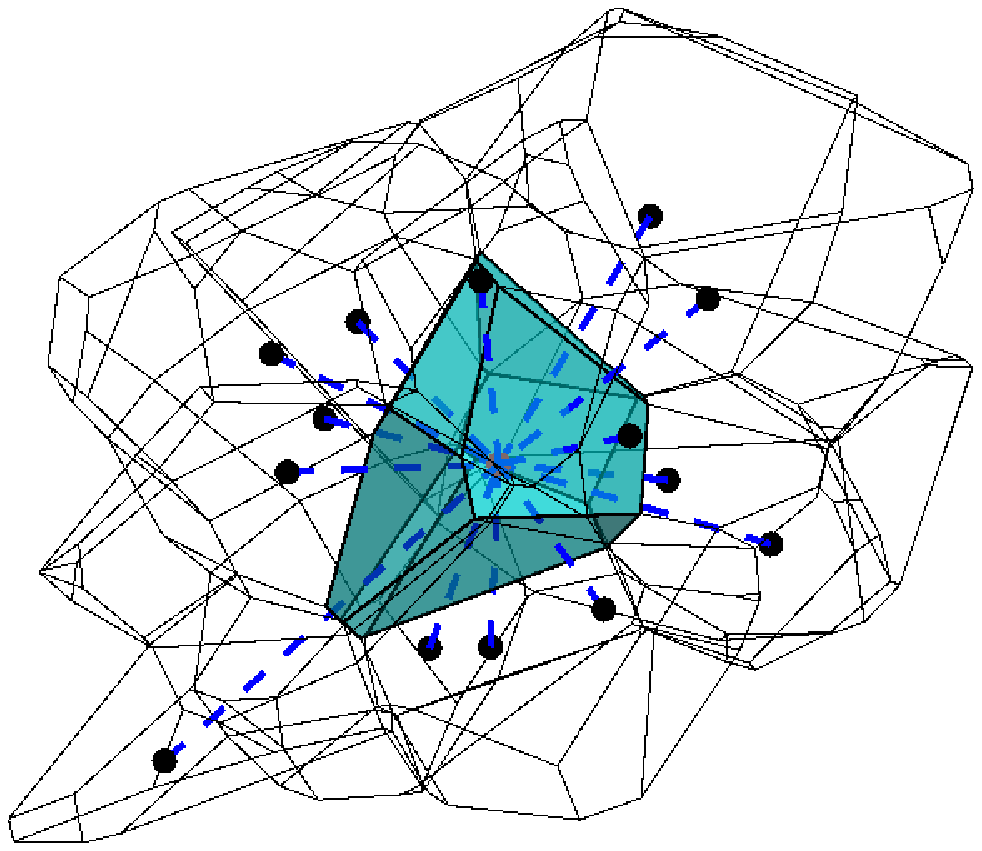}}  
  \caption{Two kinds of support of a point in 3D. } 
  \label{fig:Support} 
\end{figure}

The gradient of temperature $\nabla u$ at point $P_0$ remains unknown. In this paper, we employ the Generalized Finite Difference (GFD) method \cite{Liszka1980} to estimate $\nabla u \big| _{P_0}$ in terms of the value $u_h$ at several neighboring points of $P_0$. Unlike the common definition of the support of $P_0$ which includes all the points $P \in \left\{ P(\mathbf{x}) \mid (\mathbf{x} - \mathbf{x}_0) \leq r \right\}$ (as shown in Fig.~\ref{fig:Support_3D_01}, where $r$ is a constant radius), in the present work, the support of $P_0$ is defined to involve all the nearest neighboring points of $P_0$ in subdomains sharing boundaries with $E_0$ in the Voronoi partition (shown in Fig.~\ref{fig:Support_3D_02}). The points are named as $P_1$, $P_2$, $\cdots$, $P_m$.

In order to estimate the gradient $\nabla u \big| _{P_0}$, we minimize the following weighted discrete $L^2 $norm $J$:
\begin{align}
\begin{split}
J = \sum_{i=1}^m \left[ (\mathbf{x_i} - \mathbf{x}_0) \cdot \nabla u \Big| _{P_0} - (u_i - u_0) \right]^2 w_i,
\end{split}
\end{align}
where $\mathbf{x_i}$ donates the coordinate vector of $P_i$, $u_i$ is the value of $u_h$ at $P_i$, and $w_i$ is the value of weight function at $P_i$ ($i=1, 2, \cdots, m$). For convenience, we assume constant weight functions in this paper. Hence, the temperature gradient at $P_0$ is solved as:
\begin{align}
\begin{split}
\nabla u \Big| _{P_0} = \mathbf{B} \mathbf{u}_E,
\end{split}
\end{align}
where
\begin{align}
\begin{split}
\mathbf{u}_E = \left[ \begin{matrix} u_0 & u_1 & u_2 & \cdots & u_m \end{matrix} \right] ^ {\mathrm{T}}, \\
\mathbf{B} = (\mathbf{A}^\mathrm{T} \mathbf{A})^{-1} \mathbf{A}^\mathrm{T} \left[  \begin{matrix} \mathbf{I}_1 & \mathbf{I}_2 \end{matrix} \right], \\
\mathbf{I}_1 = \left( \left[ \begin{matrix} -1 & -1 & \cdots & -1 \end{matrix} \right] _{1 \times m} \right)^ \mathrm{T}, \\
\mathbf{I}_2 = \left[ \begin{matrix} 1 & 0 & \cdots & 0 \\ 0 & 1 & \ddots & \vdots \\ \vdots & \ddots & \ddots & 0 \\ 0 & \cdots & 0 & 1  \end{matrix} \right] _{m \times m}, \quad \mathbf{A} = \left[ \begin{matrix} \mathbf{x}_1 - \mathbf{x}_0 \\ \mathbf{x}_2 - \mathbf{x}_0 \\ \cdots \\ \mathbf{x}_m - \mathbf{x}_0 \end{matrix} \right]. \\
\end{split}  \notag
\end{align}

Therefore, the relation between $u_h$ and $\mathbf{u}_E$ can be obtained:
\begin{align}
\begin{split}
u_h (\mathbf{x}) = \mathbf{N} \mathbf{u}_E, \quad \mathbf{x} \in E_0
\end{split}
\end{align}
where $\mathbf{N}$ is called the shape function of $u_h$ in $E_0$:
\begin{align}
\begin{split}
\mathbf{N} = \left[ \mathbf{x} - \mathbf{x}_0 \right] \mathbf{B} + \left[ \begin{matrix} 1 & 0 & \cdots & 0 \end{matrix} \right] _ {1 \times (m+1)}
\end{split}
\end{align}

\begin{figure}[htbp] 
  \centering 
    \subfigure[]{ 
    \label{fig:ShapeF_2D} 
    \includegraphics[width=0.48\textwidth]{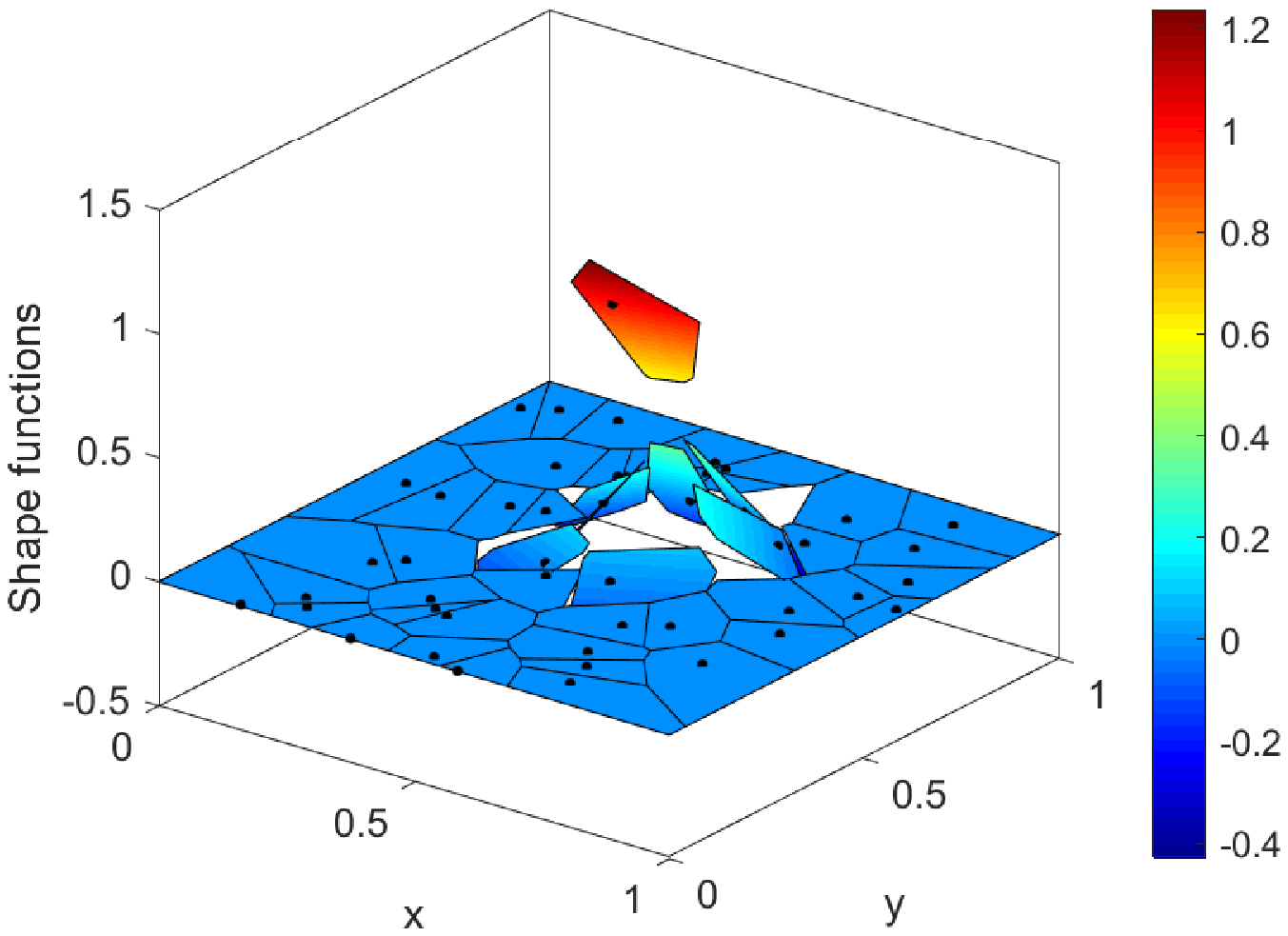}}  
    \subfigure[]{ 
    \label{fig:ShapeF_3D} 
    \includegraphics[width=0.48\textwidth]{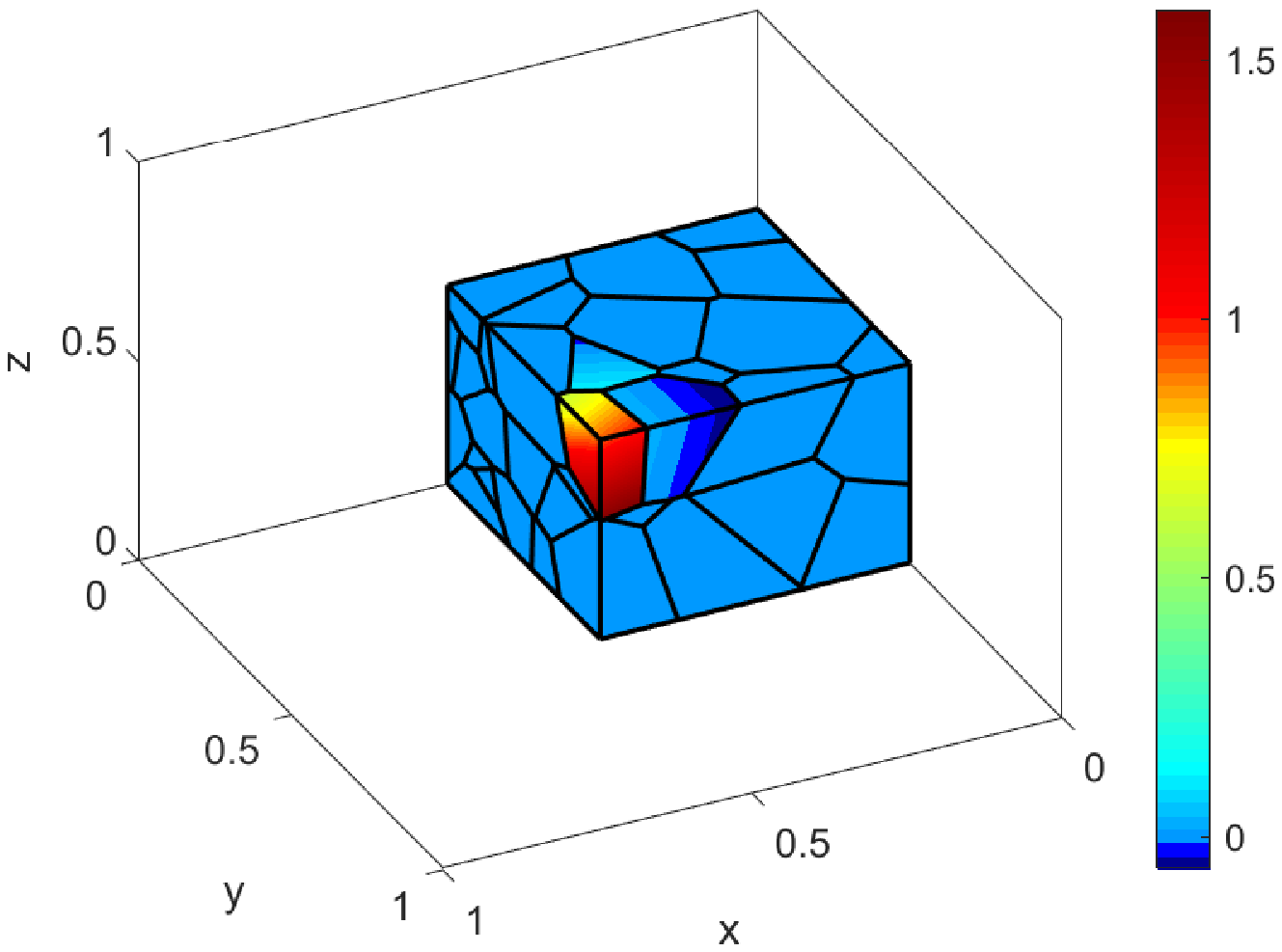}}  
  \caption{The shape functions. (a) 2D domain with 49 points. (b) 3D domain with 343 points ($x, y, z \leq 0.5$).} 
  \label{fig:ShapeF} 
\end{figure}

\begin{figure}[htbp] 
  \centering 
    \subfigure[]{ 
    \label{fig:TrialF_2D} 
    \includegraphics[width=0.48\textwidth]{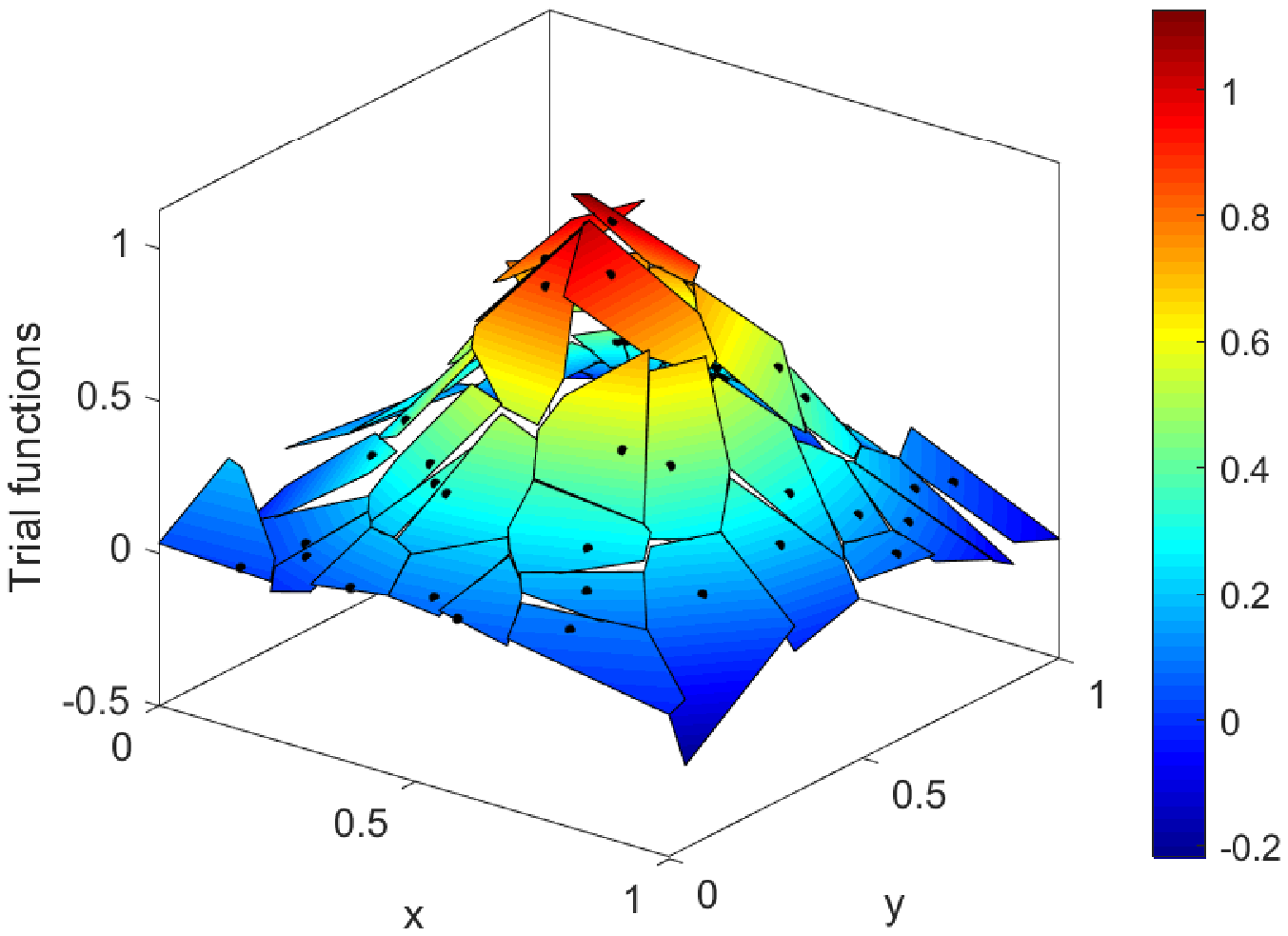}}  
    \subfigure[]{ 
    \label{fig:TrialF_3D} 
    \includegraphics[width=0.48\textwidth]{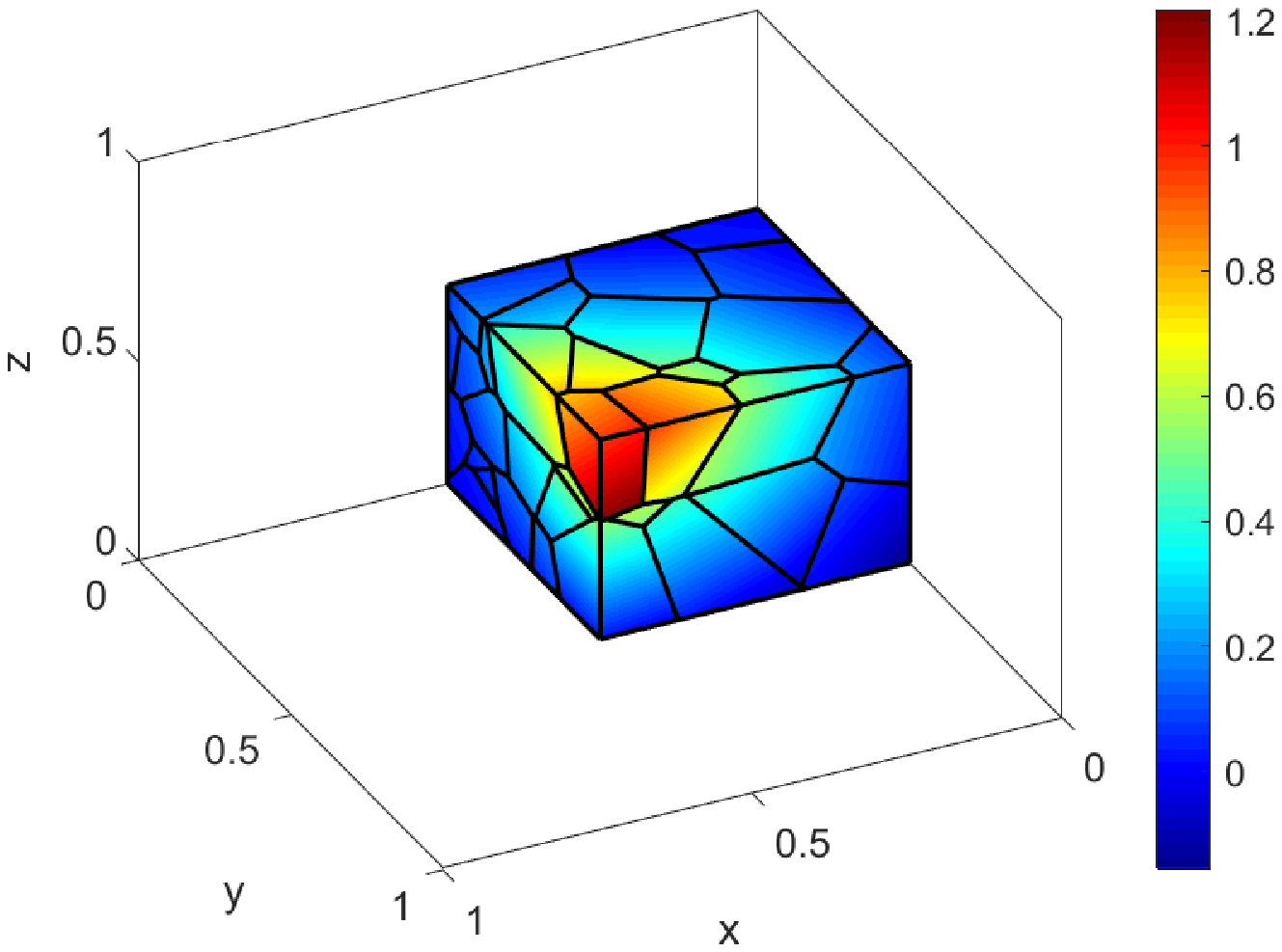}}  
  \caption{The trial functions for $u_a=e^{-10 \left| \mathbf{x} - \mathbf{x}_a \right| }$. (a) 2D domain with 49 points, $\mathbf{x}_a = [0.5 \; 0.5]$. (b) 3D domain with 343 points ($x, y, z \leq 0.5$), $\mathbf{x}_a = [0.5 \; 0.5 \; 0.5]$.} 
  \label{fig:TrialF} 
\end{figure}

Thus the shape function is defined independently in each subdomain, and no continuity requirement exists at the internal boundaries. Fig.~\ref{fig:ShapeF} shows the graphs of shape functions in 2D and 3D domains respectively. The trial function $u_h$ can be derived in each subdomain $E_i \in \Omega$ by the same process. Thus $u_h$ can also be discontinuous at the internal boundaries. For instance, the trial functions simulating an exponential function $u_a=e^{-10 \left| \mathbf{x} - \mathbf{x}_a \right| }$ is shown in Fig.~\ref{fig:TrialF}. As can be seen, the trial function is a simple local polynomial and just piecewise-continuous in the entire domain. The test function $v_h$ in the Galerkin weak-form in FPM is prescribed to have the same piecewise-continuous shape as $u_h$.

Unfortunately, the discontinuous trial and test functions will lead to an inconsistent and inaccurate result under the traditional Galerkin weak-form. To resolve that problem, we introduce Numerical Flux Corrections to the present FPM.

\subsection{Numerical flux corrections}

The Numerical Fluxes are widely used in Discontinuous Galerkin Methods \cite{Mozolevski2007} to help improving their accuracy and stability. In our work, the Interior Penalty (IP) Numerical Flux Corrections are employed.

The governing equation Eqn.~\ref{eqn:governing} can be written in the local weak-form with test function $v$ in each subdomain $E \in \Omega$. After applying the Gauss divergence theorem, we can get:
\begin{align}
\begin{split}
\int_E \rho c v \frac{\partial u}{\partial t} \mathrm{d} \Omega + \int_E \nabla v ^{\mathrm{T}} \mathbf{k} \nabla u \mathrm{d} \Omega = \int_E Qv \mathrm{d} \Omega + \int_{\partial E} v \mathbf{n}^{\mathrm{T}} \mathbf{k} \nabla u \mathrm{d} \Gamma,
\label{eqn:weakform1}
\end{split}
\end{align}
where $\partial E$ is the boundary of the subdomain, and $\mathbf{n}$ is the unit vector outward to $\partial E$.

\begin{figure}[htbp] 
  \centering 
    \subfigure[]{ 
    \label{fig:IB_2D} 
    \includegraphics[width=0.48\textwidth]{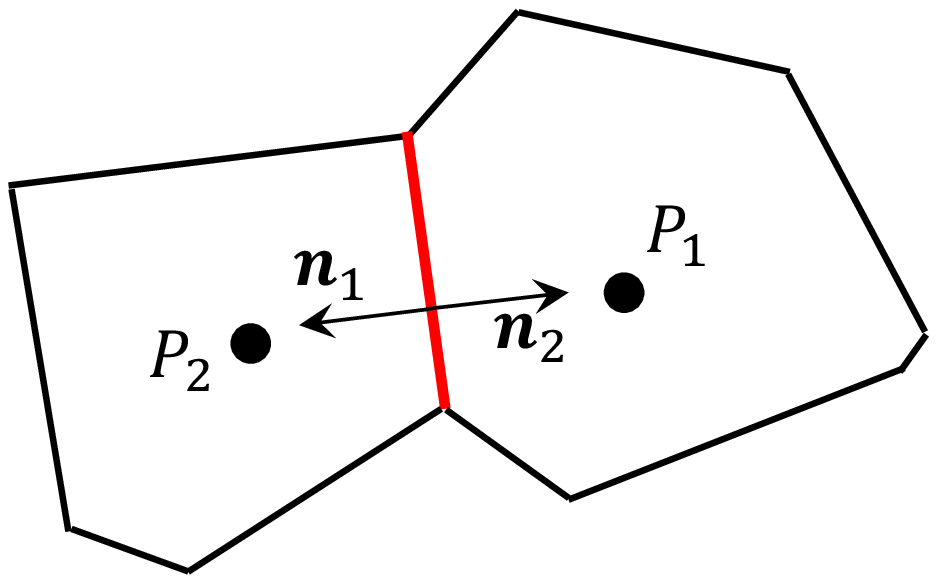}}  
    \subfigure[]{ 
    \label{fig:IB_3D} 
    \includegraphics[width=0.48\textwidth]{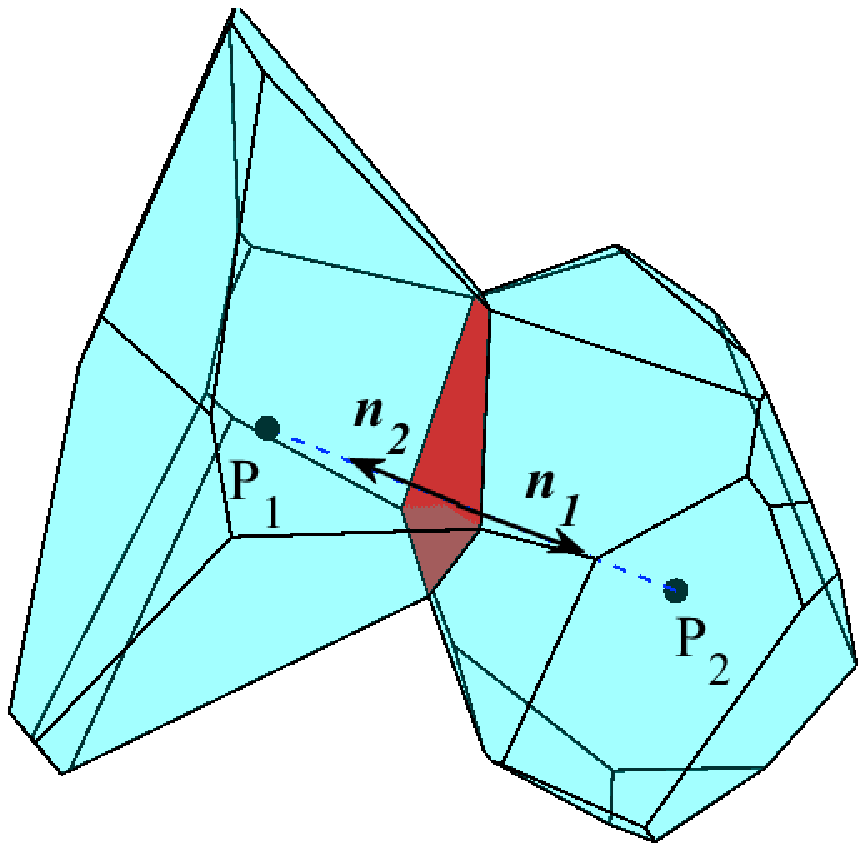}}  
  \caption{The inner boundary and normal vectors. (a) 2D case. (b) 3D case.} 
  \label{fig:IB} 
\end{figure}

Let $\Gamma$ donate the set of all internal and external boundaries, i.e., $\Gamma = \Gamma_h + \partial \Omega = \Gamma_h + \Gamma_D + \Gamma_N + \Gamma_R$, where $\Gamma_h$ is the set of all internal boundaries. We sum Eqn.~\ref{eqn:weakform1} over all subdomains and rewrite it with the jump operator $[]$ and average operator $\{\}$:
\begin{align}
\begin{split}
& \sum_{E\in \Omega} \int_E \rho c v \frac{\partial u}{\partial t} \mathrm{d} \Omega + \sum_{E\in \Omega} \int_E \nabla v ^{\mathrm{T}} \mathbf{k} \nabla u \mathrm{d} \Omega = \\
& \qquad \int_\Omega Qv \mathrm{d} \Omega + \sum_{e \in \Gamma_h} \left( \int_e \left\{ v \right\} \left[ \mathbf{n}^\mathrm{T} \mathbf{k} \nabla u \right] + \left[ v \right] \left\{ \mathbf{n}^\mathrm{T} \mathbf{k} \nabla u \right\} \right) \mathrm{d} \Gamma + \sum_{e \in \Gamma_D \cup \Gamma_N \cup \Gamma_R} \int_e \left[ v \right]  \left\{ \mathbf{n}^\mathrm{T} \mathbf{k} \nabla u \right\} \mathrm{d} \Gamma,
\label{eqn:weakform2}
\end{split}
\end{align}
where the jump operator $[]$ and average operator $\{ \}$ are defined as (for $\forall w \in \mathbb{R}$):
\begin{align}
\begin{split}
[w] = \begin{cases}
w \Big|_e^{E_1} -  w \Big|_e^{E_2} & e \in \Gamma_h\\
w \Big|_e & e \in \partial \Omega
\end{cases}, \quad
\left\{ w \right\} = \begin{cases}
\frac{1}{2} \left( w \Big|_e^{E_1} +  w \Big|_e^{E_2} \right) & e \in \Gamma_h\\
w \Big|_e & e \in \partial \Omega
\end{cases}.
\notag
\end{split}
\end{align}
When $e \in \Gamma_h$ ($e \in \partial E_1 \cap \partial E_2$), $\mathbf{n}_j^e$ is a unit vector normal to $e$ and pointing outward from $E_j$ (see Fig.~\ref{fig:IB}).

Substituting the boundary conditions (Eqn.~\ref{eqn:BC0s}) into the last term in Eqn.~\ref{eqn:weakform2}, we obtain:
\begin{align}
\begin{split}
& \sum_{e \in \Gamma_N} \int_e \left[ v \right]  \left\{ \mathbf{n}^\mathrm{T} \mathbf{k} \nabla u \right\} \mathrm{d} \Gamma = \sum_{e \in \Gamma_N} \int_e v  \widetilde{q}_N  \mathrm{d} \Gamma , \\
& \sum_{e \in \Gamma_R} \int_e \left[ v \right]  \left\{ \mathbf{n}^\mathrm{T} \mathbf{k} \nabla u \right\} \mathrm{d} \Gamma = \sum_{e \in \Gamma_R} \int_e h v  \widetilde{u}_R  \mathrm{d} \Gamma - \sum_{e \in \Gamma_R} \int_e h v  u  \mathrm{d} \Gamma.
\end{split}
\end{align}

When $u$ is the exact solution, since there is no ‘jump’ on the internal boundaries, $\left[ \mathbf{n}^\mathrm{T} \mathbf{k} \nabla u \right]= 0$. Similarly, $\left[ u \right] = 0$. This leads to $\left\{ \mathbf{n}^\mathrm{T} \mathbf{k} \nabla v \right\} \left[ u \right]=0$. Hence, we can replace the term $ \left\{ v \right\} \left[ \mathbf{n}^\mathrm{T} \mathbf{k} \nabla u \right]$ in Eqn.~\ref{eqn:weakform2} by $\left\{ \mathbf{n}^\mathrm{T} \mathbf{k} \nabla v \right\} \left[ u \right]$ without influencing the accuracy of the formula.

Two Internal Penalty Numerical Fluxes are applied on $\Gamma_h$ and $\Gamma_D$ with different penalty parameters $\eta_1$ and $\eta_2$ respectively. The formula of the FPM with  IP Numerical Flux Corrections can then be achieved: 
\begin{align}
\begin{split}
& \sum_{E \in \Omega} \int_E \rho c v \frac{\partial u}{\partial t} \mathrm{d} \Omega + \sum_{E \in \Omega} \int_E \nabla v^\mathrm{T} \mathbf{k} \nabla u \mathrm{d} \Omega - \sum_{e \in \Gamma_h \cup \Gamma_D} \int_e \left( \left\{ \mathbf{n}^\mathrm{T} \mathbf{k} \nabla u \right\} \left[ v \right] + \left\{ \mathbf{n}^\mathrm{T} \mathbf{k} \nabla v \right\} \left[ u \right] \right) \mathrm{d} \Gamma \\
& \qquad + \sum_{e \in \Gamma_R} \int_e h v u \mathrm{d} \Gamma + \sum_{e \in \Gamma_h} \frac{\eta_1}{h_e} \int_e \left[ u \right] \left[ v \right] \mathrm{d} \Gamma + \sum_{e \in \Gamma_D} \frac{\eta_2}{h_e} \int_e u v \mathrm{d} \Gamma \\
& \qquad = \sum_{E \in \Omega} \int_E Q v \mathrm{d} \Omega + \sum_{e \in \Gamma_D} \int_e \left( \frac{\eta_2}{h_e} v - \mathbf{n}^\mathrm{T} \mathbf{k} \nabla v \right) \widetilde{u}_D \mathrm{d} \Gamma + \sum_{e \in \Gamma_N} \int_e v  \widetilde{q}_N  \mathrm{d} \Gamma + \sum_{e \in \Gamma_R} \int_e h v  \widetilde{u}_R  \mathrm{d} \Gamma,
\label{eqn:FPM}
\end{split}
\end{align}
where $h_e$ is a boundary-dependent parameter with the unit of length. For instance, $h_e$ can be defined as the length of the boundary (in 2D), the square root of the boundary area (in 3D), or the distance between the points in subdomains sharing the boundary. The penalty parameters $\eta_1$, $\eta_2$ are positive numbers having the same unit of $\mathbf{k}$ and independent of the boundary size. The method is only stable when the penalty parameters are large enough. However, on the other hand, an excessively large $\eta_1$ is harmful for the accuracy and may cause a condition number problem. A discussion on recommended values of these penalty parameters is presented in section~\ref{sec:PS}. The IP Numerical Flux Correction terms vanish when $u_h$ equals to the exact solution, that is, when there is no jump on internal boundaries and the Dirichlet boundary conditions are well satisfied.

There are two ways to impose the Dirichlet boundary conditions in practice. If there are no points distributed on the boundaries (as shown in Fig.~\ref{fig:Schem_2D_01}), the Interior Penalty terms on $\Gamma_D$ are responsible for the boundary conditions. The approach is analogous to the collocation method introduced in \cite{Zhu1998}. Alternatively, if boundary points are employed (as shown in Fig.~\ref{fig:Schem_2D_02}), we can also impose $u= \widetilde{u}_D$ strongly at the boundary points and thus the corresponding Internal Penalty terms in Eqn.~\ref{eqn:FPM} vanish.

\subsection{Numerical inplementation}

The formula of the FPM can be written in the matrix form finally:
\begin{align}
\begin{split}
\mathbf{C} \dot{\mathbf{u}} + \mathbf{K} \mathbf{u} = \mathbf{q},
\label{eqn:FPM_Matrix}
\end{split}
\end{align}
where $\mathbf{C}$ and $\mathbf{K}$ are the global heat capacity and thermal conductivity matrices respectively, $\mathbf{u}$ is the unknown vector with nodal temperatures, $\mathbf{q}$ is the heat flux vector.

Substituting the shape function $\mathbf{N}$ for $u_h$ and $v$, $\mathbf{B}$ for $\nabla u$ and $\nabla v$, the point heat capacity matrix $\mathbf{C}_E$, point thermal conductivity matrix $\mathbf{K}_E$ and the boundary thermal conductivity matrices $\mathbf{K}_h$, $\mathbf{K}_D$, $\mathbf{K}_R$ can be written as:
\begin{align}
\begin{split}
\mathbf{C}_E = & \int_E \rho c \mathbf{N}^\mathrm{T} \mathbf{N} \mathrm{d} \Omega, \quad E \in \Omega, \\ 
\mathbf{K}_E = & \int_E \mathbf{B}^\mathrm{T} \mathbf{k} \mathbf{B} \mathrm{d} \Omega, \quad E \in \Omega, \\
\mathbf{K}_h = & -\frac{1}{2} \int_e \left( \mathbf{N}_1^\mathrm{T} \mathbf{n}_1^\mathrm{T} \mathbf{k} \mathbf{B}_1 + \mathbf{B}_1^\mathrm{T} \mathbf{k}^\mathrm{T} \mathbf{n}_1 \mathbf{N}_1 \right) \mathrm{d} \Gamma + \frac{\eta_1}{h_e} \int_e \mathbf{N}_1^\mathrm{T} \mathbf{N}_1 \mathrm{d} \Gamma \\ 
& -\frac{1}{2} \int_e \left( \mathbf{N}_2^\mathrm{T} \mathbf{n}_2^\mathrm{T} \mathbf{k} \mathbf{B}_2 + \mathbf{B}_2^\mathrm{T} \mathbf{k}^\mathrm{T} \mathbf{n}_2 \mathbf{N}_2 \right) \mathrm{d} \Gamma + \frac{\eta_1}{h_e} \int_e \mathbf{N}_2^\mathrm{T} \mathbf{N}_2 \mathrm{d} \Gamma \\ 
& -\frac{1}{2} \int_e \left( \mathbf{N}_1^\mathrm{T} \mathbf{n}_1^\mathrm{T} \mathbf{k} \mathbf{B}_2 + \mathbf{B}_1^\mathrm{T} \mathbf{k}^\mathrm{T} \mathbf{n}_2 \mathbf{N}_2 \right) \mathrm{d} \Gamma - \frac{\eta_1}{h_e} \int_e \mathbf{N}_1^\mathrm{T} \mathbf{N}_2 \mathrm{d} \Gamma \\
& -\frac{1}{2} \int_e \left( \mathbf{N}_2^\mathrm{T} \mathbf{n}_2^\mathrm{T} \mathbf{k} \mathbf{B}_1 + \mathbf{B}_2^\mathrm{T} \mathbf{k}^\mathrm{T} \mathbf{n}_1 \mathbf{N}_1 \right) \mathrm{d} \Gamma - \frac{\eta_1}{h_e} \int_e \mathbf{N}_2^\mathrm{T} \mathbf{N}_1 \mathrm{d} \Gamma, \quad e \in \partial E_1 \cap \partial E_2,\\
\mathbf{K}_D = & - \int_e \left( \mathbf{N}^\mathrm{T} \mathbf{n}^\mathrm{T} \mathbf{k} \mathbf{B} + \mathbf{B}^\mathrm{T} \mathbf{k}^\mathrm{T} \mathbf{n} \mathbf{N} \right) \mathrm{d} \Gamma + \frac{\eta_2}{h_e} \int_e \mathbf{N}^\mathrm{T} \mathbf{N} \mathrm{d} \Gamma, \quad e \in \Gamma_D, \\
\mathbf{K}_R = & \int_e h \mathbf{N}^\mathrm{T} \mathbf{N} \mathrm{d} \Gamma, \quad e \in \Gamma_D,
\end{split}
\end{align}

The global heat capacity and thermal conductivity matrices can be established by assembling all the submatrices. The process is the same as the FEM. Similarly, the point and boundary heat flux vectors are developed:
\begin{align}
\begin{split}
\mathbf{q}_E = & \int_E \mathbf{N}^\mathrm{T} Q \mathrm{d} \Omega, \quad E \in \Omega, \\
\mathbf{q}_D = & -\int_e \mathbf{B}^\mathrm{T} \mathbf{k}^\mathrm{T} \mathbf{n} \widetilde{u}_D \mathrm{d} \Gamma + \frac{\eta_2}{h_e} \int_e \mathbf{N}^\mathrm{T} \widetilde{u}_D \mathrm{d} \Gamma, \quad e \in \Gamma_D, \\
\mathbf{q}_N = & \int_e \mathbf{N}^\mathrm{T} \widetilde{q}_N  \mathrm{d} \Gamma, \quad e \in \Gamma_N, \\
\mathbf{q}_R = & \int_e h \mathbf{N}^\mathrm{T} \widetilde{u}_R \mathrm{d} \Gamma, \quad e \in \Gamma_R.
\label{eqn:flux_vec}
\end{split}
\end{align}

The global heat flux vector is assembled in the same way. Eventually, a set of discretized ODEs (Eqn.~\ref{eqn:FPM_Matrix}) with sparse and symmetric matrices are achieved.

\section{Local Variational Iteration Method (LVIM)} \label{sec:LVIM}
\subsection{Functional reclusive formula}

Eqn.~\ref{eqn:FPM_Matrix} can be rewritten as a system of standard first-order ODEs:
\begin{align}
\begin{split}
\dot{\mathbf{u}} = \mathbf{g} \left( \mathbf{u}, t\right) = - \mathbf{C}^\mathrm{-1} \mathbf{K} \mathbf{u} + \mathbf{C}^\mathrm{-1} \mathbf{q} (t), \qquad t \in \left[ 0, T\right].
\end{split}
\end{align}
The unknown temperature vector $\mathbf{u} = \left[ u_1, u_2, \cdots, u_L \right]^\mathrm{T}$, where $L$ is the number of points used in the FPM.

In a finitely large time interval $\left[ t_i, t_{i+1} \right] \subset \left[ 0, T \right]$, with a given initial approximation $\mathbf{u}_0 (\tau)$, the Local Variational Iteration Method (LVIM) approximates the exact solution $\mathbf{u}$ at any time $t$ with the following correctional iterative formula \cite{Wang2019}:
\begin{align}
\begin{split}
\mathbf{u}_{n+1} (t) = \mathbf{u}_n (t) + \int_{t_i} ^t \boldsymbol{\lambda} (\tau) \mathbf{R} \left( \mathbf{u}_n , \tau \right) \mathrm{d} \tau,
\label{LVIM-Iteration}
\end{split}
\end{align}
where the error residual $ \mathbf{R} \left( \mathbf{u}_n , \tau \right)$ is defined as:
\begin{align}
\begin{split}
 \mathbf{R} \left( \mathbf{u}_n , \tau \right) = \dot{\mathbf{u}}_n (\tau) -  \mathbf{g} \left( \mathbf{u}_n, \tau \right),
\end{split}
\end{align}
$\boldsymbol{\lambda} (\tau)$ is a matrix of Lagrange multipliers which are yet to be determined.

Eqn.~\ref{LVIM-Iteration} can also be regarded as a correctional iteration based on an optimally weighted error residual in time interval $\left[ t_i, t \right]$, where $\boldsymbol{\lambda} (\tau)$ is the set of optimal weighting functions. 

By making the right side of Eqn.~\ref{LVIM-Iteration} stationary, we obtain the following constraints for $\boldsymbol{\lambda} (\tau)$: 

\begin{align}
\begin{split}
\begin{cases} \mathbf{I} + \boldsymbol{\lambda} (\tau) \Big|_{\tau = t} = \mathbf{0} \\
\dot{\boldsymbol{\lambda}} (\tau) = \boldsymbol{\lambda} (\tau) \mathbf{J} \left( \mathbf{u}_n, \tau \right)
\end{cases}, \qquad \tau \in \left[ t_i, t \right], 
\end{split}
\end{align}
where
\begin{align}
\begin{split}
\mathbf{J} \left( \mathbf{u}_n, \tau \right) = \frac{\partial \mathbf{g \left( \mathbf{u}_n, \tau \right)} }{\partial \mathbf{u}_n} =  - \mathbf{C}^\mathrm{-1} \mathbf{K}
\end{split}
\end{align}
is the Jacobian matrix. $\mathbf{I}$ is the unit matrix.

Using the theory of Magnus series \cite{Blanes2009}, it can be proved that \cite{Wang2017}:
\begin{align}
\begin{split}
\begin{cases} \boldsymbol{\lambda} (t) = - \mathbf{I} \\
\frac{\partial \boldsymbol{\lambda} (t) }{\partial t} = -\mathbf{J} \left( \mathbf{u}_n, t \right)  \boldsymbol{\lambda} (t)
\end{cases}, \qquad t \in \left[ \tau, t_{i+1} \right] , 
\label{Cons-lambda}
\end{split}
\end{align}

Differentiating Eqn.~\ref{LVIM-Iteration} and substituting Eqn.~\ref{Cons-lambda} into it:
\begin{align}
\begin{split}
\dot{\mathbf{u}}_{n+1} (t) & = \dot{\mathbf{u}}_{n} (t) + \boldsymbol{\lambda} (t) \mathbf{R} \left( \mathbf{u}_n, t \right) + \int_{t_i}^t \frac{\partial \boldsymbol{\lambda} (t)}{\partial t} \mathbf{R} \left( \mathbf{u}_n, \tau \right)  \mathrm{d} \tau \\
& =  \mathbf{g} \left( \mathbf{u}_n, t \right) - \mathbf{J} \left( \mathbf{u}_n, t \right) \int_{t_i}^t  \boldsymbol{\lambda} (\tau) \mathbf{R} \left( \mathbf{u}_n, \tau \right)  \mathrm{d} \tau \\
& =  \mathbf{g} \left( \mathbf{u}_n, t \right) - \mathbf{J} \left( \mathbf{u}_n, t \right) \left[ \mathbf{u}_{n+1} (t) - \mathbf{u}_{n} (t) \right].
\end{split}
\end{align}

Therefore, we obtain the recursive formula:
\begin{align}
\begin{split}
\dot{\mathbf{u}}_{n+1} (t) + \mathbf{J} \left( \mathbf{u}_n, t \right) \mathbf{u}_{n+1} (t)  = \mathbf{g} \left( \mathbf{u}_n, t \right) +  \mathbf{J} \left( \mathbf{u}_n, t \right) \mathbf{u}_{n} (t), \qquad t \in \left[ t_i, t_{i+1} \right].
\label{eqn:LVIM-f1}
\end{split}
\end{align}

\subsection{Collocation method and numerical discretization}

Eqn.~\ref{eqn:LVIM-f1} can be written in the weak-form in the time interval $\left[ t_i, t_{i+1} \right]$ with a matrix of test functions $\mathbf{v} (t)$:
\begin{align}
\begin{split}
\int_{t_i}^{t_{i+1}} \mathbf{v} (t) \left[ \dot{\mathbf{u}}_{n+1} (t) + \mathbf{J} \left( \mathbf{u}_n, t \right) \mathbf{u}_{n+1} (t) \right] \mathrm{d} t = \int_{t_i}^{t_{i+1}} \mathbf{v} (t) \left[ \mathbf{g} \left( \mathbf{u}_n, t \right) +  \mathbf{J} \left( \mathbf{u}_n, t \right) \mathbf{u}_{n} (t) \right] \mathrm{d} t.
\end{split}
\end{align}
Let $\mathbf{v} (t) = \text{diag} \left( \left[ v, v, \cdots, v \right] \right)$, where $v$ is the Dirac Delta function for a set of collocation nodes $t_1$, $t_2$, $\cdots$, $t_M \in \left[ t_i, t_{i+1} \right]$, that is:
\begin{align}
\begin{split}
v = \delta \left( t - t_m \right), \quad t_m \in \left[ t_i, t_{i+1} \right], \quad m = 1, 2, \cdots, M.
\end{split}
\end{align}
The weak-form formula leads to:
\begin{align}
\begin{split}
\dot{\mathbf{u}}_{n+1} (t_m) + \mathbf{J} \left( \mathbf{u}_n, t_m \right) \mathbf{u}_{n+1} (t_m)  = \mathbf{g} \left( \mathbf{u}_n, t_m \right) +  \mathbf{J} \left( \mathbf{u}_n, t_m \right) \mathbf{u}_{n} (t_m), \\
\quad t_m \in \left[ t_i, t_{i+1} \right], \quad m = 1, 2, \cdots, M.
\label{eqn:LVIM-1-Col}
\end{split}
\end{align}

A set of orthogonal basis functions $\boldsymbol{\Phi} = \{ \phi_0, \phi_1, \cdots, \phi_N \}$ are used to construct the trial function $u_e$:
\begin{align}
\begin{split}
u_e (t) = \sum_{n=0}^N a_{e, n} \phi_n (t),
\label{eqn:BasisF}
\end{split}
\end{align}
where $u_e (t) (e = 1, 2, \cdots, L)$ are elements of the solution vector $\mathbf{u} (t)$. A number of types of basis functions can be used in the collocation method, including harmonics, polynomials, Radial Basis Functions (RBFs), etc. In this paper, we employed the first kind of Chebyshev polynomials \cite{MasonJohnC.2002} as an example. The collocation nodes are selected as Chebyshev-Gauss-Lobatto points. From Eqn.~\ref{eqn:BasisF}, we can get:
\begin{align}
\begin{split}
\mathbf{U}_e = \mathbf{Q} \mathbf{A}_e, \qquad \dot{\mathbf{U}}_e = (\mathbf{LQ}) \mathbf{A}_e
\end{split}
\end{align}
where
\begin{align}
\begin{split}
& \mathbf{U}_e = \left[ u_e (t_1), u_e (t_2), \cdots, u_e (t_M) \right] ^\mathrm{T}, \quad \mathbf{A}_e = \left[ a_{e,1}, a_{e,2}, \cdots, a_{e,N} \right]^\mathrm{T}, \\
& \mathbf{Q} = \left[ \begin{matrix}  \phi_0(t_1) & \phi_2(t_1) & \cdots & \phi_N (t_1) \\
\phi_0(t_2) & \phi_2(t_2) & \cdots & \phi_N (t_2) \\
\vdots        & \vdots        & \ddots & \vdots \\
\phi_0(t_M) & \phi_2(t_M) & \cdots & \phi_N (t_M) \end{matrix} \right]_{(N+1) \times M}, \quad
\mathbf{LQ} = \left[ \begin{matrix}  \dot{\phi}_0(t_1) & \dot{\phi}_2(t_1) & \cdots & \dot{\phi}_N (t_1) \\
\dot{\phi}_0(t_2) & \dot{\phi}_2(t_2) & \cdots & \dot{\phi}_N (t_2) \\
\vdots        & \vdots        & \ddots & \vdots \\
\dot{\phi}_0(t_M) & \dot{\phi}_2(t_M) & \cdots & \dot{\phi}_N (t_M) \end{matrix} \right]_{(N+1) \times M}.
\end{split} \notag
\end{align}

Normally, we set $M = N+1$. Thus, we achieve the relation between $\mathbf{U}_e$ and its derivative:
\begin{align}
\begin{split}
\dot{\mathbf{U}}_e = (\mathbf{LQ}) \mathbf{Q}^\mathrm{-1} \mathbf{U}_e.
\end{split}
\end{align}

Finally, substituting the relation into Eqn.~\ref{eqn:LVIM-1-Col} and rearranging the sequence of the collocation equations:
\begin{align}
\begin{split}
\left( \widetilde{\mathbf{E}} + \widetilde{\mathbf{J}} \right) \widetilde{\mathbf{U}}_{n+1} = \left( \widetilde{\mathbf{E}} + \widetilde{\mathbf{J}} \right) \widetilde{\mathbf{U}}_{n} - \widetilde{\mathbf{R}},
\label{eqn:CLIC-1-1}
\end{split}
\end{align}
where
\begin{align}
\begin{split}
& \widetilde{\mathbf{U}} = \left[ \mathbf{U}_1^\mathrm{T},  \mathbf{U}_2^\mathrm{T}, \cdots,  \mathbf{U}_L^\mathrm{T} \right] ^\mathrm{T},\qquad  \widetilde{\mathbf{E}} = \mathbf{I}_{L \times L} \otimes \left[ (\mathbf{LQ}) \mathbf{Q}^\mathrm{-1} \right], \\
& \widetilde{\mathbf{J}} = \mathbf{J} \left[ \text{diag} (\hat{\mathbf{t}}) \right] = \left( -\mathbf{C}^\mathrm{T} \mathbf{K} \right) \otimes \mathbf{I}_{M \times M}, \qquad \hat{\mathbf{t}} = \left[ t_1, t_2, \cdots, t_M \right]^\mathrm{T}, \\
& \widetilde{\mathbf{C}} = \mathbf{C} \otimes I_{M \times M}, \qquad \widetilde{\mathbf{K}} = \mathbf{K} \otimes I_{M \times M}, \qquad \widetilde{\mathbf{q}} = \mathbf{q} (\hat{\mathbf{t}}),  \\
& \widetilde{\mathbf{R}} = \widetilde{\mathbf{E}} \widetilde{\mathbf{U}} _n + \widetilde{\mathbf{C}}^\mathrm{-1} \widetilde{\mathbf{K}} \widetilde{\mathbf{U}}_n - \widetilde{\mathbf{C}}^\mathrm{-1} \widetilde{\mathbf{q}},
\end{split} \notag
\end{align}
here $\otimes$ denotes the Kronecker product.

The LVIM can usually achieve good estimates with very simple initial guess functions, e.g., linear functions. In the current approach, we simply assume $ \widetilde{\mathbf{U}}_0 = \mathbf{u} (t_i) \otimes \left[ 1, 1, \cdots, 1 \right]^\mathrm{T}$, that is, a constant function as the initial condition at all time steps. To apply the initial conditions, we usually select the first collocation point at the initial boundary, i.e., $t_1 = t_i$. However, this would make Eqn.~\ref{eqn:CLIC-1-1} overdetermined. To solve that problem, the collocation equations at the initial boundary need to be eliminated. Thus, the final iteration formula in LVIM can be written as:
\begin{align}
\begin{split}
\widetilde{\mathbf{U}}_{n+1}^\mathrm{r} = \widetilde{\mathbf{U}}_{n}^\mathrm{r} - \left( \widetilde{\mathbf{E}}^\mathrm{r}  +\widetilde{\mathbf{J}}^\mathrm{r} \right)^\mathrm{-1} \widetilde{\mathbf{R}}^\mathrm{r},
\label{eqn:CLIC-1-2}
\end{split}
\end{align}
where $[]^\mathrm{r}$ stands for the remained vector (or matrix) after eliminating the ($pM+1$)th rows (and columns), $p = 0, 1, \cdots, L-1$. 

There are some other modifications of LVIM in which the matrix of Lagrange multipliers $\boldsymbol{\lambda} (\tau)$ is approximated in different ways \cite{Wang2019a}. Some of these modifications have potentials in further improving the computing efficiency by avoiding the inversion of the Jacobian matrix, especially for systems dominated by a few eigenvalues. Yet in this paper, we just concentrate on the basic LVIM scheme shown in Eqn.~\ref{eqn:CLIC-1-2}.

\section{Numerical results and discussion}  \label{sec:NR}

In this section, a number of 2D and 3D numerical examples are carried out to illustrate the implementation and effectiveness of our approach. Both steady-state and transient heat conduction problems are presented. The FPM is employed for spatial discretization in all the examples with either uniform or random points. The LVIM is applied in transient examples and the results are compared with explicit and implicit Euler schemes. Anisotropic nonhomogeneous materials are considered. Some complex and practical examples are solved after that, followed by a discussion on the penalty parameters in the FPM and the number of collocation nodes in LVIM. The relative errors $r_0$ and $r_1$ used in this section are defined as:
\begin{align}
\begin{split}
r_0 = \frac{\left\| u^h - u \right\| _{L^2}}{\left\| u \right\| _{L^2}},  \qquad r_1 = \frac{\left\| \nabla u^h - \nabla u \right\| _{L^2}}{\left\| \nabla u \right\| _{L^2}}
\end{split}
\end{align}
where
\begin{align}
\begin{split}
\left\| u \right\| _{L^2} = \left( \int_{\Omega} u^2 \mathrm{d} \Omega \right)^{1/2}, \qquad \left\| \nabla u \right\| _{L^2} = \left( \int_{\Omega} \left| \nabla u \right|^2 \mathrm{d} \Omega \right)^{1/2}.
\end{split}
\end{align}

\subsection{2D examples}

\subsubsection{Isotropic homogeneous benchmark examples}

\begin{figure}[htbp] 
  \centering 
    \subfigure[]{ 
    \label{fig:EX1-U601} 
    \includegraphics[width=0.48\textwidth]{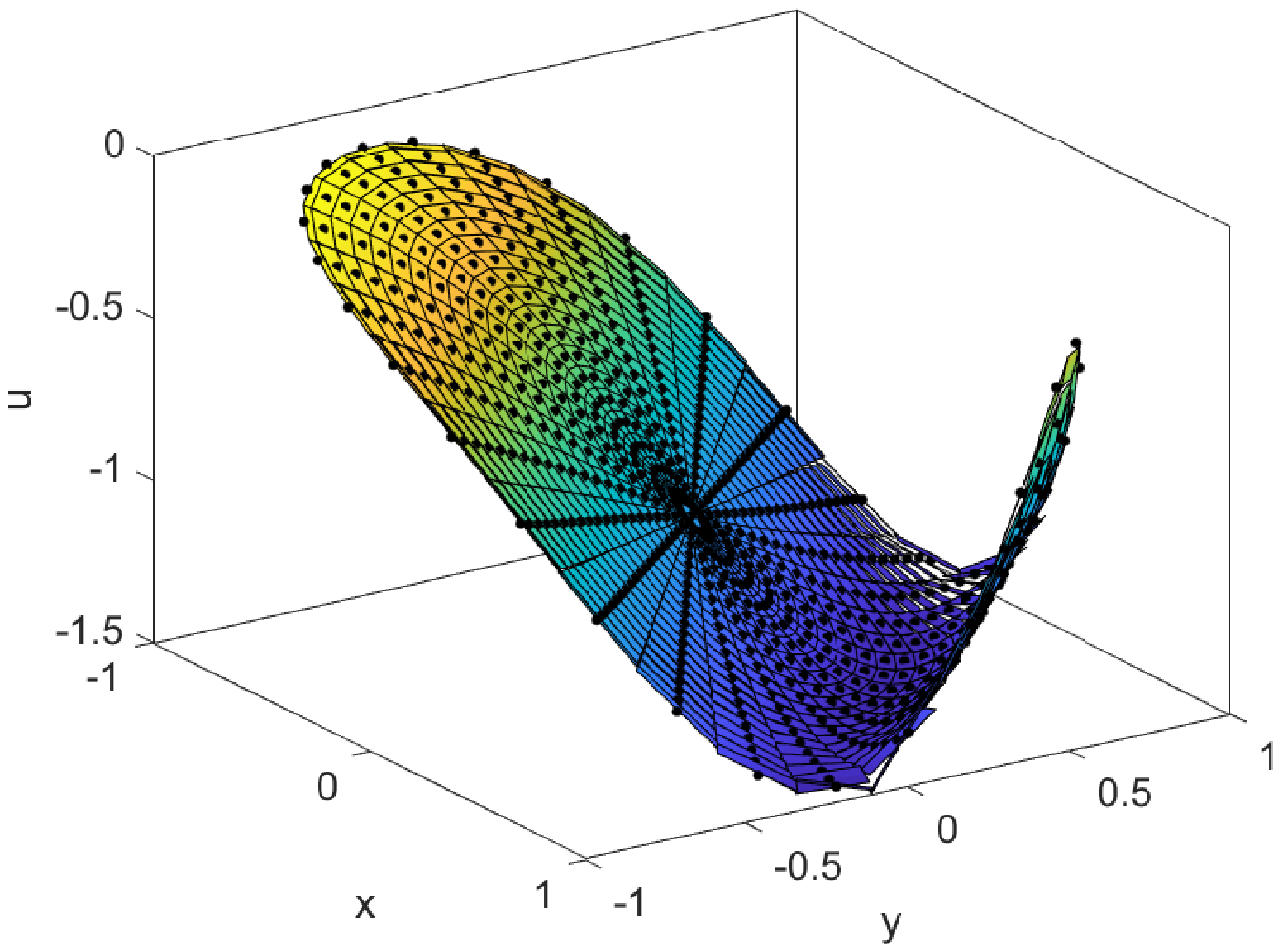}}  
    \subfigure[]{ 
    \label{fig:EX1-R601} 
    \includegraphics[width=0.48\textwidth]{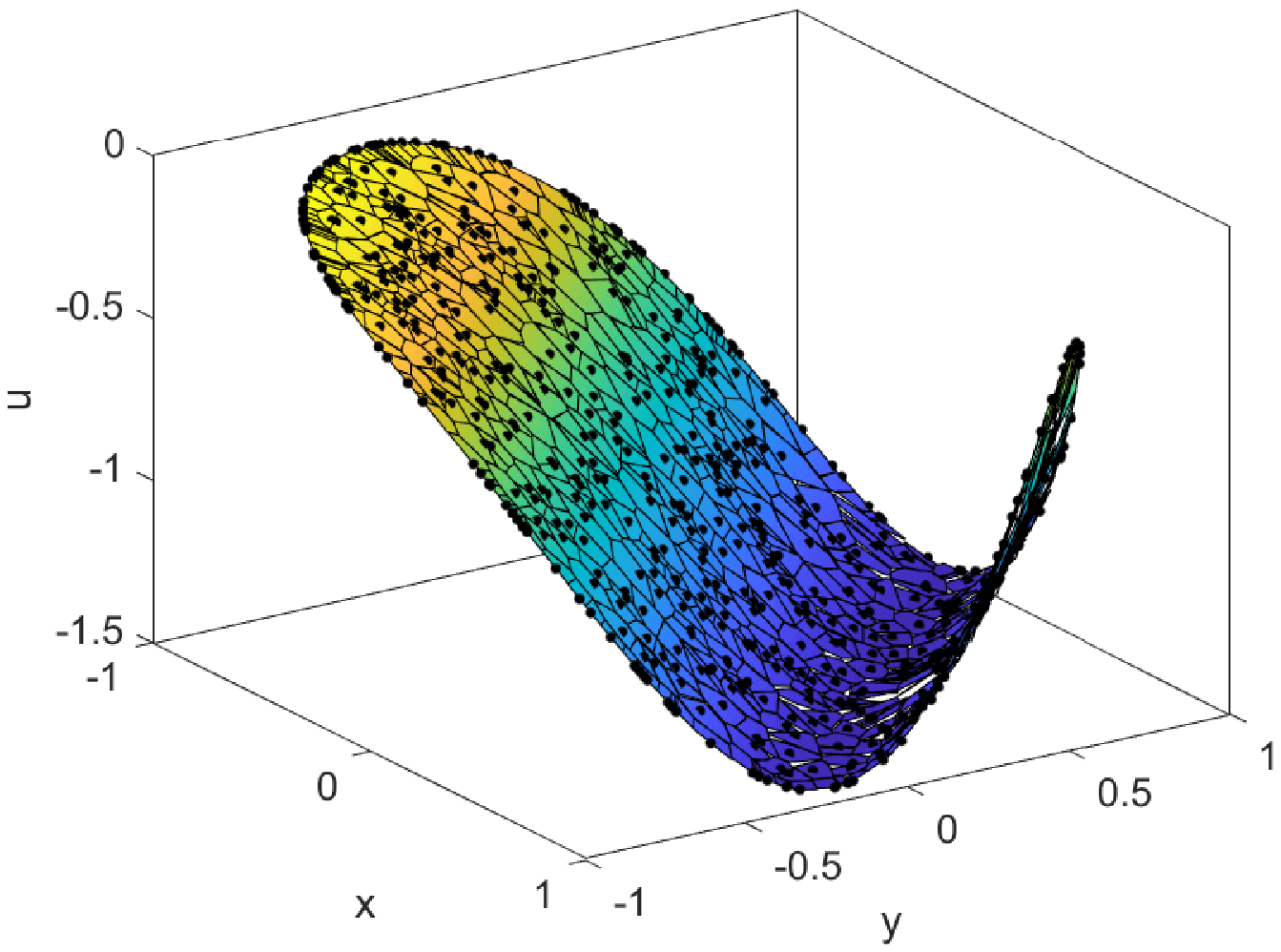}}  
  \caption{Ex. (1.1) - The computed solution when $t = 0.8$. (a) 601 uniform points. (b) 601 random points.} 
  \label{fig:EX1} 
\end{figure}

In the first example (Ex.~(1.1)), a circular isotropic and homogenous domain is considered. Without loss of generality, we assume the material properties $\rho = 1$, $c=1$, thermal conductivity tensor components $k_{11} = k_{22}=1$, $k_{12} = k_{21}=0$. The body source density $Q$ is absent. The simplified governing equation can be written as:
\begin{align}
\begin{split}
\dot{u} (x,y,t) =\nabla u (x,y,t). \\
\end{split}
\end{align}
We consider a postulated analytical solution:
\begin{align}
\begin{split}
u(x,y,t) = e^{x+y} \mathrm{cos} (x+y+4t), \quad (x,y) \in \left\{ (x,y) \mid x^2 + y^2 \leq 1 \right\},
\end{split}
\end{align}
Dirichlet boundary conditions are prescribed on the circumference, corresponding to the given postulated solution. A total of 601 points are distributed uniformly or randomly in the domain, 30 of which are on the boundary ($x^2+y^2=1$). The Dirichlet boundary condition is applied directly. Hence the penalty parameter $\eta_2$ is eliminated. The solutions based on the FPM + LVIM / Backward Euler scheme and their relative errors at $t = 0.8$ are presented in Fig.~\ref{fig:EX1} and Table~\ref{table:Ex1}, in which $\eta_1$ donates the first penalty parameter in the FPM, $M$ is the number of collocation points in each time interval, and $tol$ is the error tolerance in stopping criteria in the LVIM. As can be seen, the FPM can be incorporated with different ODE solvers and achieve highly accurate solutions. Whereas the LVIM in the time domain reduces the computational cost significantly.

\begin{table}[htbp]
\caption{Relative errors and computational time of FPM + LVIM / backward Euler approach in solving Ex.~(1.1).}
\centering
{
\begin{tabular}{ c c c c c }
\toprule[2pt]
Method & \tabincell{c}{Computational \\ parameters} & Time step & Relative errors & \tabincell{c}{Computational \\ time (s)} \\
\toprule[2pt]
\tabincell{c}{FPM + LVIM\\ (601 uniform points)} & \tabincell{c}{$\eta_1 = 2$, \\$M=5$, $tol = 10^{-8}$}  & $\Delta t = 0.4$ & \tabincell{c}{$r_0 = 6.9 \times 10^{-3}$\\$r_1 = 1.7 \times 10^{-1}$} & 2.5 \\
\hline
\tabincell{c}{FPM + LVIM\\ (601 random points)} & \tabincell{c}{$\eta_1 = 2$, \\$M=5$, $tol = 10^{-8}$}  & $\Delta t = 0.4$ & \tabincell{c}{$r_0 = 5.9 \times 10^{-3}$\\$r_1 = 1.5 \times 10^{-1}$} & 1.4 \\
\hline
\tabincell{c}{FPM + backward Euler \\ (601 uniform points)} & $\eta_1 = 2$  & $\Delta t = 0.0016$ & \tabincell{c}{$r_0 =7.1 \times 10^{-3}$\\$r_1 = 1.7 \times 10^{-1}$} & 11 \\
\hline
\tabincell{c}{FPM + backward Euler \\ (601 random points)} & $\eta_1 = 2$  & $\Delta t = 0.0016$ & \tabincell{c}{$r_0 =5.4 \times 10^{-3}$\\$r_1 = 1.5 \times 10^{-1}$} & 6.2 \\
\toprule[2pt]
\end{tabular}}
\label{table:Ex1}
\end{table}

\begin{figure}[htbp] 
  \centering 
    \subfigure[]{ 
    \label{fig:EX2-U144} 
    \includegraphics[width=0.48\textwidth]{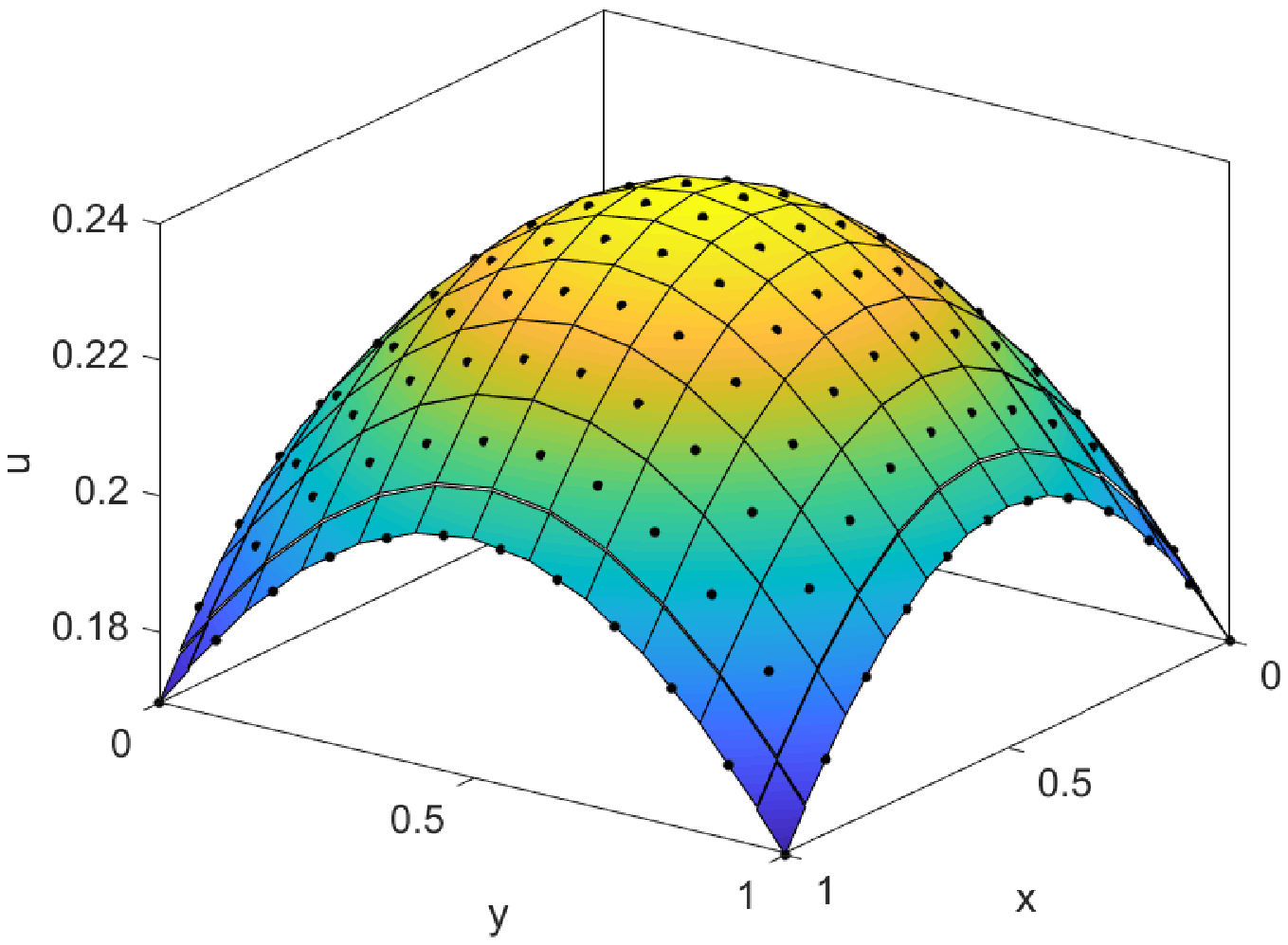}}  
    \subfigure[]{ 
    \label{fig:EX2-R144} 
    \includegraphics[width=0.48\textwidth]{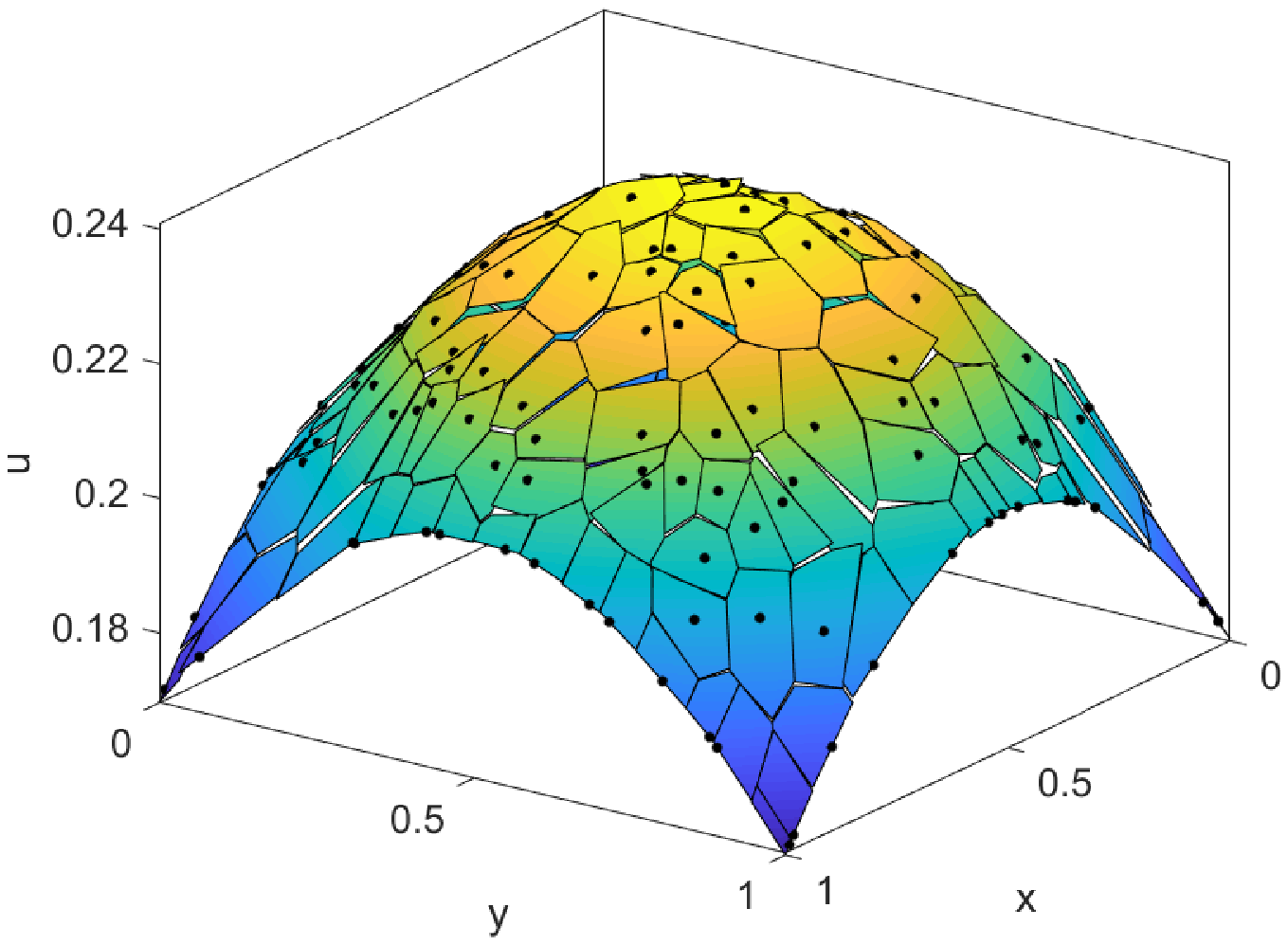}}  
  \caption{Ex. (1.2) - The computed solution when $t = 1$. (a) 144 uniform points. (b) 144 random points.} 
  \label{fig:EX2} 
\end{figure}

The second numerical example (Ex.~(1.2)) is in a square domain. The material properties are the same as Ex.~(1.1). The following postulated analytical solution is considered:
\begin{align}
\begin{split}
u(x,y,t) = \sqrt{2} e^{-\pi^2 t /4} \left[ \mathrm{cos} (\frac{\pi x}{2} - \frac{\pi}{4}) + \mathrm{cos} (\frac{\pi y}{2} - \frac{\pi}{4}) \right], \quad (x,y) \in \left\{ (x,y) \mid x \in \left[ 0, 1 \right], \; y \in \left[ 0, 1 \right] \right\}.
\end{split}
\end{align}
Neumann boundary condition consistent with the postulated solution is applied on $x = 1$, while the other sides are under Dirichlet boundary conditions. 144 uniform or random points are utilized, of which 44 points are on the boundaries. The computed solutions are shown in Table~\ref{table:Ex2} and Fig.~\ref{fig:EX2}. Our current FPM + LVIM approach presents significantly high accuracy for the mixed boundary value problem. The computational speed is ten times higher than the forward and backward Euler schemes. While the forward Euler scheme may become unstable and result in divergent results with a large time step, the LVIM shows its reliability under relatively large time intervals.

\begin{table}[htbp]
\caption{Relative errors and computational time of FPM + LVIM / forward Euler / backward Euler approach in solving Ex.~(1.2).}
\centering
{
\begin{tabular}{ c c c c c }
\toprule[2pt]
Method & \tabincell{c}{Computational \\ parameters} & Time step & Relative errors & \tabincell{c}{Computational \\ time (s)} \\
\toprule[2pt]
\tabincell{c}{FPM + LVIM\\ (144 uniform points)} & \tabincell{c}{$\eta_1 = 2$, \\ $M=5$, $tol = 10^{-8}$}  & $\Delta t = 0.5$ & \tabincell{c}{$r_0 = 5.8 \times 10^{-3}$\\$r_1 = 1.5 \times 10^{-1}$} & 0.09 \\
\hline
\tabincell{c}{FPM + LVIM\\ (144 random points)} & \tabincell{c}{$\eta_1 = 2$, \\ $M=5$, $tol = 10^{-8}$}  & $\Delta t = 0.5$ & \tabincell{c}{$r_0 = 1.1 \times 10^{-2}$\\$r_1 = 2.5 \times 10^{-1}$} & 0.1 \\
\hline
\tabincell{c}{FPM + backward Euler \\ (144 uniform points)} & $\eta_1 = 2$  & $\Delta t = 1 \times 10^{-3}$ & \tabincell{c}{$r_0 =6.1 \times 10^{-3}$\\$r_1 = 1.6 \times 10^{-1}$} & 1.9 \\
\hline
\tabincell{c}{FPM + backward Euler \\ (144 random points)} & $\eta_1 = 2$  & $\Delta t = 1 \times 10^{-3}$ & \tabincell{c}{$r_0 =1.1 \times 10^{-2}$\\$r_1 = 2.5 \times 10^{-1}$} & 2.3 \\
\hline
\tabincell{c}{FPM + forward Euler \\ (144 uniform points)} & $\eta_1 = 2$  & $\Delta t = 5 \times 10^{-4} $ & \tabincell{c}{$r_0 =6.0 \times 10^{-3}$\\$r_1 = 1.5 \times 10^{-1}$} & 3.3 \\
\hline
\tabincell{c}{FPM + forward Euler \\ (144 random points)} & $\eta_1 = 2$  & $\Delta t = 1 \times 10^{-4}$ & \tabincell{c}{$r_0 =1.1 \times 10^{-2}$\\$r_1 = 2.5 \times 10^{-1}$} & 19 \\
\toprule[2pt]
\end{tabular}}
\label{table:Ex2}
\end{table}

\subsubsection{Anisotropic nonhomogeneous examples in a square domain}

In the following four examples, a benchmark mixed boundary value problem in anisotropic nonhomogeneous materials is considered. The tested domain is a $L \times L$ square with Dirichlet boundary conditions on $y = 0$ and $y = L$. Symmetric boundary conditions are applied on the lateral sides. For isotropic problems, symmetry is equivalent to Neumann boundary condition with $\widetilde{q}_N = 0$. Whereas for anisotropic problems, an additional boundary thermal conductivity matrix has to be employed:
\begin{align}
\begin{split}
\mathbf{K}_S = - \int_e \left[ \mathbf{N}^\mathrm{T} \mathbf{n}^\mathrm{T} \left( \mathbf{k} - {k}_{11} \mathbf{I} \right) \mathbf{B} \right] \mathrm{d} \Gamma, \qquad e \in \Gamma_S,
\end{split}
\end{align}
where ${k}_{11}$ is the first diagonal element of the thermal conductivity tensor $\mathbf{k}$. $\Gamma_S$ stands for the symmetric boundaries. Clearly, the matrix vanishes in isotropic domain. The initial, boundary conditions and material properties are given as:
\begin{align}
\begin{split}
& u(x, 0, t) = u_0, \qquad  u(x, L, t) =  u_L, \qquad u(x, y ,0) =  u_0,\\
&\rho (x, y) =  1, \qquad c(x, y) = f(y), \qquad \mathbf{k} (x, y) = f(y) \left[ \begin{matrix} \hat{k}_{11} & \hat{k}_{12} \\ \hat{k}_{21} & \hat{k}_{22} \end{matrix} \right],
\end{split}
\end{align}
where $u_0$, $u_L$, $\hat{k}_{ij} (i,j = 1, 2)$ are constant. In isotropic case, $\hat{k}_{ij} = \delta_{ij}$. Whereas in anisotropic case, $\hat{k}_{11} = \hat{k}_{22} = 2, \hat{k}_{12} = \hat{k}_{21} = 1$. The body source density $Q = \mathrm{const} = 0$. It turns out that the resulting temperature distribution is not dependent on $x$, i.e., the example can be equivalent to a 1D heat conduction problem.

\begin{figure}[htbp] 
  \centering 
    \subfigure[]{ 
    \label{fig:EX3-Trans} 
    \includegraphics[width=0.48\textwidth]{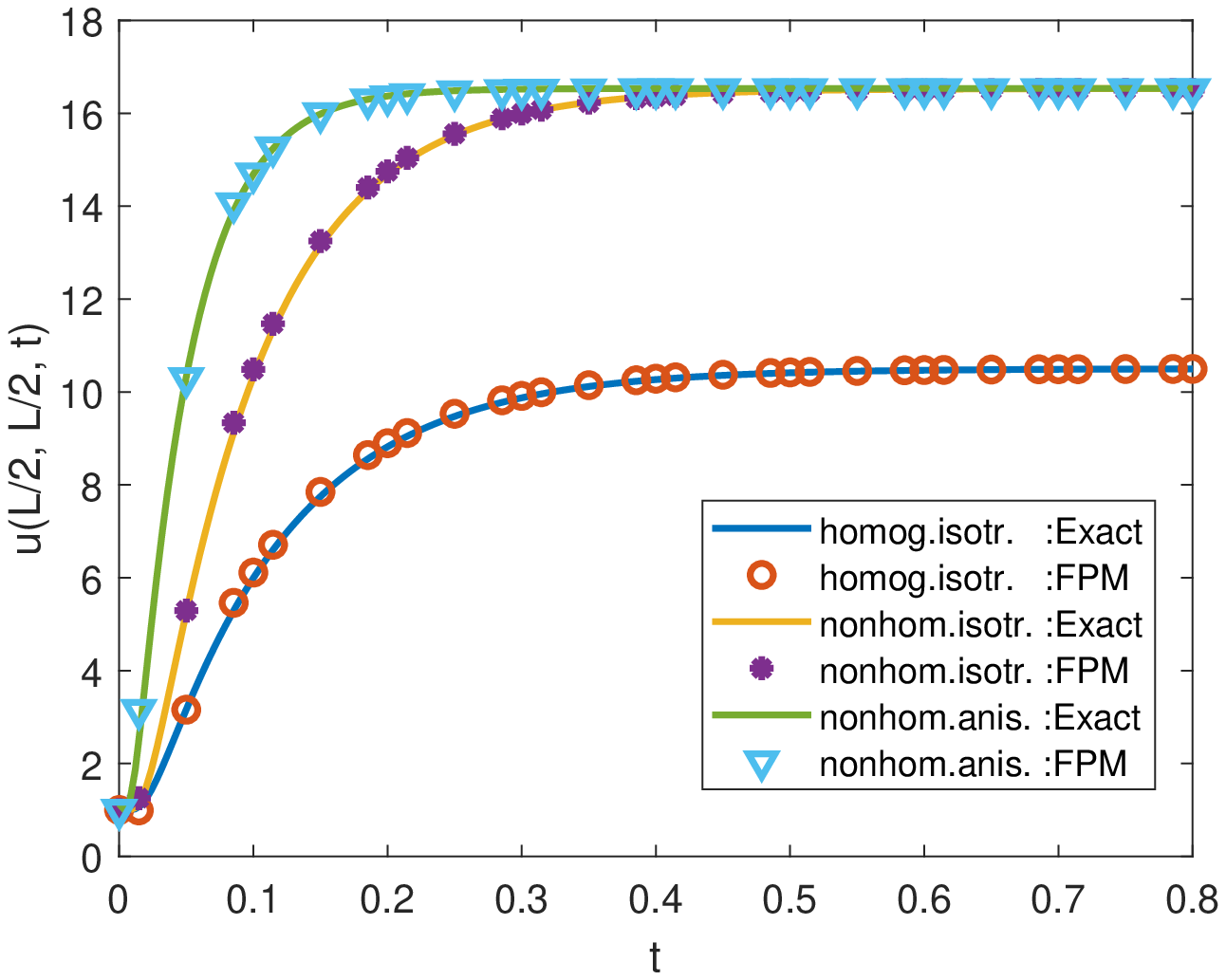}}  
    \subfigure[]{ 
    \label{fig:EX3-Conf} 
    \includegraphics[width=0.48\textwidth]{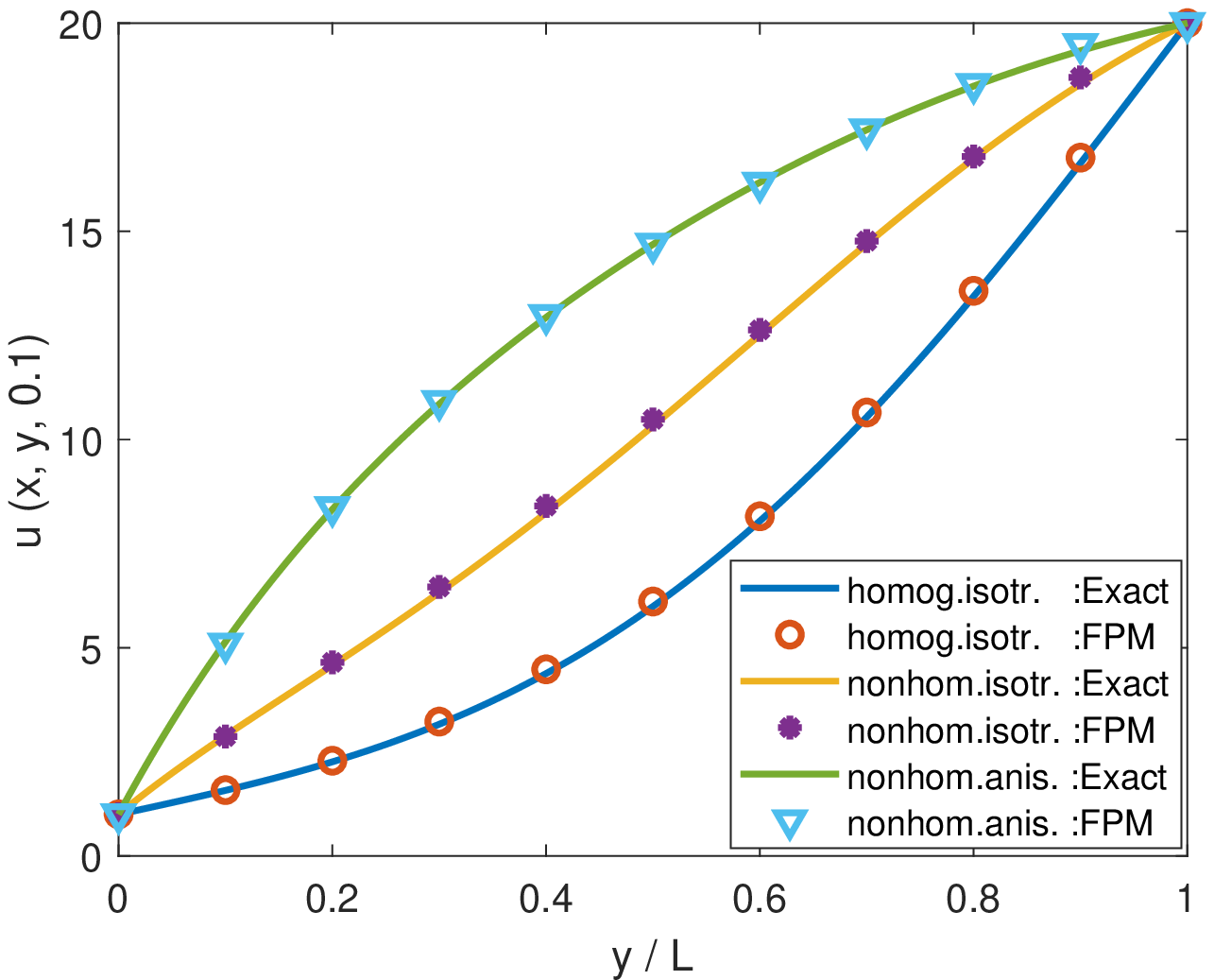}}  
  \caption{Ex. (1.3) - The computed solution with different material properties. (a) transient temperature solution at the midpoint of the domain in time scope $[0, 0.8]$. (b) vertical temperature distribution when $t=0.1$.} 
  \label{fig:EX3} 
\end{figure}

In Ex.~(1.3), $u_0 = 1$, $u_L = 20$, the gradation function $f(y) = \mathrm{exp}(\delta y / L)$. The exact solution is obtained and given in \cite{Sladek2005}. When $\delta = 0$, the material is homogenous. The computed solution for isotropic homogenous, isotropic nonhomogeneous, and anisotropic nonhomogeneous materials are presented and compared with exact solutions in Fig.~\ref{fig:EX3}. With only 121 ($11 \times 11$) uniform points in the domain, the result shows great agreement with the exact solution. It is also consistent with the results shown in \cite{Sladek2005} based on meshless point interpolation method (PIM) and Laplace-transform (LT) approach. The time cost and average errors of the present FPM + LVIM approach is listed in Table~\ref{table:EX3}, as well as the backward Euler scheme. The average error $\overline{r}_0$ is defined as the average value of $r_0$ in time interval $[0, 0.8]$. It should be pointed out that in order to get a continuous solution in the entire domain, the FPM with random points usually requires a larger penalty parameter $\eta_1$. Unfortunately, the accuracy drops down as $\eta_1$ increases. 

As can be seen from Fig.~\ref{fig:EX3} and Table~\ref{table:EX3}, while the nonhomogeneity and anisotropy of the material have a significant influence on the temperature distribution, they do not give rise to any difficulties in the present computing method. As the solution achieves steady state before $t = 0.8$, the advantage of LVIM approach in computational time is not distinct, especially when comparing with Ex.~(1.1) and (1.2) in which the temperature solution varies violently. Yet the LVIM approach still saves approximately one half of the computing time.

\begin{table}[htbp]
\caption{Relative errors and computational time of FPM + LVIM / backward Euler approach in solving Ex.~(1.3).}
\centering
{
\begin{tabular}{ c c c c c }
\toprule[2pt]
Method & \tabincell{c}{Computational \\ parameters} & Time step & Average errors & \tabincell{c}{Computational \\ time (s)} \\
\toprule[2pt]
\multicolumn{5}{c}{Homogenous isotropic ($\delta = 0$; $\hat{k}_{11} = \hat{k}_{22} = 1, \hat{k}_{12} = \hat{k}_{21} = 0$)} \\
\hline
\tabincell{c}{FPM + LVIM\\ (144 uniform points)} & \tabincell{c}{$\eta_1 = 10$, \\ $M=5$, $tol = 10^{-8}$}  & $\Delta t = 0.1$ & $\overline{r}_0 = 5.2 \times 10^{-3}$ & 0.6 \\
\hline
\tabincell{c}{FPM + LVIM\\ (144 random points)} & \tabincell{c}{$\eta_1 = 20$, \\ $M=5$, $tol = 10^{-8}$}  & $\Delta t = 0.5$ & $\overline{r}_0 = 4.8 \times 10^{-2}$ & 0.6 \\
\hline
\tabincell{c}{FPM + backward Euler \\ (144 uniform points)} & $\eta_1 = 10$  & $\Delta t = 0.005$ & $\overline{r}_0 = 2.7 \times 10^{-3}$ & 1.4 \\
\hline
\tabincell{c}{FPM + backward Euler \\ (144 random points)} & $\eta_1 = 20$  & $\Delta t = 0.005$ & $\overline{r}_0 = 2.8 \times 10^{-2}$ & 1.2 \\
\toprule[2pt]
\multicolumn{5}{c}{Nonhomogenous isotropic ($\delta = 3$; $\hat{k}_{11} = \hat{k}_{22} = 1, \hat{k}_{12} = \hat{k}_{21} = 0$)} \\
\hline
\tabincell{c}{FPM + LVIM\\ (144 uniform points)} & \tabincell{c}{$\eta_1 = 10$, \\ $M=5$, $tol = 10^{-8}$}  & $\Delta t = 0.1$ & $\overline{r}_0 = 7.7 \times 10^{-3}$ & 0.6 \\
\hline
\tabincell{c}{FPM + LVIM\\ (144 random points)} & \tabincell{c}{$\eta_1 = 20$, \\ $M=5$, $tol = 10^{-8}$}  & $\Delta t = 0.5$ & $\overline{r}_0 = 4.2 \times 10^{-2}$ & 0.7 \\
\hline
\tabincell{c}{FPM + backward Euler \\ (144 uniform points)} & $\eta_1 = 10$  & $\Delta t = 0.005$ & $\overline{r}_0 = 7.1 \times 10^{-3}$ & 1.2 \\
\hline
\tabincell{c}{FPM + backward Euler \\ (144 random points)} & $\eta_1 = 20$  & $\Delta t = 0.005$ & $\overline{r}_0 = 2.8 \times 10^{-2}$ & 1.3 \\
\toprule[2pt]
\multicolumn{5}{c}{Nonhomogenous anisotropic ($\delta = 3$; $\hat{k}_{11} = \hat{k}_{22} = 2, \hat{k}_{12} = \hat{k}_{21} = 1$)} \\
\hline
\tabincell{c}{FPM + LVIM\\ (144 uniform points)} & \tabincell{c}{$\eta_1 = 10$, \\ $M=5$, $tol = 10^{-8}$}  & $\Delta t = 0.1$ & $\overline{r}_0 = 7.3 \times 10^{-3}$ & 0.6 \\
\hline
\tabincell{c}{FPM + LVIM\\ (144 random points)} & \tabincell{c}{$\eta_1 = 20$, \\ $M=5$, $tol = 10^{-8}$}  & $\Delta t = 0.5$ & $\overline{r}_0 = 4.9 \times 10^{-2}$ & 0.7 \\
\hline
\tabincell{c}{FPM + backward Euler \\ (144 uniform points)} & $\eta_1 = 10$  & $\Delta t = 0.005$ & $\overline{r}_0 = 7.8 \times 10^{-3}$ & 1.2 \\
\hline
\tabincell{c}{FPM + backward Euler \\ (144 random points)} & $\eta_1 = 20$  & $\Delta t = 0.005$ & $\overline{r}_0 = 4.2 \times 10^{-2}$ & 1.5 \\
\toprule[2pt]
\end{tabular}}
\label{table:EX3}
\end{table}

In Ex.~(1.4) – (1.6), we consider the same initial boundary value problem as shown in Ex.~(1.3). The material gradation function $f(y)$ and boundary values are given as:
\begin{align}
 \text{Ex.~(1.4):} \quad  & \text{exponential}:  & f(y) = \left[ \mathrm{exp} (\delta y / L) + \mathrm{exp} (-\delta y / L) \right] ^2 &, \delta = 2, u_0 = 1, u_L =20; & \notag \\
 \text{Ex.~(1.5):} \quad  & \text{trigonometric}: & f(y) = \left[ \mathrm{cos} (\delta y / L) + 5 \mathrm{sin} (\delta y /L) \right]^2 &, \delta = 2, u_0 = 0, u_L =100; & \notag \\
 \text{Ex.~(1.6):} \quad & \text{power-law}: & f(y) = \left( 1 + \delta y / L \right)^2 &, \delta = 3, u_0 = 1, u_L =20; & \notag
 \label{eqn:BCs}
\end{align}

\begin{figure}[htbp] 
  \centering 
    \subfigure[]{ 
    \label{fig:EX4-Trans} 
    \includegraphics[width=0.48\textwidth]{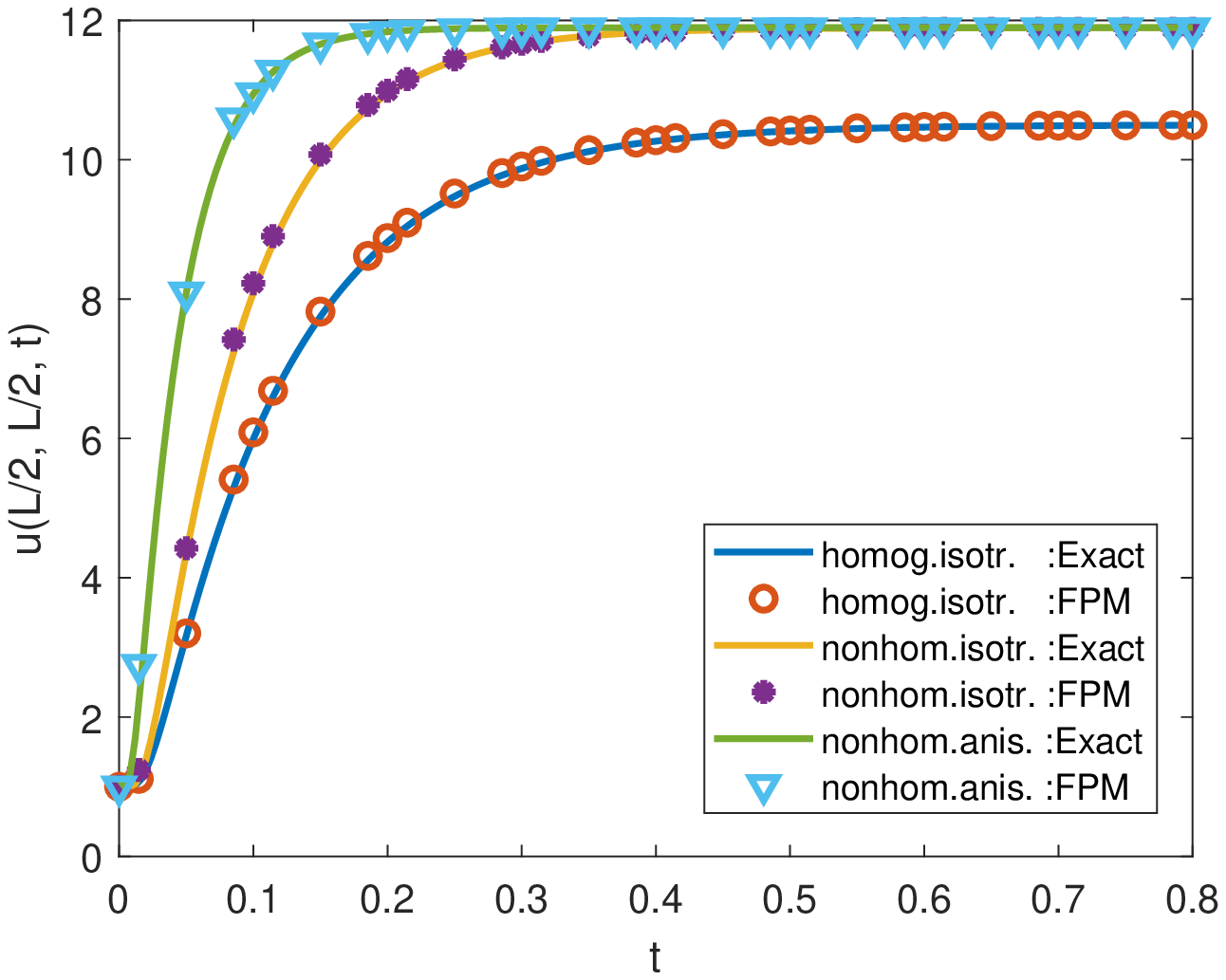}}  
    \subfigure[]{ 
    \label{fig:EX4-Conf} 
    \includegraphics[width=0.48\textwidth]{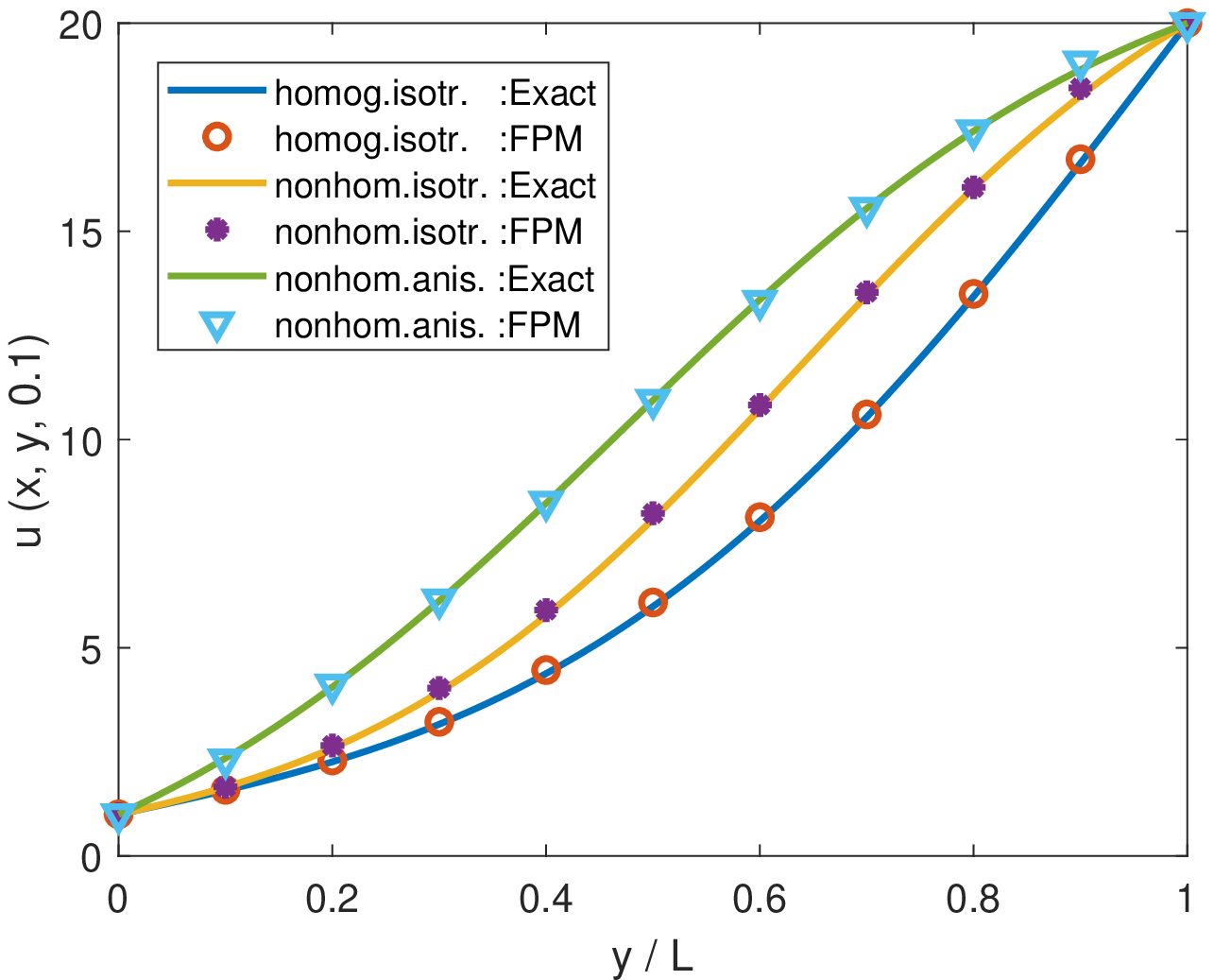}}  
  \caption{Ex. (1.4) - The computed solution with different material properties. (a) transient temperature solution in time scope $[0, 0.8]$. (b) vertical temperature distribution when $t=0.1$.} 
  \label{fig:EX4} 
\end{figure}

\begin{figure}[htbp] 
  \centering 
    \subfigure[]{ 
    \label{fig:EX5-Trans} 
    \includegraphics[width=0.48\textwidth]{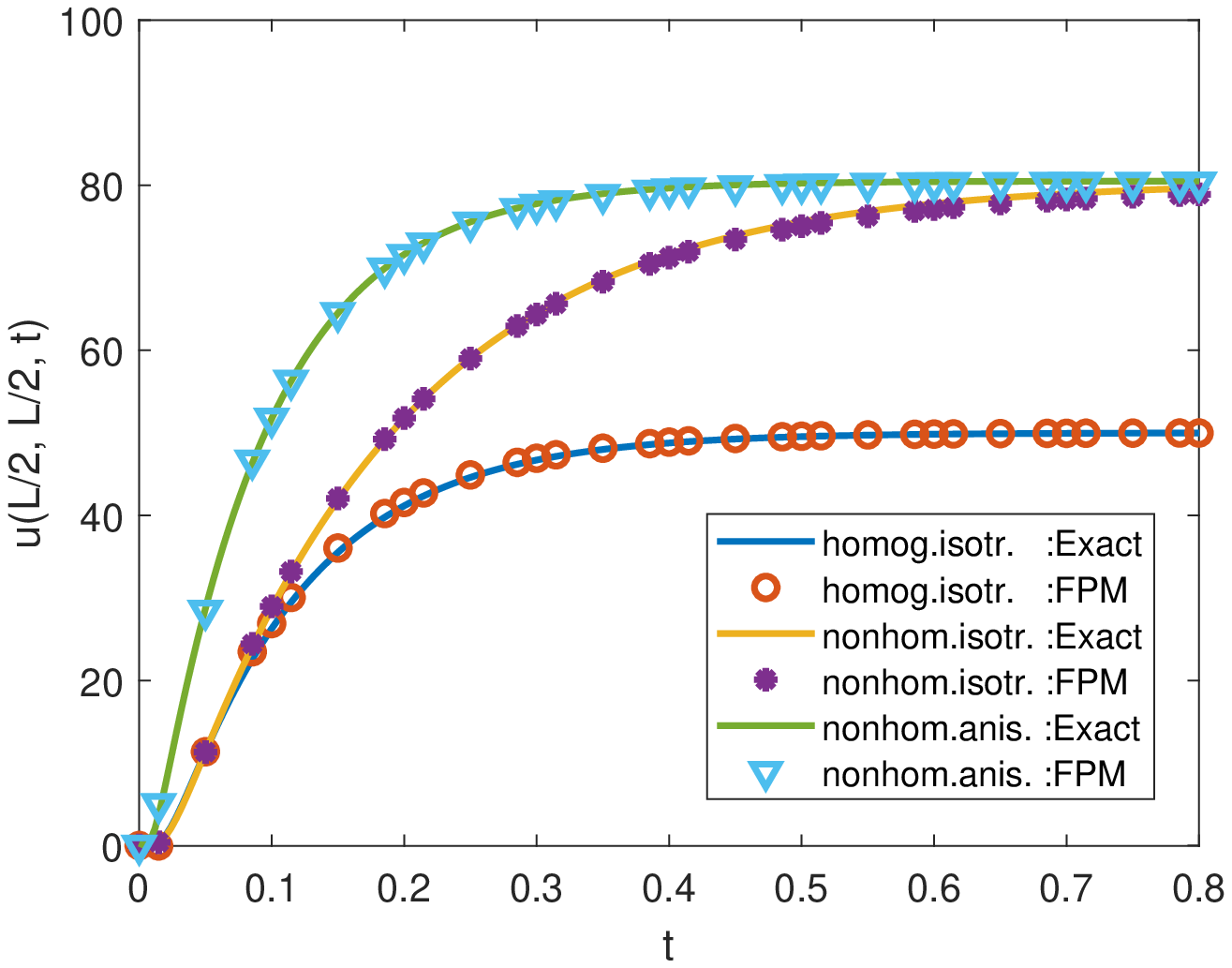}}  
    \subfigure[]{ 
    \label{fig:EX5-Conf} 
    \includegraphics[width=0.48\textwidth]{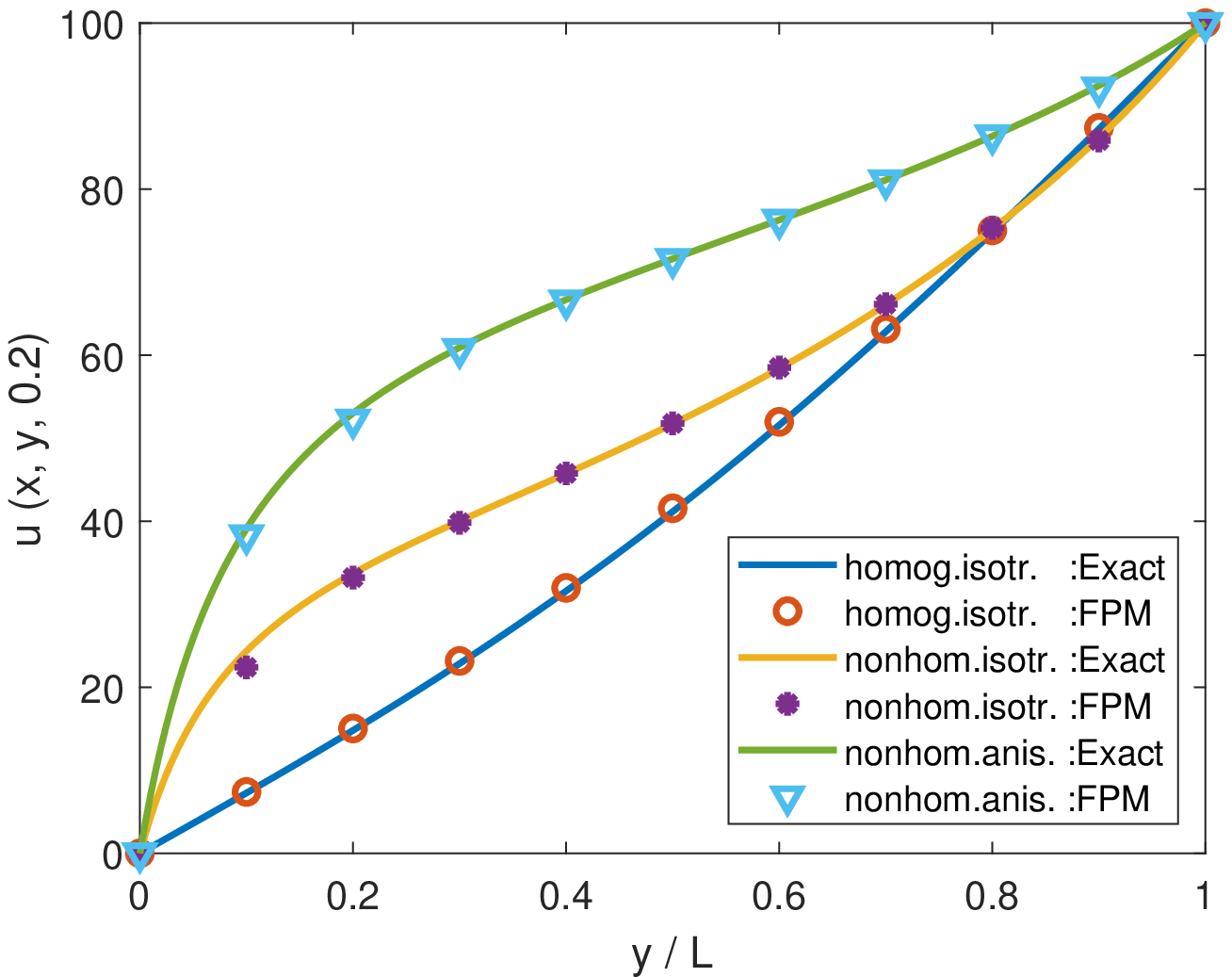}}  
  \caption{Ex. (1.5) - The computed solution with different material properties. (a) transient temperature solution in time scope $[0, 0.8]$. (b) vertical temperature distribution when $t=0.2$.} 
  \label{fig:EX5} 
\end{figure}

\begin{figure}[htbp] 
  \centering 
    \subfigure[]{ 
    \label{fig:EX6-Trans} 
    \includegraphics[width=0.48\textwidth]{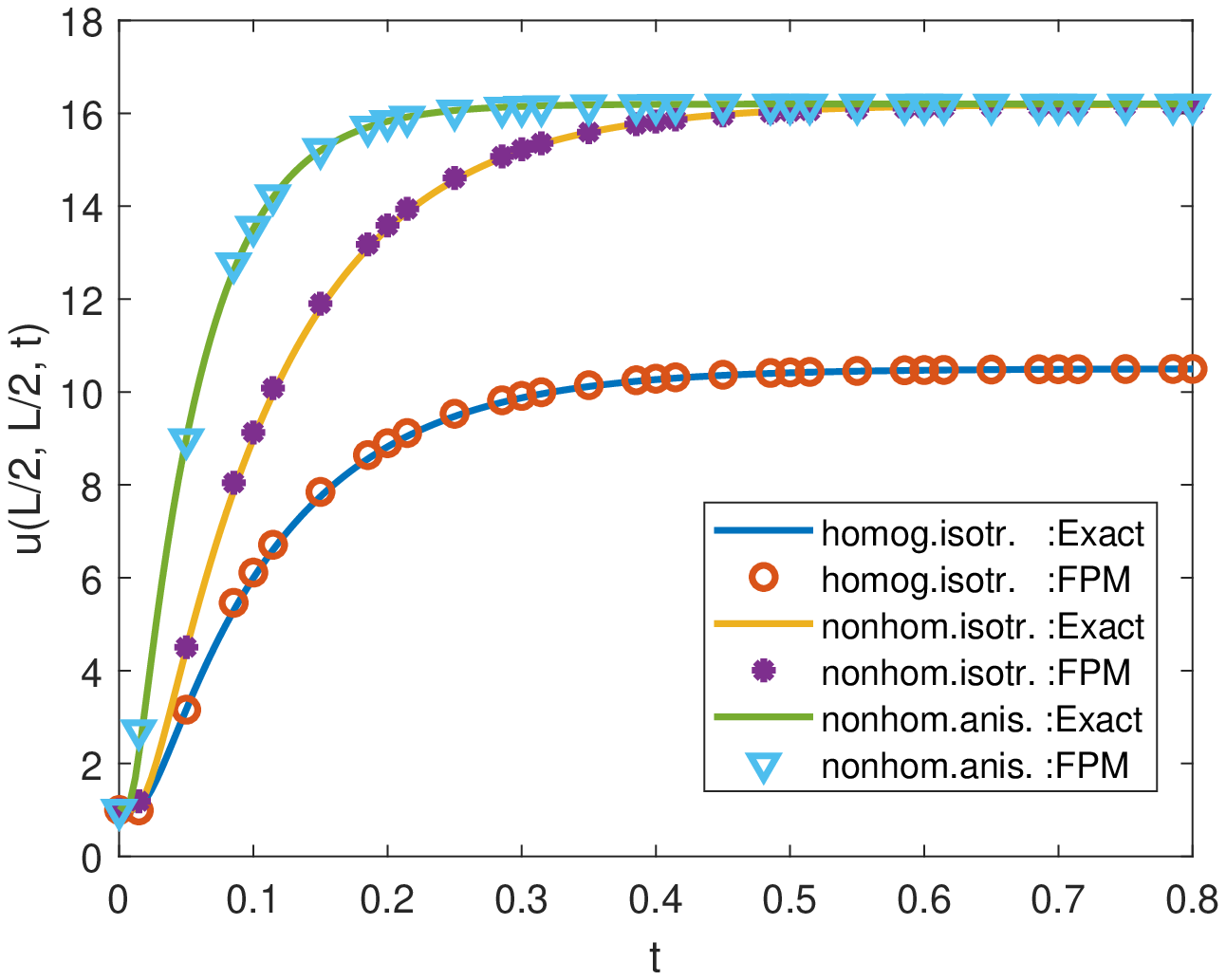}}  
    \subfigure[]{ 
    \label{fig:EX6-Conf} 
    \includegraphics[width=0.48\textwidth]{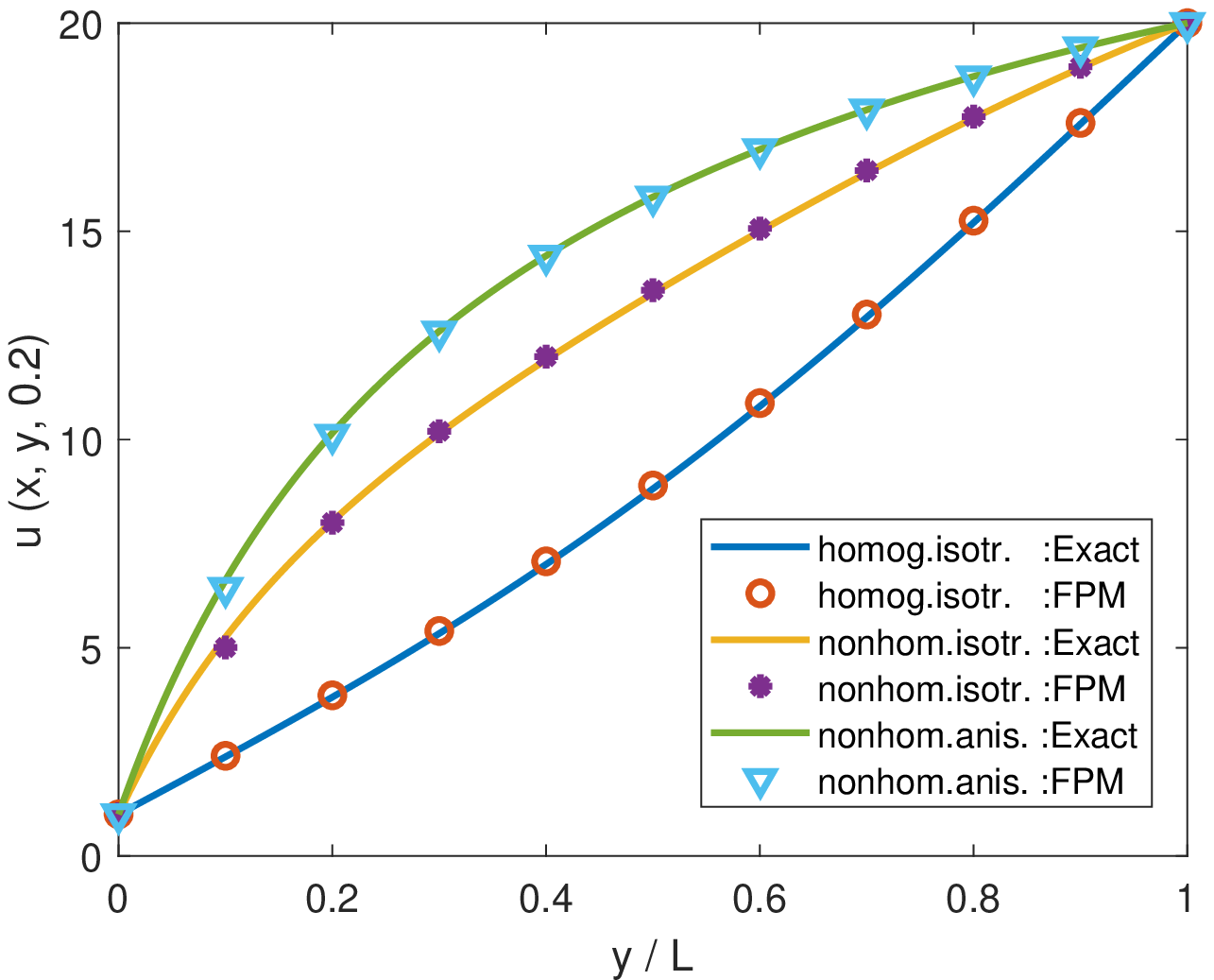}}  
  \caption{Ex. (1.6) - The computed solution with different material properties. (a) transient temperature solution in time scope $[0, 0.8]$. (b) vertical temperature distribution when $t=0.2$.} 
  \label{fig:EX6} 
\end{figure}

The computed solutions of these three examples are shown in Fig.~\ref{fig:EX4}, \ref{fig:EX5} and \ref{fig:EX6} respectively. 121 uniform points are utilized. The results achieve great agreement with the analytical solutions, confirming that the nonhomogeneity and anisotropy do not cause any difficulties in the FPM + LVIM approach. The corresponding relative errors and computational times are shown in Table~\ref{table:EX4}, \ref{table:EX5} and \ref{table:EX6}. The LVIM approach cuts the computing time approximately by a half and does not cause any stability problems. All these results are consistent with the numerical example solutions in \cite{Sladek2005}.

\begin{table}[htbp]
\caption{Relative errors and computational time of FPM + LVIM / backward Euler approach (121 uniform points) in solving Ex.~(1.4).}
\centering
{
\begin{tabular}{ c c c c c }
\toprule[2pt]
Method & \tabincell{c}{Computational \\ parameters} & Time step & Average errors & \tabincell{c}{Computational \\ time (s)} \\
\toprule[2pt]
\multicolumn{5}{c}{Homogenous isotropic ($\delta = 0$; $\hat{k}_{11} = \hat{k}_{22} = 1, \hat{k}_{12} = \hat{k}_{21} = 0$)} \\
\hline
FPM + LVIM & $\eta_1 = 10$, $M=5$, $tol = 10^{-8}$ & $\Delta t = 0.1$ & $\overline{r}_0 = 6.8 \times 10^{-3}$ & 0.5 \\
\hline
FPM + backward Euler & $\eta_1 = 10$  & $\Delta t = 0.005$ & $\overline{r}_0 = 5.8 \times 10^{-3}$ & 1.1 \\
\toprule[2pt]
\multicolumn{5}{c}{Nonhomogenous isotropic ($\delta = 2$; $\hat{k}_{11} = \hat{k}_{22} = 1, \hat{k}_{12} = \hat{k}_{21} = 0$)} \\
\hline
FPM + LVIM & $\eta_1 = 10$, $M=5$, $tol = 10^{-8}$  & $\Delta t = 0.1$ & $\overline{r}_0 = 7.1 \times 10^{-3}$ & 0.5 \\
\hline
FPM + backward Euler & $\eta_1 = 10$  & $\Delta t = 0.005$ & $\overline{r}_0 = 5.7 \times 10^{-3}$ & 1.2 \\
\toprule[2pt]
\multicolumn{5}{c}{Nonhomogenous anisotropic ($\delta = 2$; $\hat{k}_{11} = \hat{k}_{22} = 2, \hat{k}_{12} = \hat{k}_{21} = 1$)} \\
\hline
FPM + LVIM & $\eta_1 = 10$, $M=5$, $tol = 10^{-8}$ & $\Delta t = 0.1$ & $\overline{r}_0 = 8.2 \times 10^{-3}$ & 0.5 \\
\hline
FPM + backward Euler & $\eta_1 = 10$  & $\Delta t = 0.005$ & $\overline{r}_0 = 8.1 \times 10^{-3}$ & 1.3 \\
\toprule[2pt]
\end{tabular}}
\label{table:EX4}
\end{table}

\begin{table}[htbp]
\caption{Relative errors and computational time of FPM + LVIM / backward Euler approach (121 uniform points) in solving Ex.~(1.5).}
\centering
{
\begin{tabular}{ c c c c c }
\toprule[2pt]
Method & \tabincell{c}{Computational \\ parameters} & Time step & Average errors & \tabincell{c}{Computational \\ time (s)} \\
\toprule[2pt]
\multicolumn{5}{c}{Homogenous isotropic ($\delta = 0$; $\hat{k}_{11} = \hat{k}_{22} = 1, \hat{k}_{12} = \hat{k}_{21} = 0$)} \\
\hline
FPM + LVIM & $\eta_1 = 10$, $M=5$, $tol = 10^{-8}$ & $\Delta t = 0.1$ & $\overline{r}_0 = 6.9 \times 10^{-3}$ & 0.4 \\
\hline
FPM + backward Euler & $\eta_1 = 10$  & $\Delta t = 0.005$ & $\overline{r}_0 = 6.1 \times 10^{-3}$ & 1.0 \\
\toprule[2pt]
\multicolumn{5}{c}{Nonhomogenous isotropic ($\delta = 2$; $\hat{k}_{11} = \hat{k}_{22} = 1, \hat{k}_{12} = \hat{k}_{21} = 0$)} \\
\hline
FPM + LVIM & $\eta_1 = 10$, $M=5$, $tol = 10^{-8}$  & $\Delta t = 0.1$ & $\overline{r}_0 = 2.0 \times 10^{-2}$ & 0.4 \\
\hline
FPM + backward Euler & $\eta_1 = 10$  & $\Delta t = 0.005$ & $\overline{r}_0 = 2.1 \times 10^{-2}$ & 1.0 \\
\toprule[2pt]
\multicolumn{5}{c}{Nonhomogenous anisotropic ($\delta = 2$; $\hat{k}_{11} = \hat{k}_{22} = 2, \hat{k}_{12} = \hat{k}_{21} = 1$)} \\
\hline
FPM + LVIM & $\eta_1 = 10$, $M=5$, $tol = 10^{-8}$ & $\Delta t = 0.1$ & $\overline{r}_0 = 9.3 \times 10^{-3}$ & 0.4 \\
\hline
FPM + backward Euler & $\eta_1 = 10$  & $\Delta t = 0.005$ & $\overline{r}_0 = 1.0 \times 10^{-2}$ & 1.0 \\
\toprule[2pt]
\end{tabular}}
\label{table:EX5}
\end{table}

\begin{table}[htbp]
\caption{Relative errors and computational time of FPM + LVIM / backward Euler approach (121 uniform points) in solving Ex.~(1.6).}
\centering
{
\begin{tabular}{ c c c c c }
\toprule[2pt]
Method & \tabincell{c}{Computational \\ parameters} & Time step & Average errors & \tabincell{c}{Computational \\ time (s)} \\
\toprule[2pt]
\multicolumn{5}{c}{Homogenous isotropic ($\delta = 0$; $\hat{k}_{11} = \hat{k}_{22} = 1, \hat{k}_{12} = \hat{k}_{21} = 0$)} \\
\hline
FPM + LVIM & $\eta_1 = 10$, $M=5$, $tol = 10^{-8}$ & $\Delta t = 0.1$ & $\overline{r}_0 = 6.4 \times 10^{-3}$ & 0.4 \\
\hline
FPM + backward Euler & $\eta_1 = 10$  & $\Delta t = 0.005$ & $\overline{r}_0 = 5.6 \times 10^{-3}$ & 1.0 \\
\toprule[2pt]
\multicolumn{5}{c}{Nonhomogenous isotropic ($\delta = 2$; $\hat{k}_{11} = \hat{k}_{22} = 1, \hat{k}_{12} = \hat{k}_{21} = 0$)} \\
\hline
FPM + LVIM & $\eta_1 = 10$, $M=5$, $tol = 10^{-8}$  & $\Delta t = 0.1$ & $\overline{r}_0 = 1.1 \times 10^{-2}$ & 0.4 \\
\hline
FPM + backward Euler & $\eta_1 = 10$  & $\Delta t = 0.005$ & $\overline{r}_0 = 1.1 \times 10^{-2}$ & 0.9 \\
\toprule[2pt]
\multicolumn{5}{c}{Nonhomogenous anisotropic ($\delta = 2$; $\hat{k}_{11} = \hat{k}_{22} = 2, \hat{k}_{12} = \hat{k}_{21} = 1$)} \\
\hline
FPM + LVIM & $\eta_1 = 10$, $M=5$, $tol = 10^{-8}$ & $\Delta t = 0.1$ & $\overline{r}_0 = 7.9 \times 10^{-3}$ & 0.4 \\
\hline
FPM + backward Euler & $\eta_1 = 10$  & $\Delta t = 0.005$ & $\overline{r}_0 = 7.8 \times 10^{-3}$ & 0.9 \\
\toprule[2pt]
\end{tabular}}
\label{table:EX6}
\end{table}

\begin{figure}[htbp] 
  \centering 
    \subfigure[]{ 
    \label{fig:EX6-R144} 
    \includegraphics[width=0.48\textwidth]{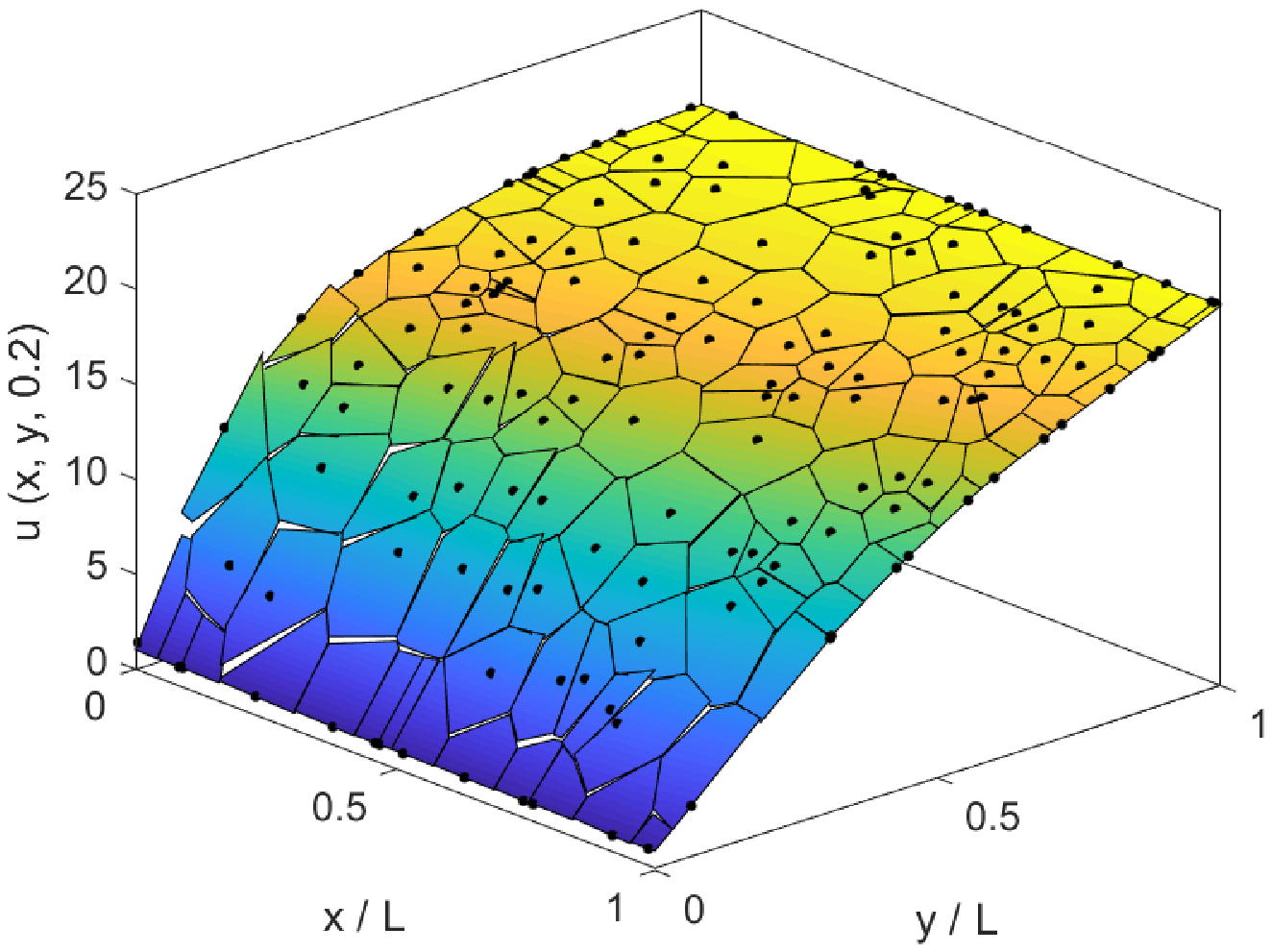}}  
    \subfigure[]{ 
    \label{fig:EX6-R144-NBC} 
    \includegraphics[width=0.48\textwidth]{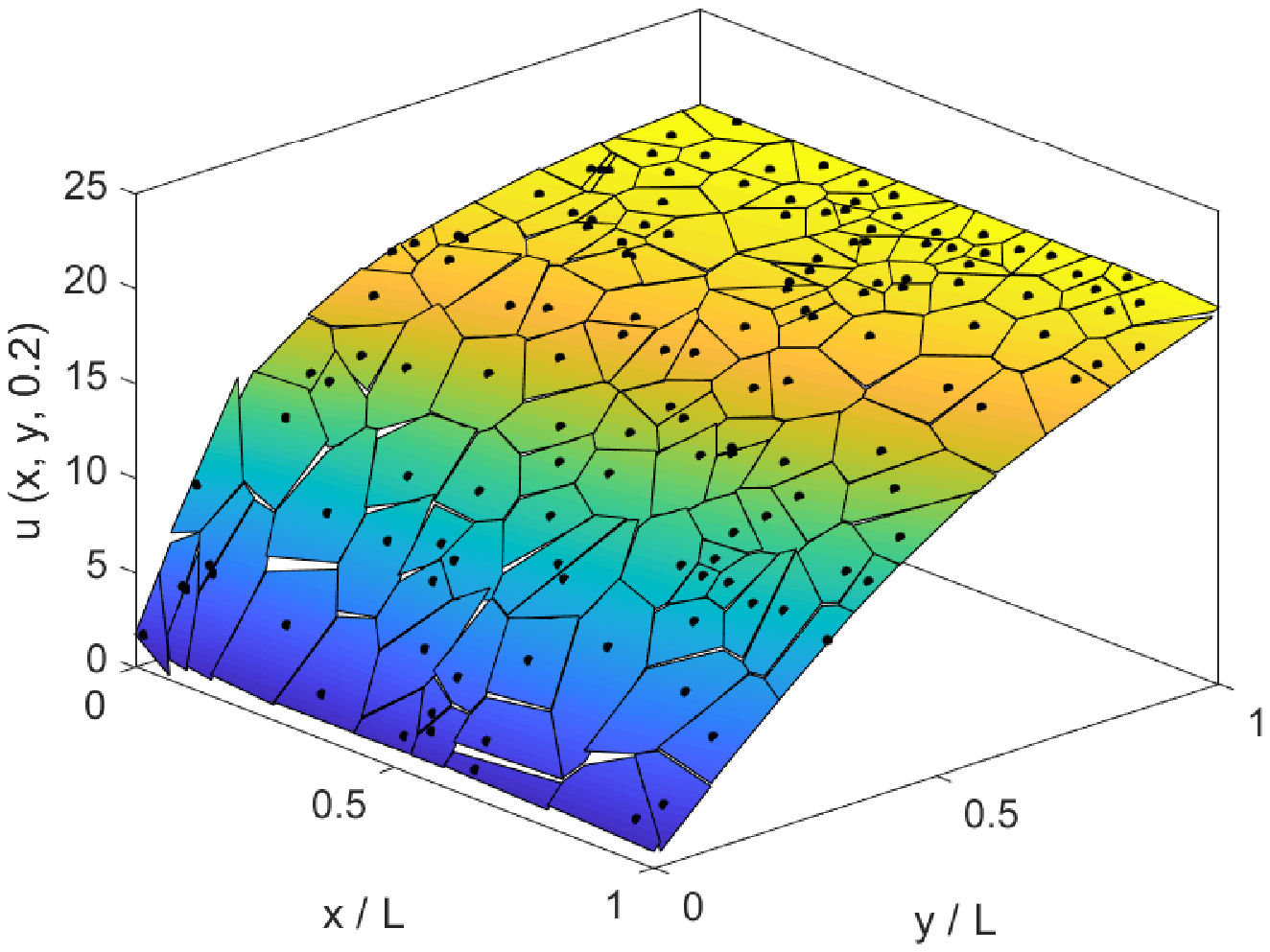}}  
  \caption{Ex. (1.6) - The computed solution with vanishing heat fluxes on the lateral sides. (a) 44 points on the boundaries, $\eta_1 = 10$. (b) no points on the boundaries, $\eta_1 = 10$, $\eta_2 = 20$.} 
  \label{fig:EX6-02} 
\end{figure}

In Ex.~(1.6), we can also replace the symmetric boundary conditions by Neumann boundary conditions with heat flux vanishing on the sides. In the anisotropic case, as a result, the temperature variation in $x$ - direction is no longer constant. The computed 2D temperature distribution based on 144 random points is shown in Fig.~\ref{fig:EX6-02}. In Fig.~\ref{fig:EX6-R144}, 44 of the points are distributed on the boundaries, hence the Dirichlet boundary condition is imposed directly. Whereas in Fig.~\ref{fig:EX6-R144-NBC}, no points are on the boundary. A collocation method based on integral terms on the boundaries is employed to enforce the essential boundary conditions. That is, in Eqn.~\ref{eqn:FPM}, penalty parameter $\eta_2$ is utilized. The result presents a good consistency between different domain partitions, as well as the direct and collocation methods in imposing the essential boundary conditions.

\subsubsection{Some practical examples}

\begin{figure}[htbp] 
  \centering 
    \subfigure[]{ 
    \label{fig:EX7-Partition} 
    \includegraphics[width=0.48\textwidth]{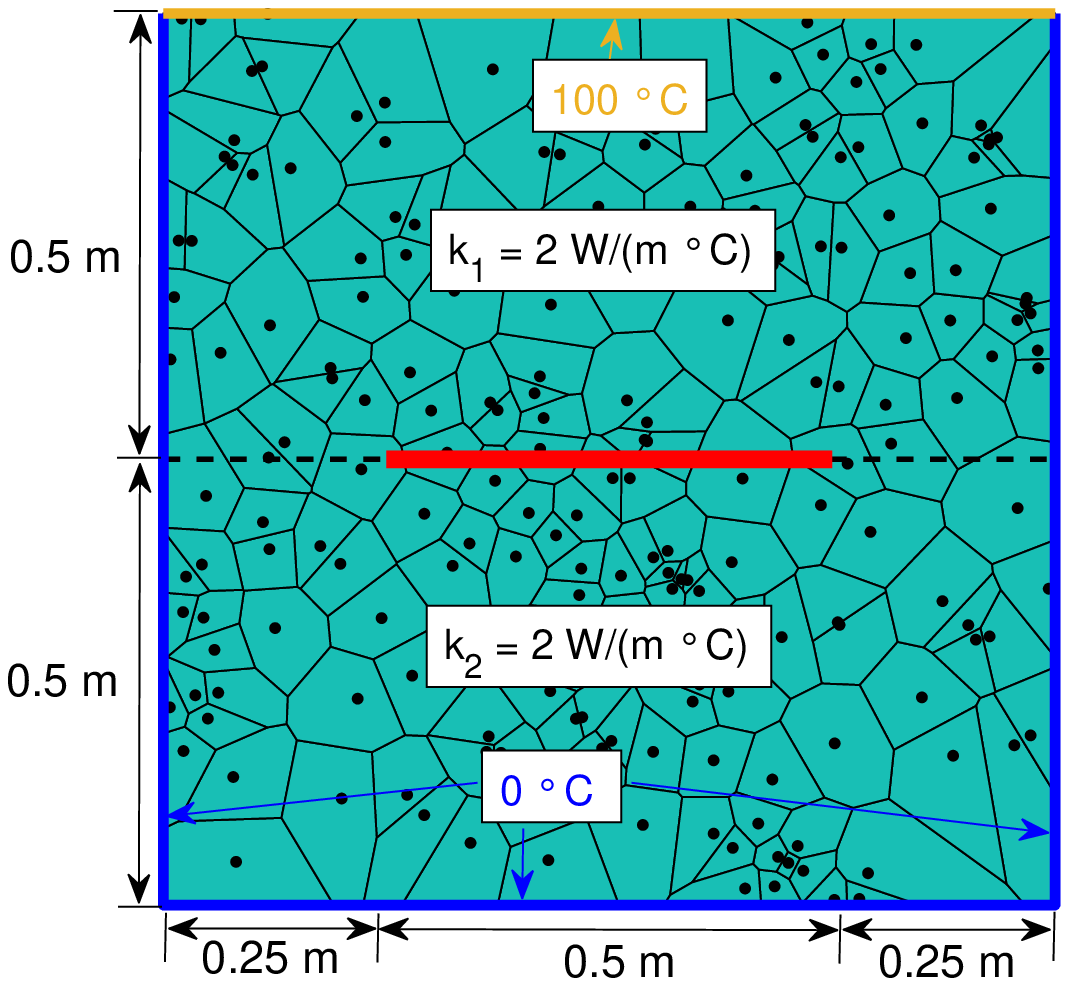}}  
    \subfigure[]{ 
    \label{fig:EX7-U200} 
    \includegraphics[width=0.48\textwidth]{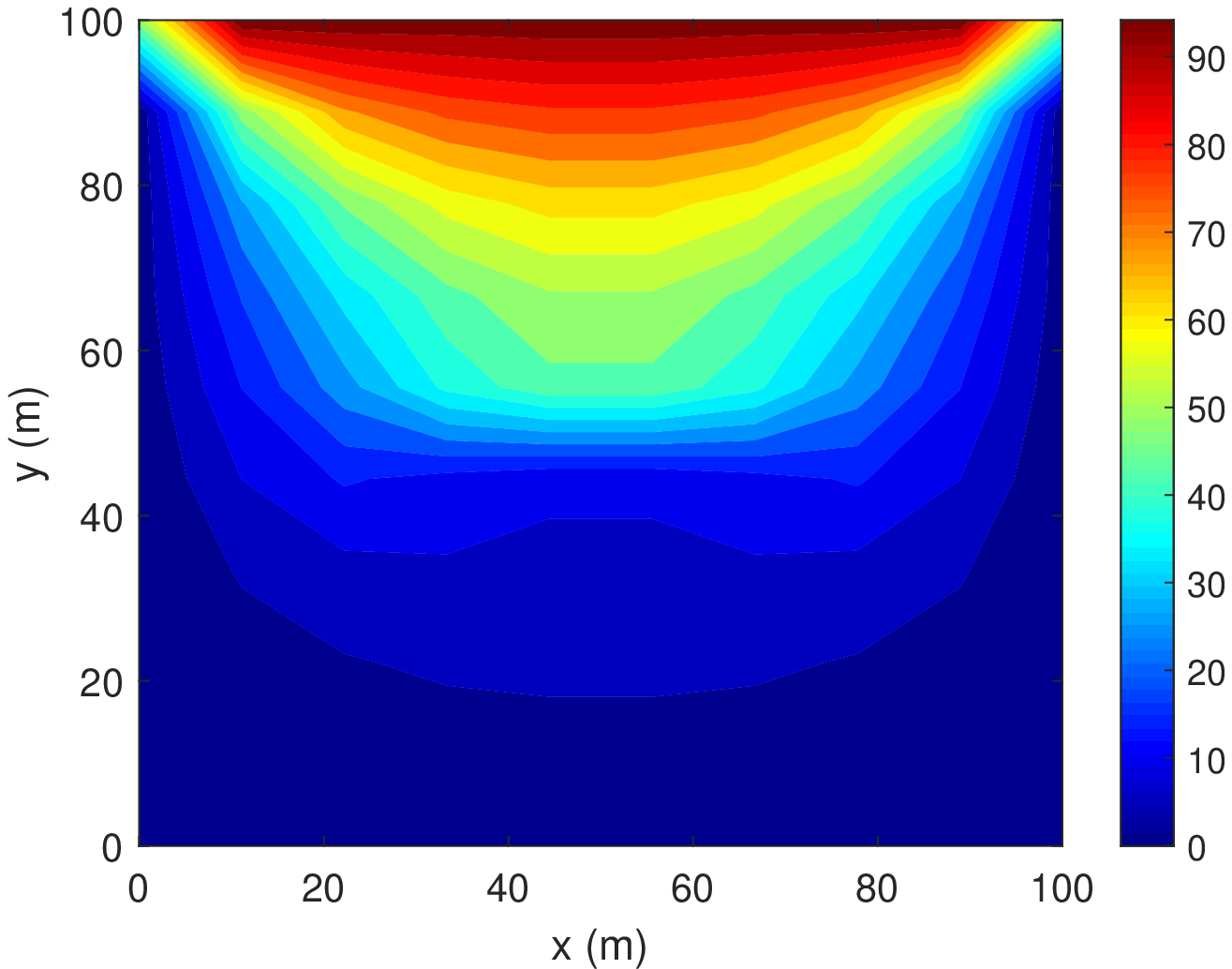}}  
        \subfigure[]{ 
    \label{fig:EX7-R200} 
    \includegraphics[width=0.48\textwidth]{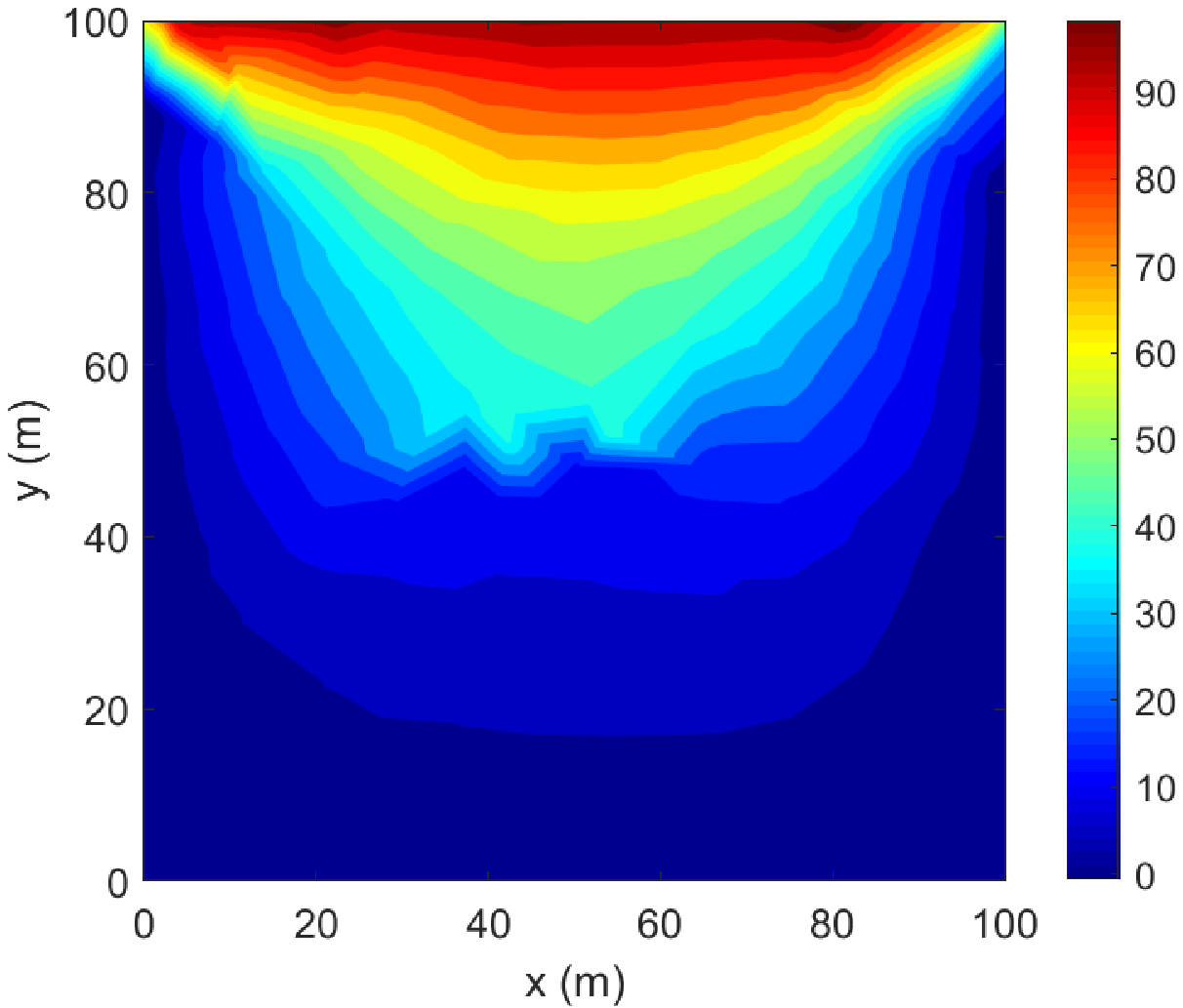}}  
        \subfigure[]{ 
    \label{fig:EX7-R2000} 
    \includegraphics[width=0.48\textwidth]{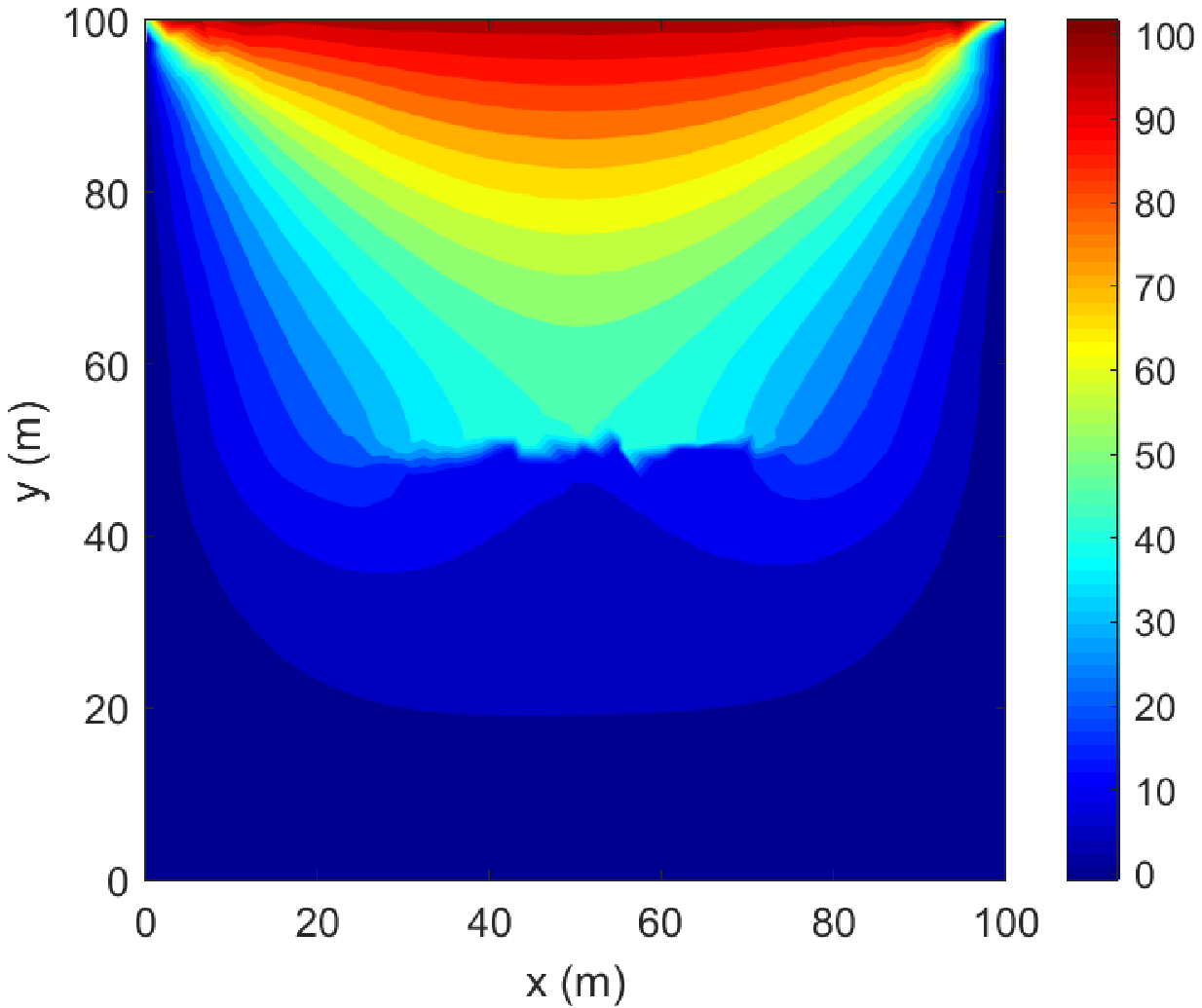}}  
  \caption{Ex. (1.7) – The adiabatic crack and computed solutions ($\eta_1 = 5 \sqrt{k_1 k_1}$, $\eta_2 = 20 \sqrt{k_1 k_1}$). (a) the points, partition and adiabatic crack. (b) 100 uniform points. (c) 100 random points. (d) 2000 random points.} 
  \label{fig:EX7} 
\end{figure}

Ex.~(1.7) is still in a square domain. However, the material property is no longer continuous. As shown in Fig.~\ref{fig:EX7-Partition}, in the top half of the domain ($ y > 50~\mathrm{m}$), the medium is isotropic and has a thermal conductivity $ k_1 = 2 \mathrm{W / (m ^\circ C)}$, while in the bottom half ($ y < 50~\mathrm{m}$), the isotropic thermal conductivity is $ k_2 = 1 \mathrm{W / (m ^\circ C)}$. An adiabatic crack emanates on the midline of the domain ($25~\mathrm{m} < x < 75~\mathrm{m}, y = 50~\mathrm{m}$). Dirichlet boundary condition is applied on all the external boundaries. On the bottom and lateral sides, $\widetilde{u}_D = 0 \mathrm{^\circ C}$, while on the top side, $\widetilde{u}_D = 100 \mathrm{^\circ C}$. For simplicity, only the steady-state solution is considered in this example.

In FPM, the subdomain boundaries shared by two points on either side of the crack are regarded as external boundaries. In this example, Neumann (adiabatic) boundary condition is applied. The computed steady-state temperature distribution is presented in Fig.~\ref{fig:EX7}. The result based on 100 uniform points (Fig.~\ref{fig:EX7-U200}) shows a very good accuracy and is consistent with the numerical result given in \cite{Liu2019}, since the crack is just on top of some subdomain boundaries. However, in a partition with random distributed points, the crack may not coincide with the internal boundaries (as can be seen in Fig.~\ref{fig:EX7-Partition}). Yet the FPM can still get a good approximation of the temperature distribution in the entire domain, especially outside the vicinity of the crack. When the number of points increases (see Fig.~\ref{fig:EX7-R2000}), the computed result approaches the exact solution gradually.

Such a result shows the potential of the FPM in solving thermal-shock problems with crack propagation in brittle materials. Without knowing the exact geometry of the cracks, an approximate solution can be obtained by simply shifting some internal subdomain boundaries from $\Gamma_h$ to $\partial \Omega$ in where the thermal stress is above the yield stress. Other than the adiabatic crack, multiple thermal crack models can be incorporated with the FPM.

\begin{figure}[htbp] 
  \centering 
   \subfigure[]{ 
    \label{fig:EX8-BC} 
    \includegraphics[width=0.48\textwidth]{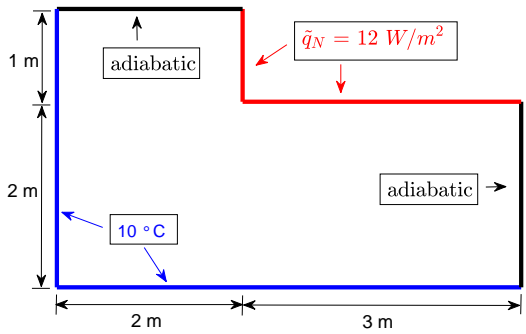}}  
    \subfigure[]{ 
    \label{fig:EX8-Abaqus} 
    \includegraphics[width=0.48\textwidth]{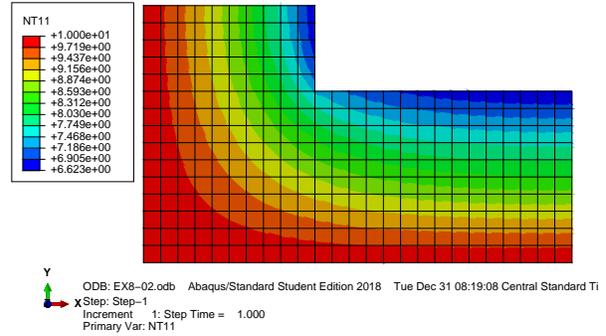}}  
    \subfigure[]{ 
    \label{fig:EX8-U48} 
    \includegraphics[width=0.48\textwidth]{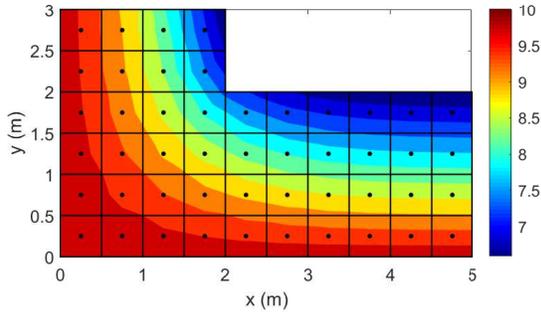}}  
    \subfigure[]{ 
    \label{fig:EX8-R100} 
    \includegraphics[width=0.48\textwidth]{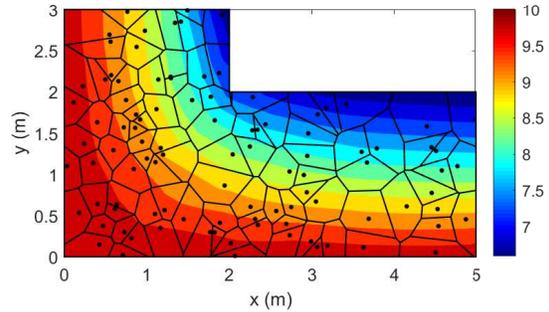}}  
  \caption{Ex. (1.8) - The boundary conditions and computed solutions. (a) the problem domain and boundary conditions. (b) ABAQUS solution with 310 DC2D4 elements (341 nodes) (c) FPM solution with 48 uniform points. (d) FPM solution with 100 random points.} 
  \label{fig:EX8} 
\end{figure}

In Ex.~(1.8), a L-shaped orthotropic domain is considered. As shown in Fig.~\ref{fig:EX8-BC}, the temperature is fixed to $10 \mathrm{^\circ C}$ on the left and bottom sides. The other sides are Neumann boundaries with $\widetilde{q}_N = 0$ on the black sides and $\widetilde{q}_N = 12 \mathrm{W / m^2}$ on the red sides. The orthotropic material has thermal conductivity coefficients $k_{11} = 4~\mathrm{W / m ^\circ C}$, $k_{22} = 7~\mathrm{W / m ^\circ C}$, and $k_{12} = k_{21} = 0$. Dirichlet boundary conditions are applied by the collocation method. The penalty parameters $\eta_1 = 5 \sqrt{k_{11} k_{22}}$, $\eta_2 = 20 \sqrt{k_{11} k_{22}}$. The steady-state results are shown in Fig.~\ref{fig:EX8}. The FPM solution agrees well with the FEM solution achieved by ABAQUS with 310 linear quadrilateral elements (341 nodes).

\begin{figure}[htbp] 
  \centering 
   \subfigure[]{ 
    \label{fig:EX9-BC} 
    \includegraphics[width=0.48\textwidth]{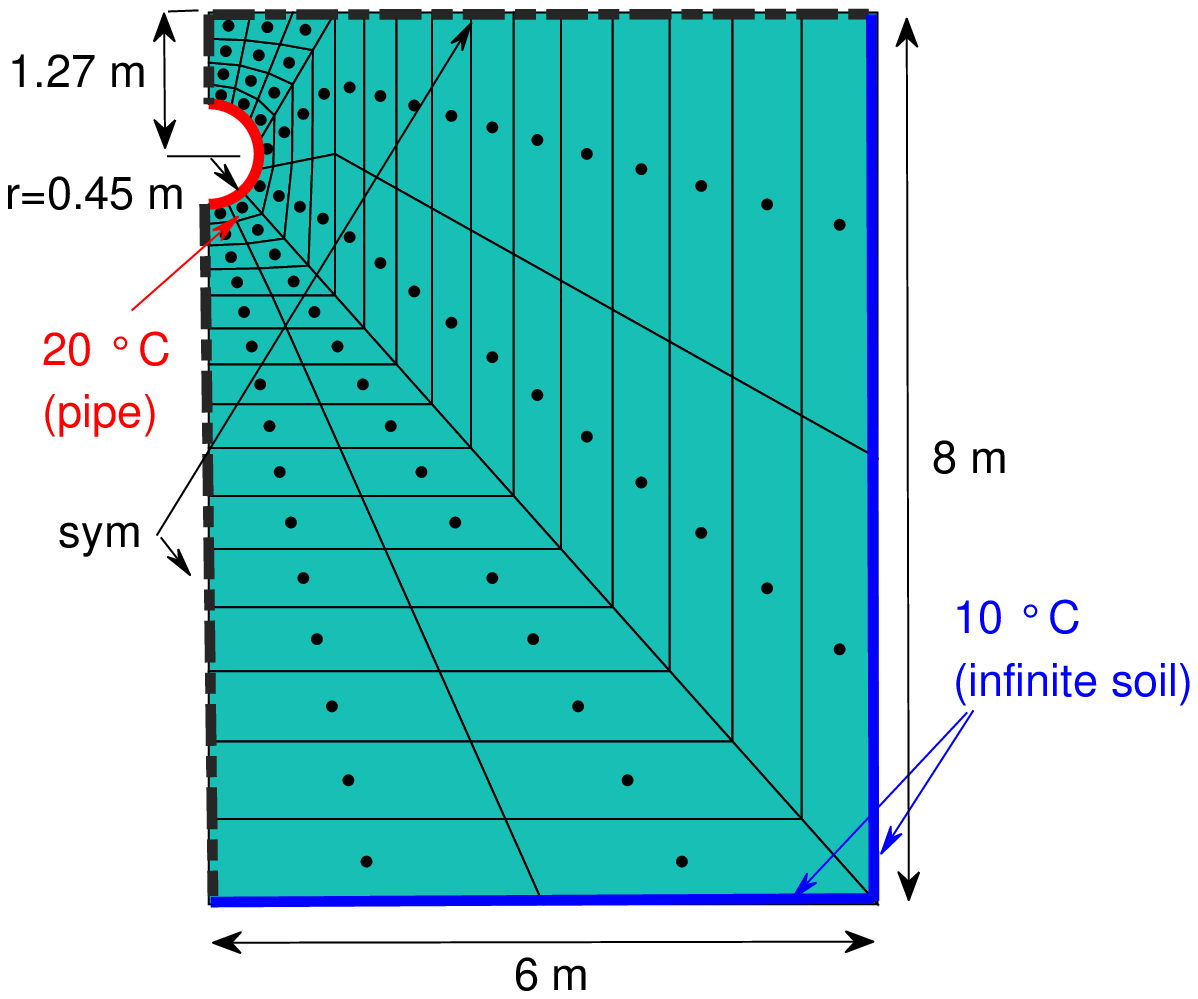}}  
    \subfigure[]{ 
    \label{fig:EX9-Trans} 
    \includegraphics[width=0.48\textwidth]{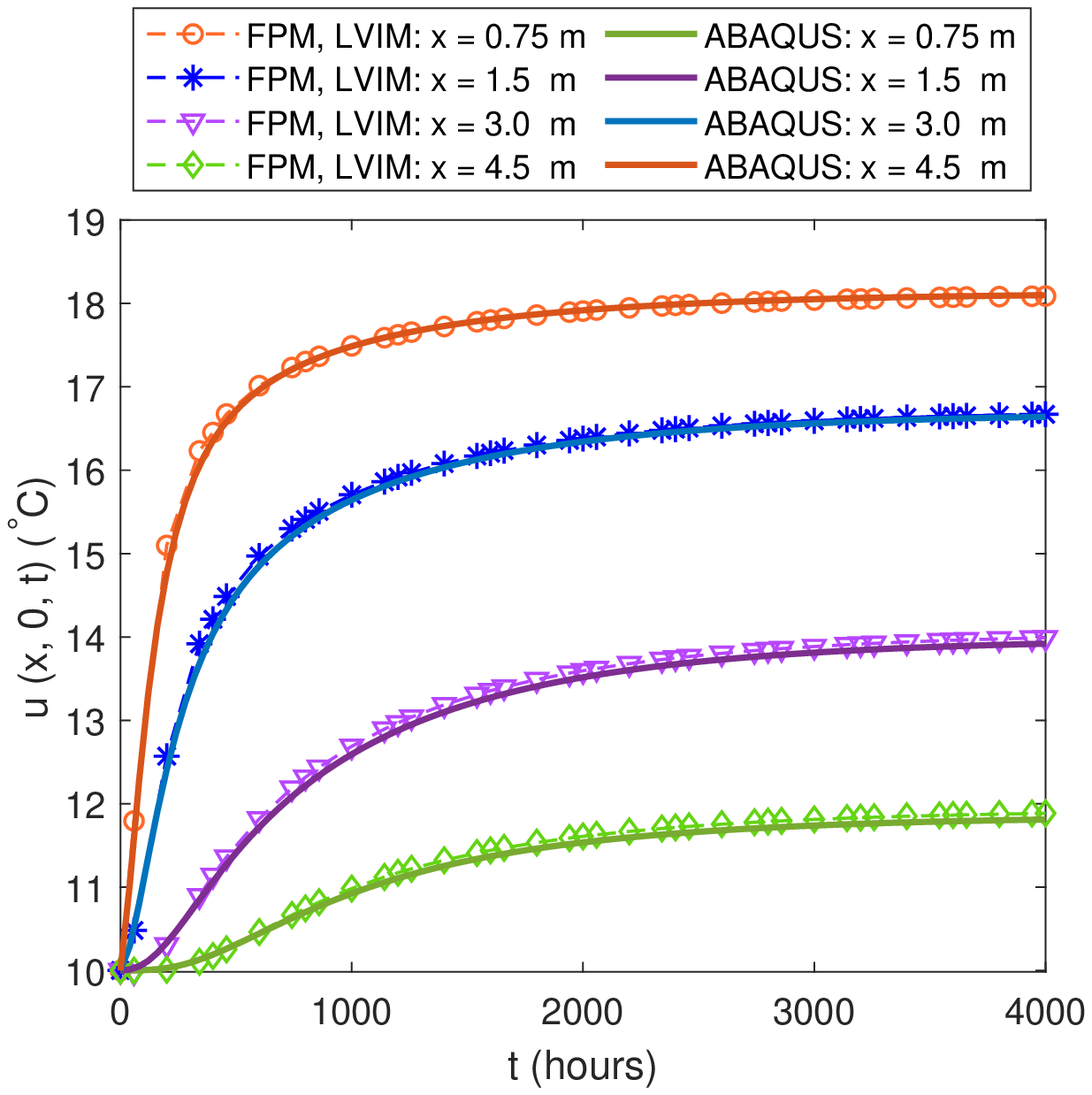}}  
    \subfigure[]{ 
    \label{fig:EX9-U360} 
    \includegraphics[width=0.98\textwidth]{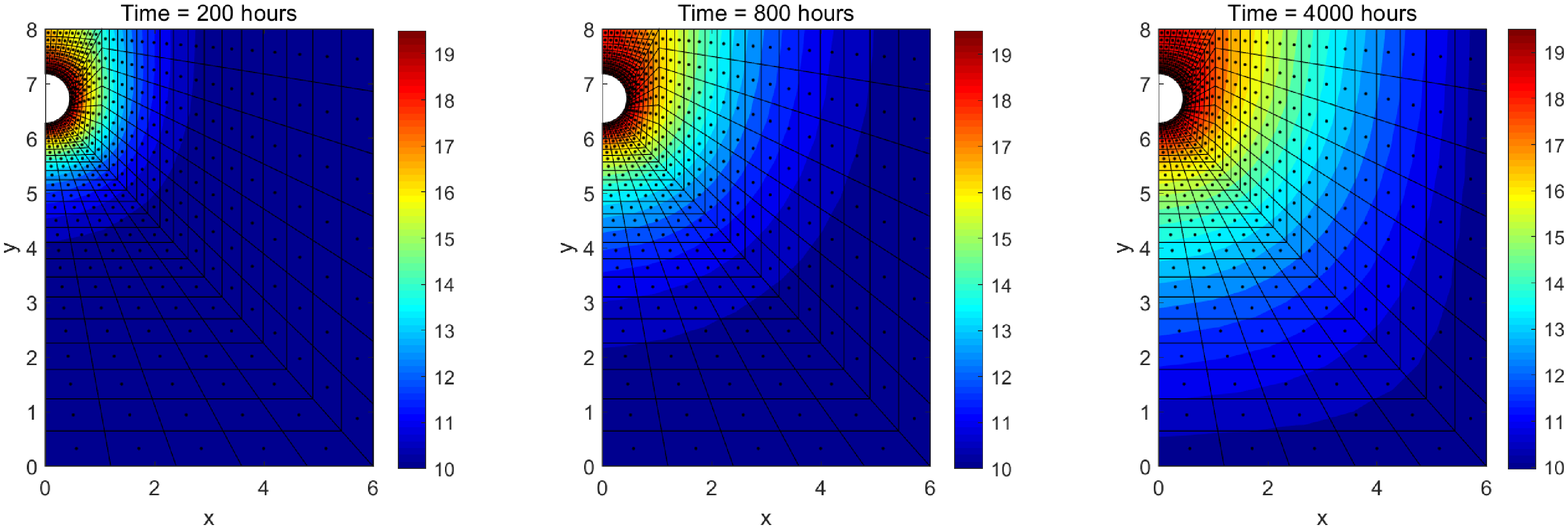}}  
    \subfigure[]{ 
    \label{fig:EX9-R360} 
    \includegraphics[width=0.98\textwidth]{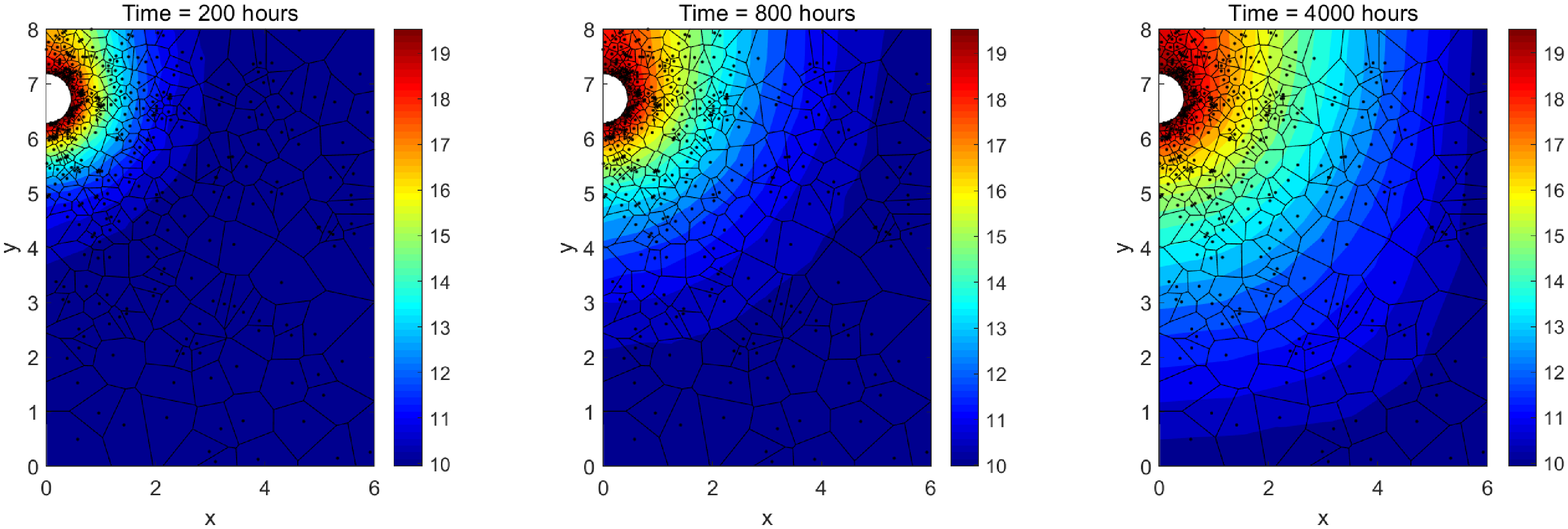}}  
  \caption{Ex. (1.9) - The problem and computed solutions. (a) the problem domain and boundary conditions. (b) transient temperature solution. (c) temperature distribution when $t=$ 200, 800, 400 hours (360 organized points). (d) temperature distribution when $t=$ 200, 800, 400 hours (360 random points).} 
  \label{fig:EX9} 
\end{figure}

In the last 2D example (Ex.~(1.9)), we consider the transient heat conduction in a semi-infinite isotropic soil medium caused by an oil pipe. A $12~\mathrm{m} \times 8~\mathrm{m}$ domain is considered. According to the symmetry, we only compute one half of the domain. As shown in Fig.~\ref{fig:EX9-BC}, the pipe wall with a radius of $0.45~\mathrm{m}$ is modeled as a Dirichlet boundary with $\widetilde{u}_D = 20 \mathrm{^\circ C}$. The infinite boundary is applied as $\widetilde{u}_D = 10 \mathrm{^\circ C}$ on the right and bottom sides ($x = 6~\mathrm{m}$ and $y = -8~\mathrm{m}$). The left side ($x = 0$) is symmetric, and the top side ($y = 0$) is adiabatic. The two boundary conditions are equivalent here since the material is isotropic. A number of points are distributed in the domain. As the variation of temperature is more violent, more points are distributed in the vicinity of the pipe wall. Generally, the number of points in a unit area decreases exponentially with the distance from the pipe center. The material properties are given as: $\rho = 2620~\mathrm{kg / m^2}$, $c = 900~\mathrm{J / kg ^\circ C}$, $k = 2.92~\mathrm{W / m ^\circ C}$. The initial condition is $u (x, y, 0) = \mathrm{const} = 10~\mathrm{^\circ C}$.

In FPM, the essential boundary conditions are imposed by the collocation method. The penalty parameters are set as: $\eta_1 = 5k$, $\eta_2 = 20k$. The computed time-variation of temperature at four representative points on the adiabatic side is shown in Fig.~\ref{fig:EX9-Trans}. The results present great consistency with the FEM result achieved by ABAQUS with 661 DC2D4 elements (715 nodes) and an explicit solver. The number of time steps is 100 in ABAQUS and 10 in the LVIM approach. The temperature distribution results at $t=$ 200, 800 and 4000~hours are presented in Fig.~\ref{fig:EX9-U360} and \ref{fig:EX9-R360} with 360 organized and random points respectively. The results agree well with the corresponding ABAQUS solutions. This example confirms the high accuracy and efficiency of the FPM + LVIM approach in solving complex 2D heat conductivity problems with unevenly distributed points.

\subsection{3D examples}

\subsubsection{Anisotropic nonhomogeneous examples in a cubic domain}

In this section, we consider a number of 3D heat conduction examples in a cubic domain $\Omega = \left\{ (x,y,z) \mid x,y,z \in [0, L] \right\}$. Various boundary conditions and material properties are tested. The heat source density $Q$ vanishes in all the following examples.

\begin{figure}[htbp] 
  \centering 
    \subfigure[]{ 
    \label{fig:EX21-U1000} 
    \includegraphics[width=0.48\textwidth]{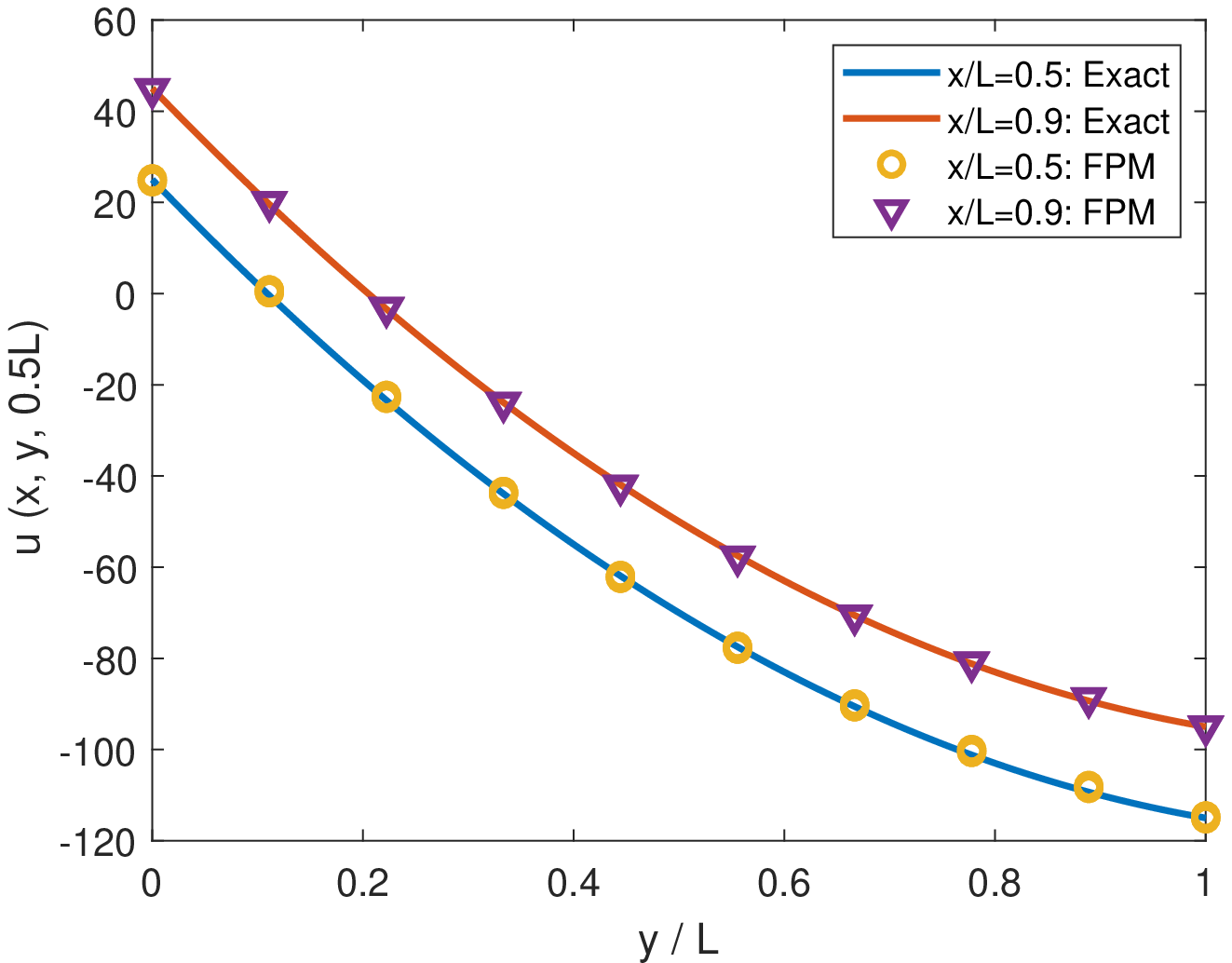}}  
    \subfigure[]{ 
    \label{fig:EX21-R1000} 
    \includegraphics[width=0.48\textwidth]{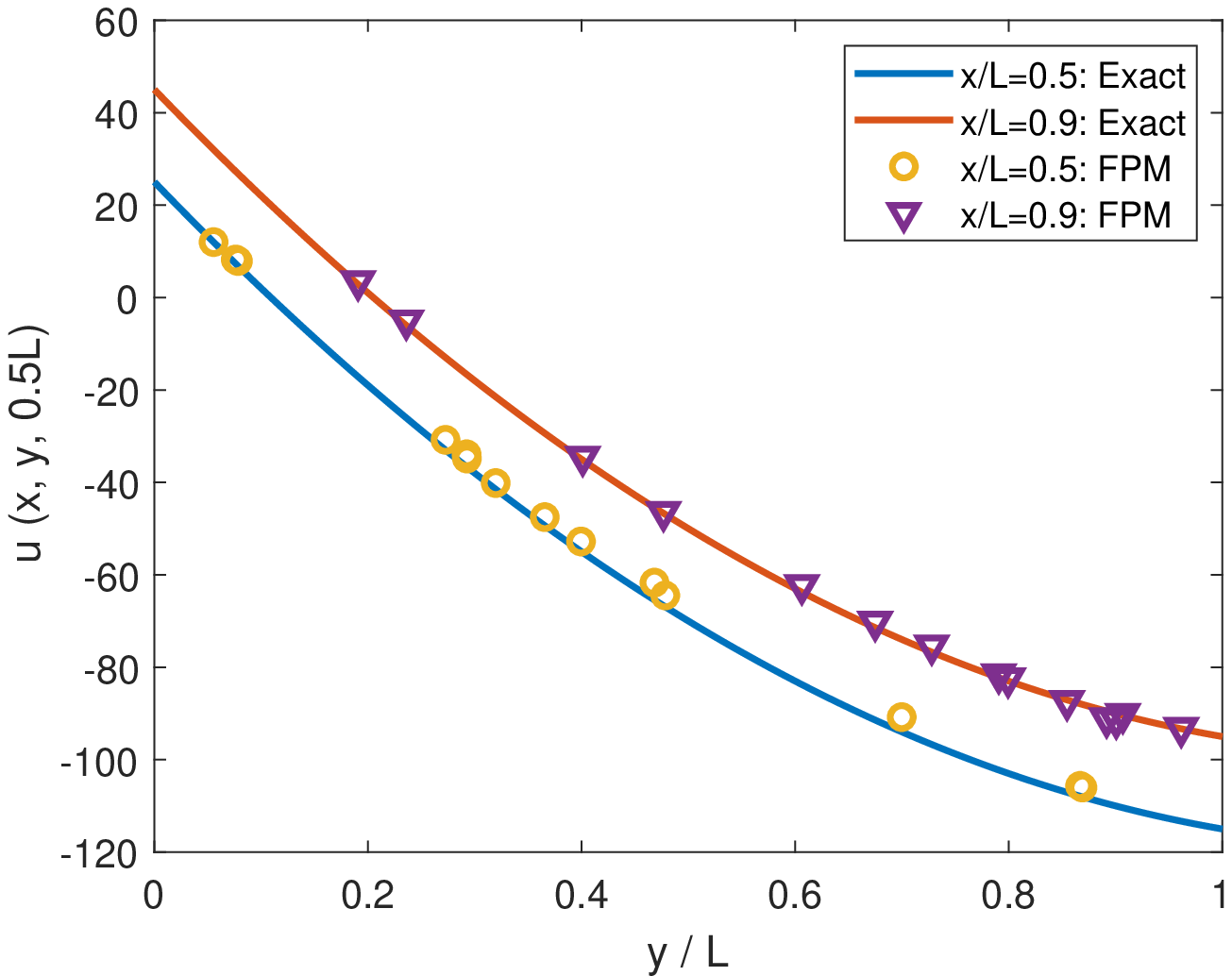}}  
  \caption{Ex. (2.1) - The computed solutions at $z = 0.5L$. (a) 1000 uniform points. (b) 1000 random points.} 
  \label{fig:EX21} 
\end{figure}

First, a steady-state problem with homogenous anisotropic material is considered. The thermal conductivity tensor components $k_{11} = k_{22} = k_{33} = 1 \times 10^{-4}$, $k_{23} = 0.2 \times 10^{-4}$, $k_{12} = k_{13} = 0$. A postulated analytical solution is considered:
\begin{align}
\begin{split}
u (x, t, z) = y^2+ y -5yz +xz. 
\end{split}
\end{align}
Dirichlet boundary conditions satisfying the postulated solution are prescribed on all the faces of the cube. The FPM is employed to solve the anisotropic example. The computed temperature distribution at $z = 0.5 L$ is shown in Fig.~\ref{fig:EX21}. In Fig.~\ref{fig:EX21-U1000}, $10 \times 10 \times 10$ points are distributed uniformly in the cube, while in Fig.~\ref{fig:EX21-R1000} the points are distributed randomly. The penalty parameters are: $\eta_1 = 5 k_{11}$ for the uniform points, and $\eta_1 = 10 k_{11}$, $\eta_2 = 20 k_{11}$ for the random points. Both results match well with the exact solution. The relative errors $r_0$ are $9.6 \times 10^{-3}$ and $1.1 \times 10^{-2}$ respectively.

\begin{figure}[htbp]
    \centering
    \begin{minipage}[t]{0.48\textwidth}
        \centering
        \includegraphics[width=1\textwidth]{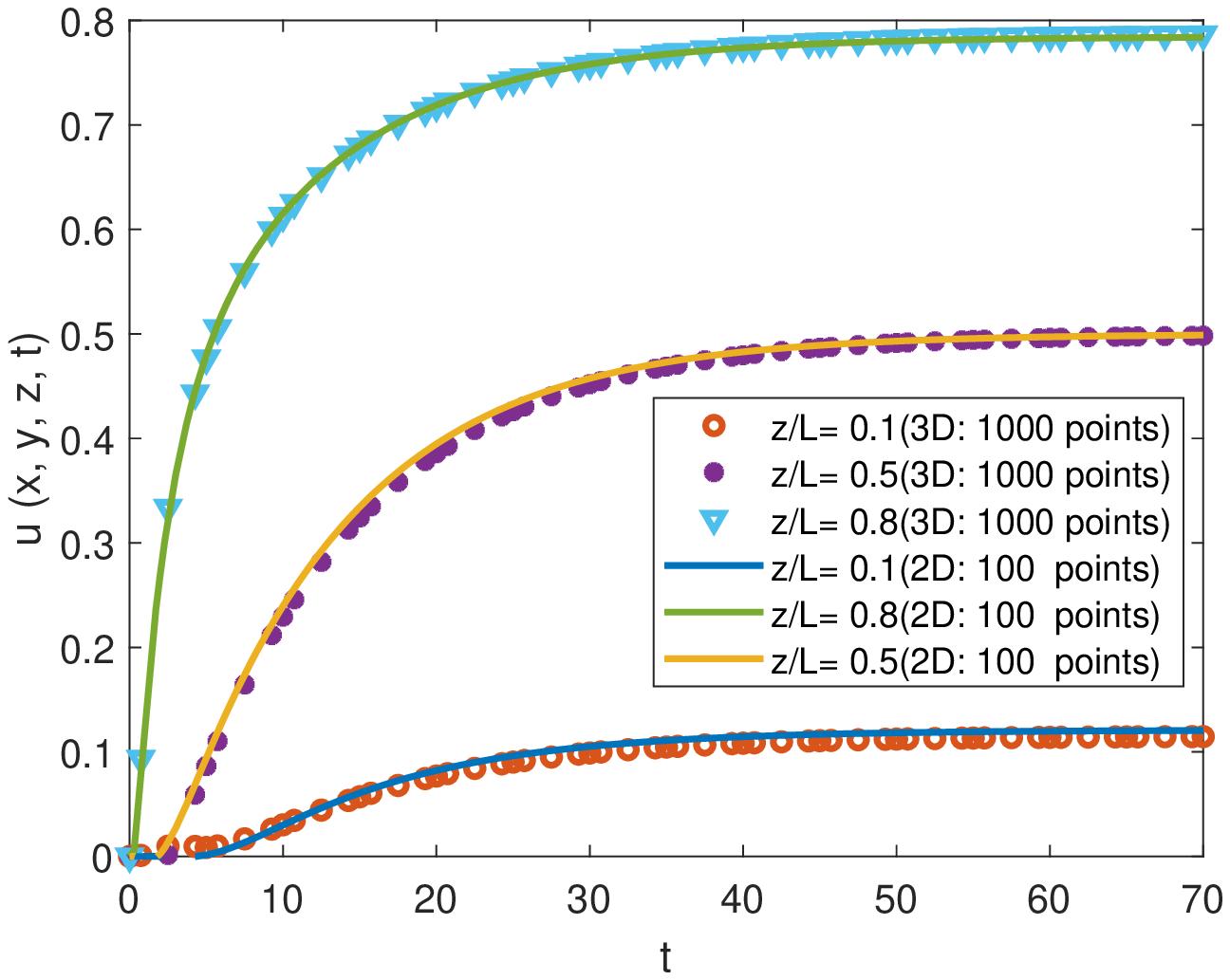}
        \caption{Ex. (2.2) - The computed transient temperature solution.}
        \label{fig:EX22}
    \end{minipage}
    \begin{minipage}[t]{0.48\textwidth}
        \centering
        \includegraphics[width=1\textwidth]{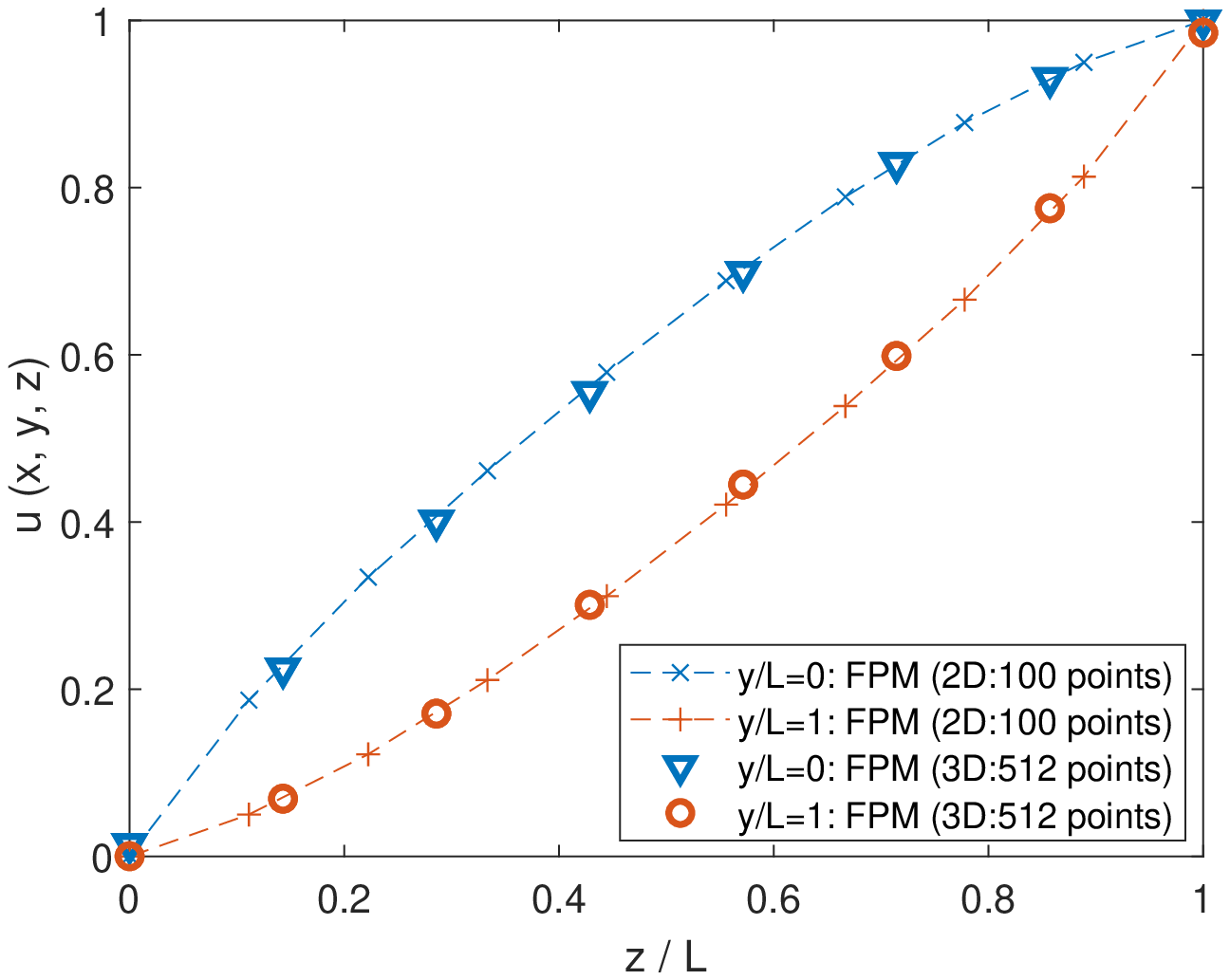}
        \caption{Ex. (2.3) - The computed steady-state result.}
        \label{fig:EX23}
    \end{minipage}
\end{figure}

Next, a transient heat conduction example is considered. The material properties are given as: $\rho = 1$, $c = 1$, and $k_{11} = k_{22} = k_{33} = 1$, $k_{12} = k_{13} = k_{23} = 0$. The boundary condition on the top surface ($z = L$) is prescribed as a thermal shock $\widetilde{u}_D = H (t-0)$, where $H$ is the Heaviside time step function. The bottom boundary condition on $z = 0$ is given as $\widetilde{u}_D = 0$. And all the lateral surfaces ($x, y = 0, L$) have vanishing heat fluxes. The initial condition is $u(x, y, z, 0) = 0$. The side length $L = 10$. It turns out that the temperature distribution in this example is not dependent on $x$ and $y$ coordinates. As a result, the problem can be analyzed equivalently in 2D. The transient temperatures at $z = 0.1L$, $z = 0.5 L$ and $z = 0.8L$ are computed by the 2D and 3D FPM and presented in Fig.~\ref{fig:EX22} respectively. The computational times cost by the LVIM approach and the backward Euler scheme are shown in Table~\ref{table:EX22}. As the time-variation of temperature is smooth in this case, the LVIM approach can only improve the computing efficiency slightly.

In Ex.~(2.3), we consider a similar initial boundary condition problem as Ex.~(2.2) in an isotropic medium. The thermal conductivity tensor components are: $k_{11} = k_{33} = 1$, $k_{22} = 1.5$, $k_{23} = 0.5$, $k_{12} = k_{13} = 0$. Symmetric boundary conditions are given on the left and right surfaces ($x = 0, L$) instead of the Neumann boundary conditions. The temperature distribution is then independent of $x$ coordinate, i.e., the example can also be equivalent to a 2D problem. Fig.~\ref{fig:EX23} compares the computed steady-state temperature distribution on $y = 0$ and $y = L$ analyzed by 2D and 3D FPM ($\eta_1 = 10 k_{11}$, $\eta_2 = 20 k_{11}$ in both cases). Very good agreement can be observed. The transient result is shown as a comparison of Ex.~(2.4) in the following Fig.~\ref{fig:EX24-Trans}.

\begin{table}[htbp]
\caption{Computational time of 3D FPM + LVIM / backward Euler approach (1000 points) in solving Ex.~(2.2).}
\centering
{
\begin{tabular}{ c c c c c }
\toprule[2pt]
Method & \tabincell{c}{Computational \\ parameters} & Time step & \ \tabincell{c}{Computational \\ time (s)} \\
\hline
FPM + LVIM & $\eta_1 = 10$, $\eta_2 = 20$, $M=3$, $tol = 10^{-8}$ & $\Delta t = 5.5 $ & 5.2 \\
\hline
FPM + backward Euler & $\eta_1 = 10$, $\eta_2 = 20$  & $\Delta t = 0.5 $ & 7.0 \\
\toprule[2pt]
\end{tabular}}
\label{table:EX22}
\end{table}

Ex.~(2.4) is a nonhomogeneous anisotropic problem with the same initial and boundary conditions as Ex.~(2.3). The material density $\rho$ and heat capacity $c$ remain constant in the whole domain. Whereas the thermal conductivity tensor is prescribed as: $k_{33} (z) = 1 + z/L$, $k_{11} = 1$, $k_{22} = 1.5$, $k_{23} = 0.5$, $k_{12} = k_{13} = 0$. The side length $L = 10$. The example can also be analyzed in 2D. A comparison of the transient 2D and 3D computed temperatures on $z = 0.2 L$ is presented in Fig.~\ref{fig:EX24-Trans}. The homogenous result (Ex.~(2.3)) is also shown as a comparison. Table~\ref{table:EX24} shows the computational times for the LVIM approach and backward Euler scheme when achieving the same accuracy. As can be seen, the nonhomogeneity has a considerable influence on the temperature distribution, while it has no influence on the accuracy or efficiency of the FPM + LVIM approach. The transient temperature solution approaches the steady-state result, as shown in Fig.~\ref{fig:EX24-SS}, gradually.

\begin{figure}[htbp] 
  \centering 
    \subfigure[]{ 
    \label{fig:EX24-Trans} 
    \includegraphics[width=0.48\textwidth]{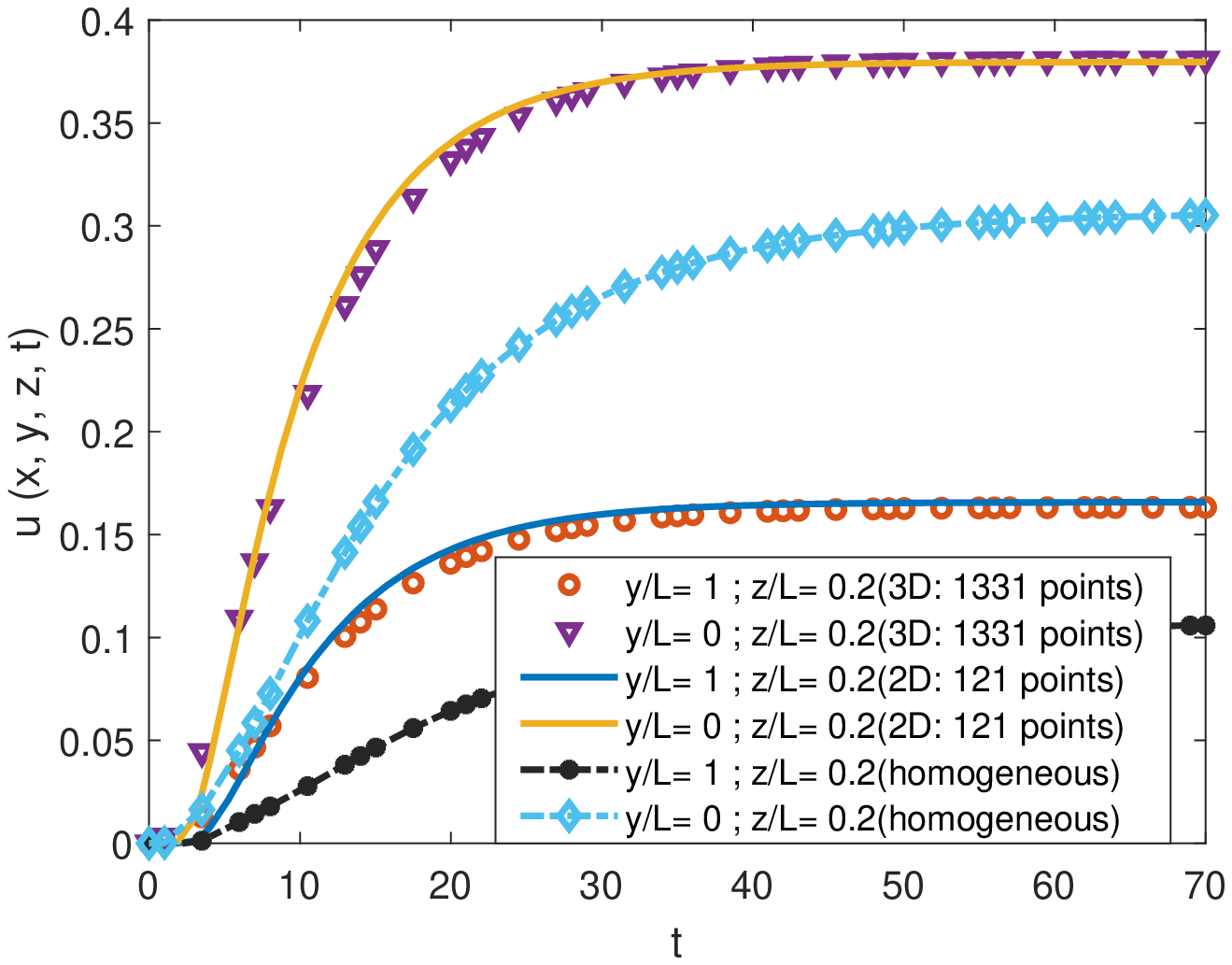}}  
    \subfigure[]{ 
    \label{fig:EX24-SS} 
    \includegraphics[width=0.48\textwidth]{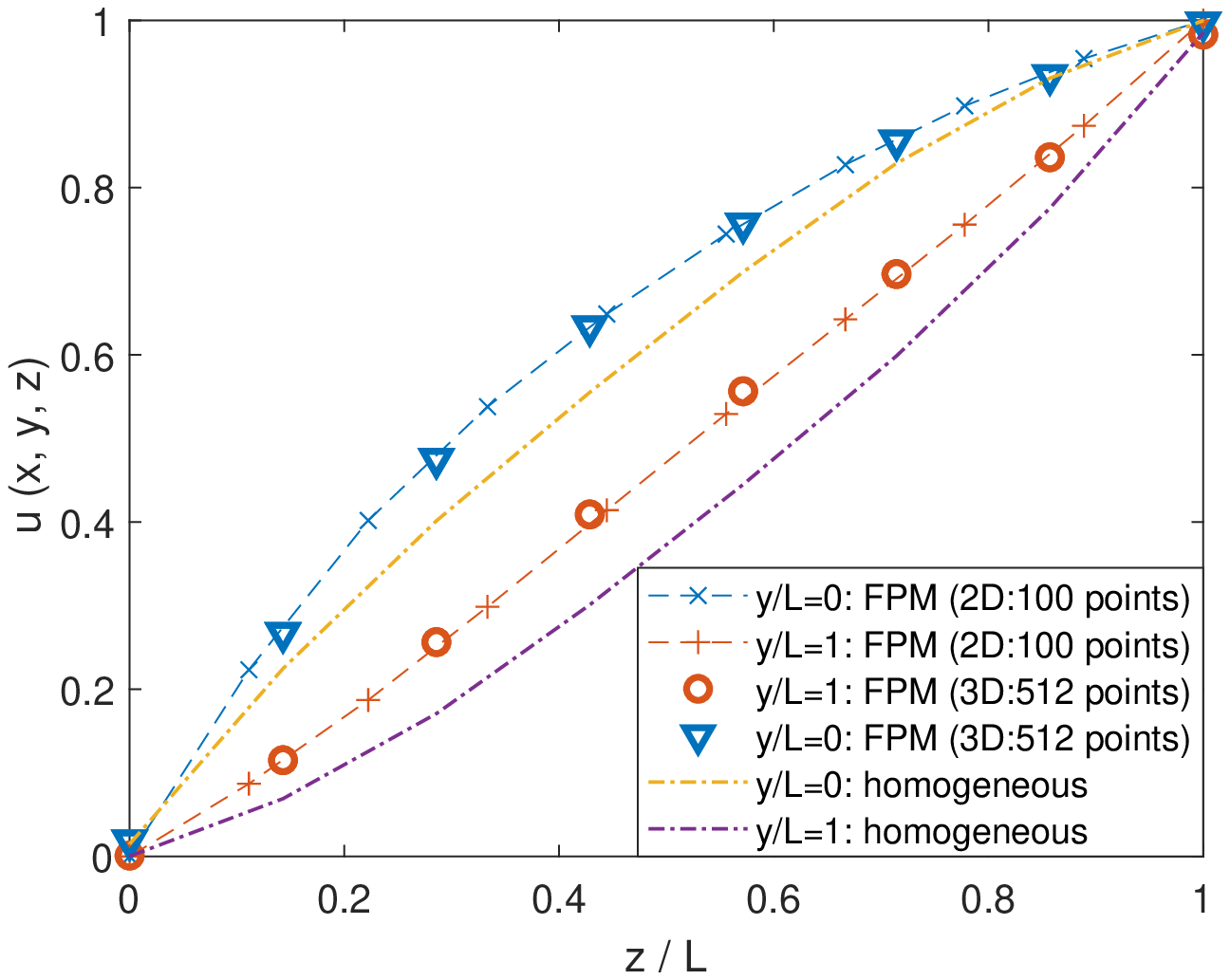}}  
  \caption{Ex. (2.4) - The computed solutions. (a) transient temperature solution. (b) steady-state result.} 
  \label{fig:EX24} 
\end{figure}

\begin{table}[htbp]
\caption{Computational time of 3D FPM + LVIM / backward Euler approach (1331 points) in solving Ex.~(2.4).}
\centering
{
\begin{tabular}{ c c c c c }
\toprule[2pt]
Method & \tabincell{c}{Computational \\ parameters} & Time step & \ \tabincell{c}{Computational \\ time (s)} \\
\hline
FPM + LVIM & $\eta_1 = 10$, $\eta_2 = 20$, $M=4$, $tol = 10^{-6}$ & $\Delta t = 25 $ & 3.4 \\
\hline
FPM + backward Euler & $\eta_1 = 10$, $\eta_2 = 20$  & $\Delta t = 0.5 $ & 7.9 \\
\toprule[2pt]
\end{tabular}}
\label{table:EX24}
\end{table}

\begin{figure}[htbp] 
  \centering 
    \subfigure[]{ 
    \label{fig:EX25-Trans} 
    \includegraphics[width=0.48\textwidth]{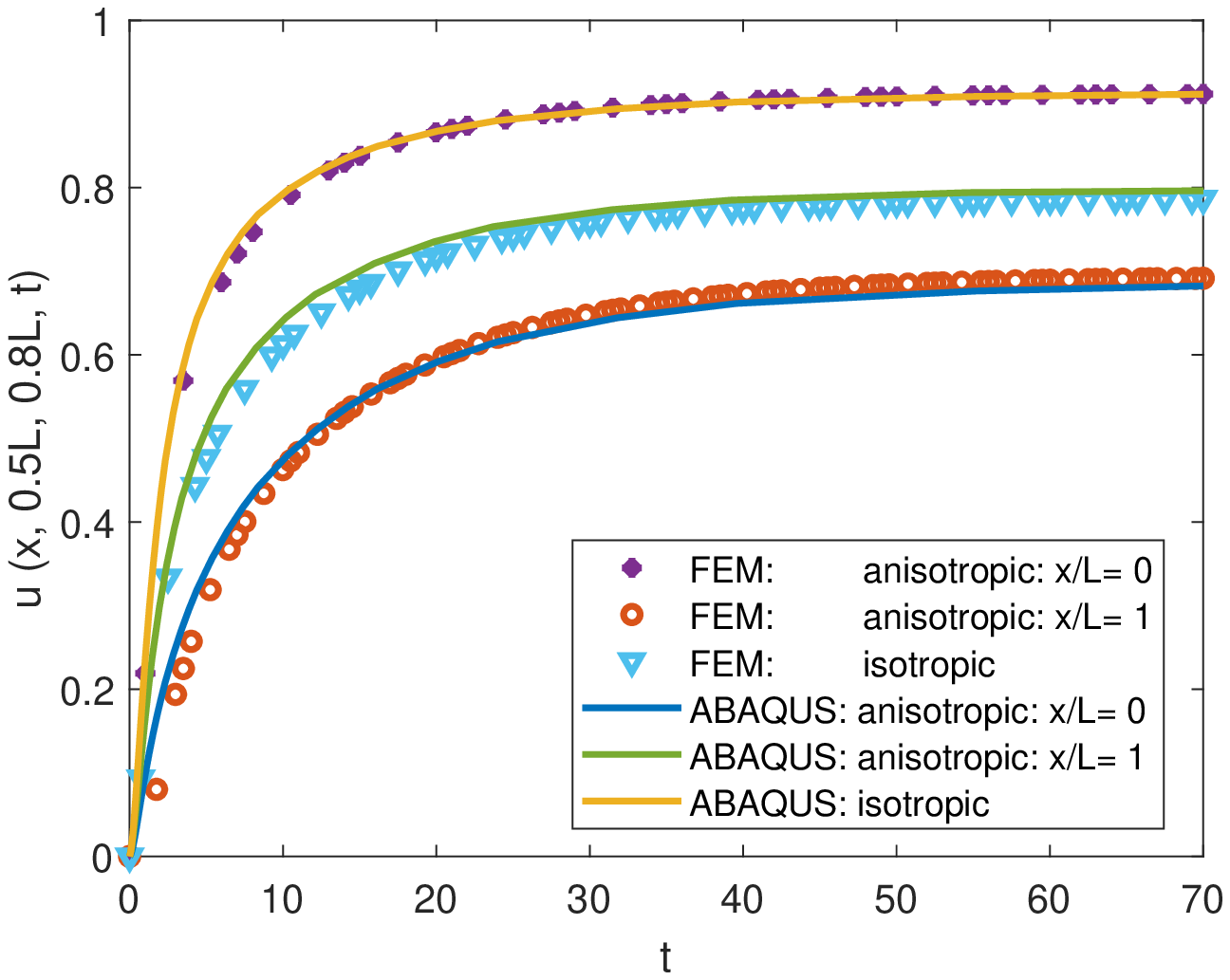}}  
    \subfigure[]{ 
    \label{fig:EX25-SS} 
    \includegraphics[width=0.48\textwidth]{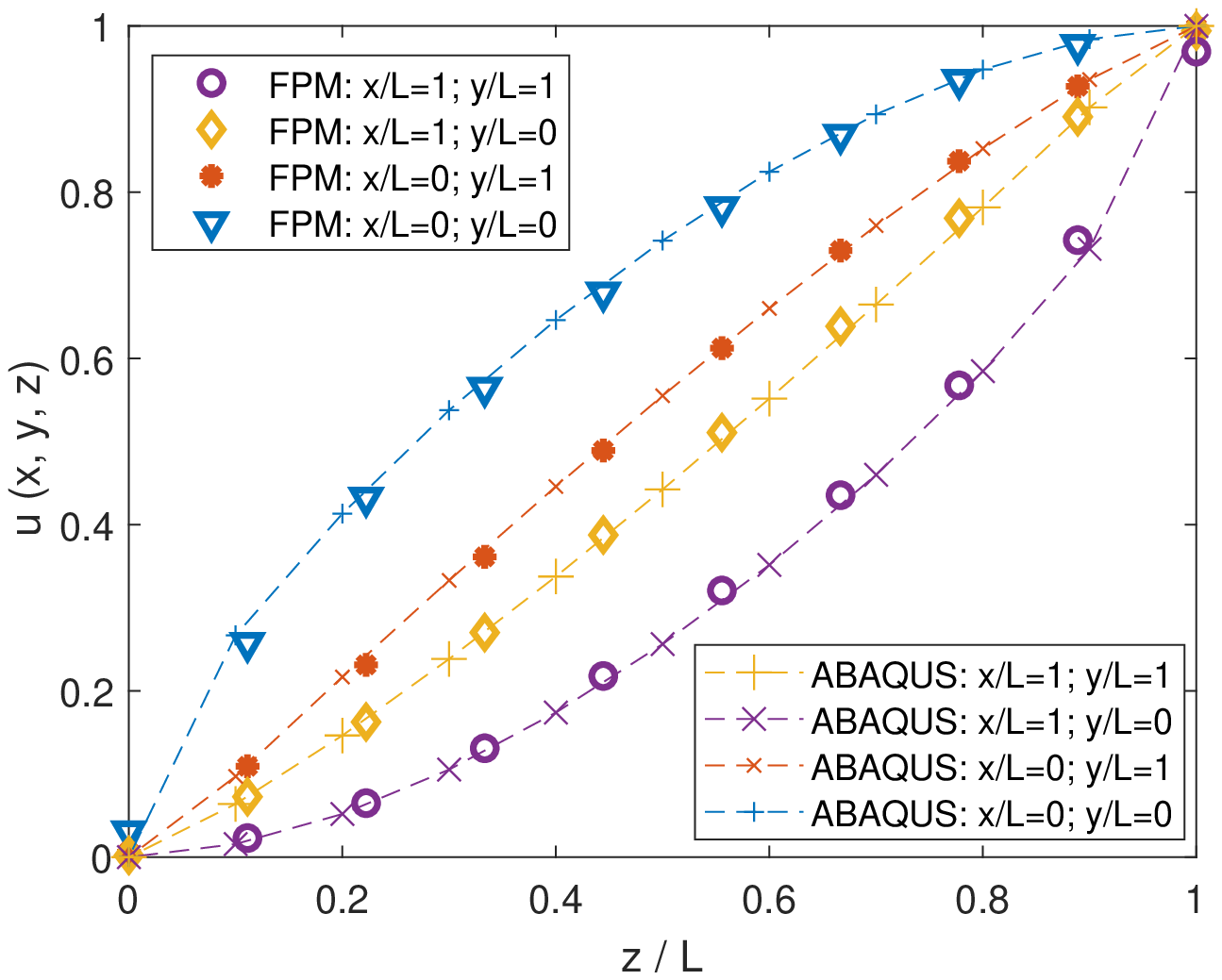}}  
  \caption{Ex. (2.5) - The computed solutions. (a) transient temperature solution. (b) steady-state result.} 
  \label{fig:EX25} 
\end{figure}

In Ex.~(2.5), we consider a 3D example that can no longer be analyzed in 2D. The problem domain is still a $L \times L \times L$ cube with vanishing flux on all the lateral surfaces. The boundary conditions on the top and bottom surfaces are given as: $\widetilde{u}_D = H(t-0)$, for $z = L$; and $\widetilde{u}_D = 0$, for $z = 0$. The homogenous anisotropic thermal conductivity coefficients: $k_{11} = k_{33} = 1$, $k_{22} = 1.5$, $k_{12}=k_{13} = k_{23} = 0.5$. The other conditions are the same as the previous examples. The computed solution is compared with FEM result achieved by ABAQUS with 1000 linear heat transfer elements (DC3D8) and shown in Fig.~\ref{fig:EX25-Trans}. The homogenous solution (Ex.~(2.2)) is also shown for comparison. As can be seen, a good agreement is observed between the FPM + LVIM and ABAQUS results. As time goes on, the transient solution keeps approaching the steady state. The computed temperature distributions on the four lateral sides of the domain ($x, y = 0, L$) are shown in Fig.~\ref{fig:EX25-SS}, as well as the ABAQUS results. It is clear that the solution is dependent on all $x$, $y$ and $z$ coordinates, and cannot be simplified as a 2D problem. The penalty parameters and computational times are listed in Table~\ref{table:EX25}, confirming that the FPM + LVIM approach can work with considerable large time intervals and achieving accurate solutions. 

\begin{table}[htbp]
\caption{Computational time of 3D FPM + LVIM / backward Euler approach (1000 points) in solving Ex.~(2.5).}
\centering
{
\begin{tabular}{ c c c c c }
\toprule[2pt]
Method & \tabincell{c}{Computational \\ parameters} & Time step & \ \tabincell{c}{Computational \\ time (s)} \\
\hline
FPM + LVIM & $\eta_1 = 5$, $\eta_2 = 10$, $M=5$, $tol = 10^{-6}$ & $\Delta t = 25 $ & 6.9 \\
\hline
FPM + backward Euler & $\eta_1 = 5$, $\eta_2 = 10$  & $\Delta t = 0.5 $ & 11 \\
\toprule[2pt]
\end{tabular}}
\label{table:EX25}
\end{table}

\begin{figure}[htbp]
    \centering
    \begin{minipage}[t]{0.48\textwidth}
        \centering
        \includegraphics[width=1\textwidth]{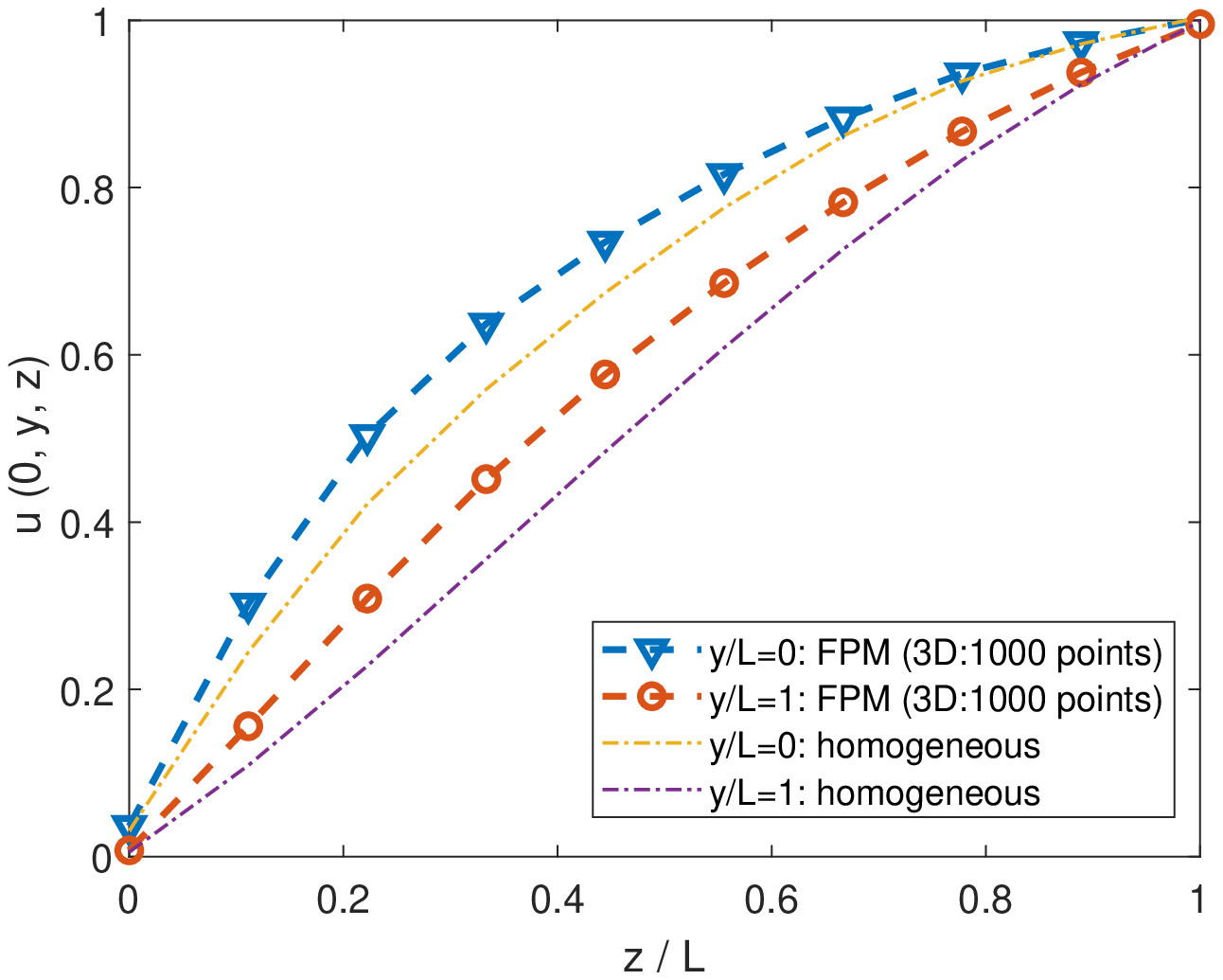}
        \caption{Ex. (2.6) - The computed steady-state result.}
        \label{fig:EX26}
    \end{minipage}
    \begin{minipage}[t]{0.48\textwidth}
        \centering
        \includegraphics[width=1\textwidth]{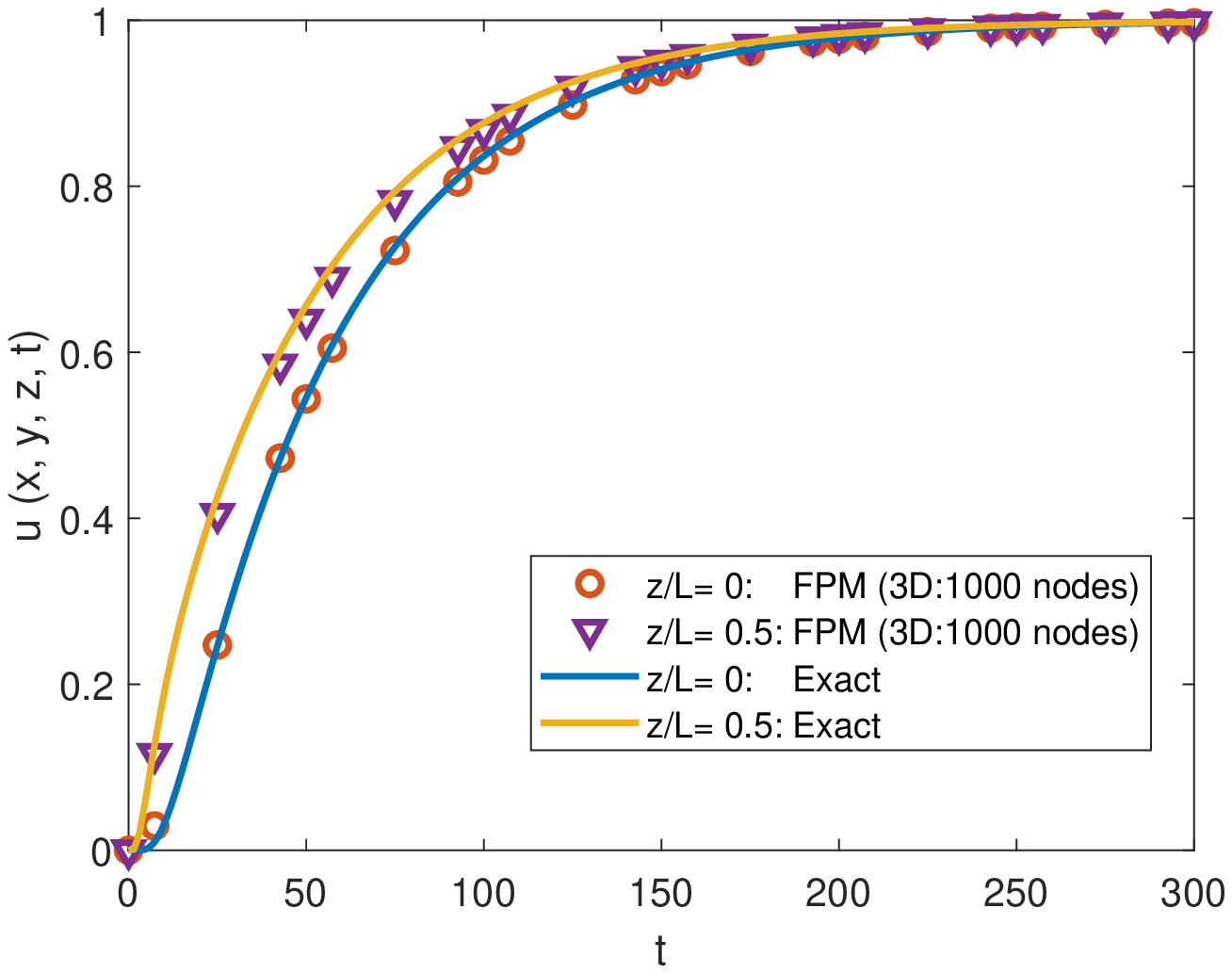}
        \caption{Ex. (2.7) - The computed transient temperature solution.}
        \label{fig:EX27}
    \end{minipage}
\end{figure}

Furthermore, a nonhomogeneous anisotropic problem is considered in Ex.~(2.6). The coordinate-dependent thermal conductivity tensor components: $k_{33} (z) = 1 + z/L$, $k_{11} = 1$, $k_{22} = 1.5$, $k_{12}=k_{13} = k_{23} = 0.5$. All the boundary conditions are the same as Ex.~(2.5). Fig.~\ref{fig:EX26} presents the computed steady-state solution obtained by the FPM. The solution, as well as all the previous solutions in Ex.~(2.1) – Ex.~(2.5), are consistent with the computed solutions achieved by \citet{Sladek2008} with the Meshless Local Petrov-Galerkin (MLPG) method and Laplace-transform technique.

In Ex.~(2.7), a transient heat conduction example with Robin boundary condition is tested. The material is homogenous and isotropic: $\rho = 1$, $c = 1$, $k_{11} = k_{22} = k_{33} = 1$, $k_{12} = k_{13} = k_{23} = 0$. The top surface has a heat transfer coefficient $h = 1.0$. And the temperature outside the top surface is prescribed as $\widetilde{u}_R = H(t-0)$. All the lateral surfaces and bottom surface have heat fluxes $\widetilde{q}_N = 0$. Started from an initial condition $u (x,y,z,0)=0$, the temperature distribution depends only on $z$ coordinate and the time. The analytical solution can be written as \cite{Sladek2008}:
\begin{align}
\begin{split}
u (x,y,z,t) = u(z, t) = 1 - 2m \sum_{i = 1} ^{\infty} \frac{\mathrm{sin} \beta_i \mathrm{cos} \left( \frac{\beta_i z}{L} \right) \mathrm{exp} \left( - \frac{\beta_i^2 k_{33} t}{\rho c L^2} \right)}{\beta_i \left( m + \mathrm{sin}^2 \beta_i \right)}, \notag
\end{split}
\end{align}
where $\beta_i$ are roots of the transcendental equation:
\begin{align}
\begin{split}
\frac{\beta \mathrm{sin} \beta}{\mathrm{cos} \beta} - m =0, \quad \text{where} \; m=\frac{hL}{k_{33}}.
\end{split}
\end{align}
Let $L = 10$. The computed time-variations of temperature on the bottom and midsurface of the cube ($z = 0, 0.5L$) are shown in Fig.~\ref{fig:EX27}, in which an excellent agreement is observed between the FPM + LVIM solution and the analytical result.

\subsubsection{Some practical examples}

\begin{figure}[htbp] 
  \centering 
   \subfigure[]{ 
    \label{fig:EX28-BC} 
    \includegraphics[width=0.48\textwidth]{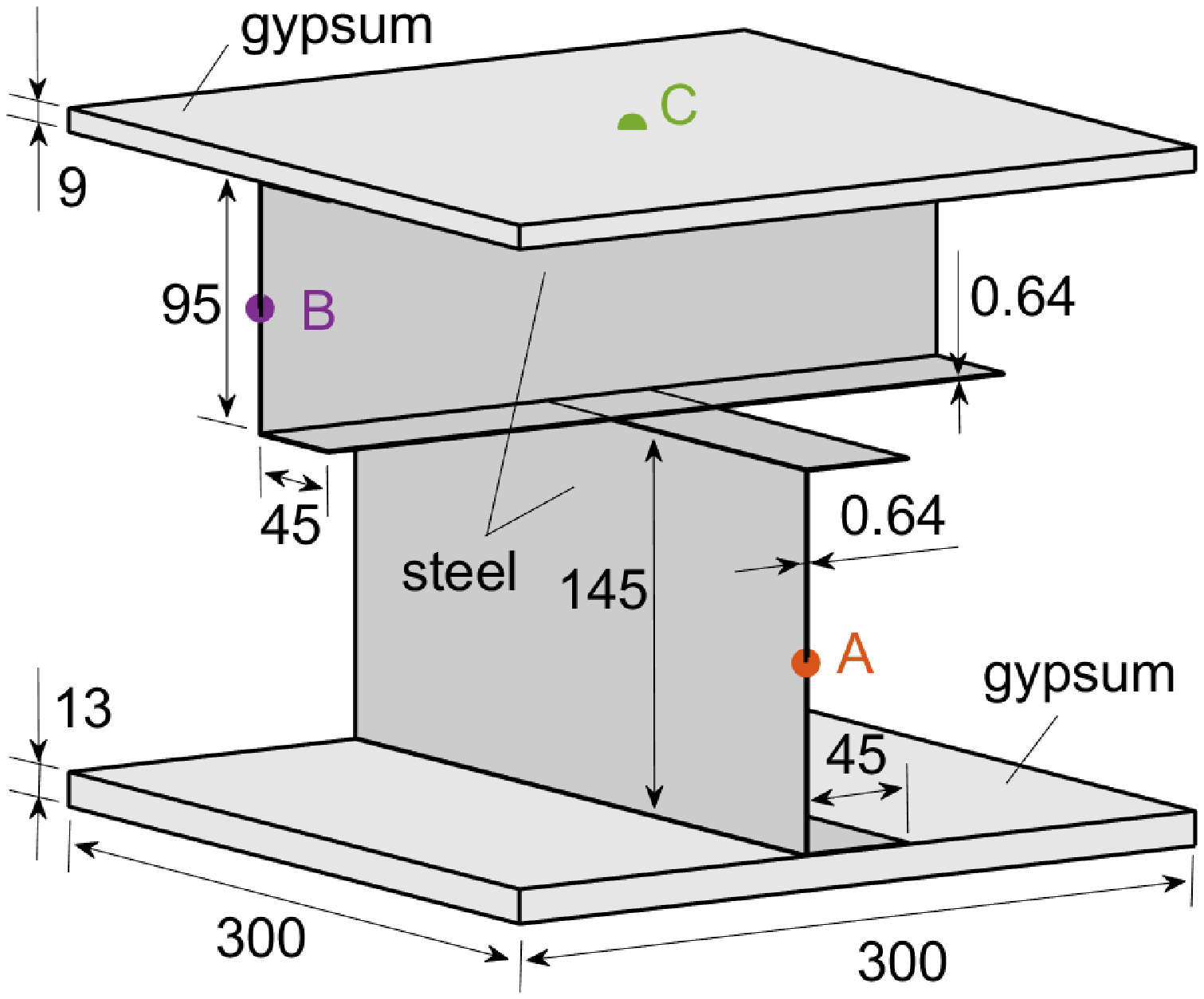}}  
   \subfigure[]{ 
    \label{fig:EX28-Trans} 
    \includegraphics[width=0.48\textwidth]{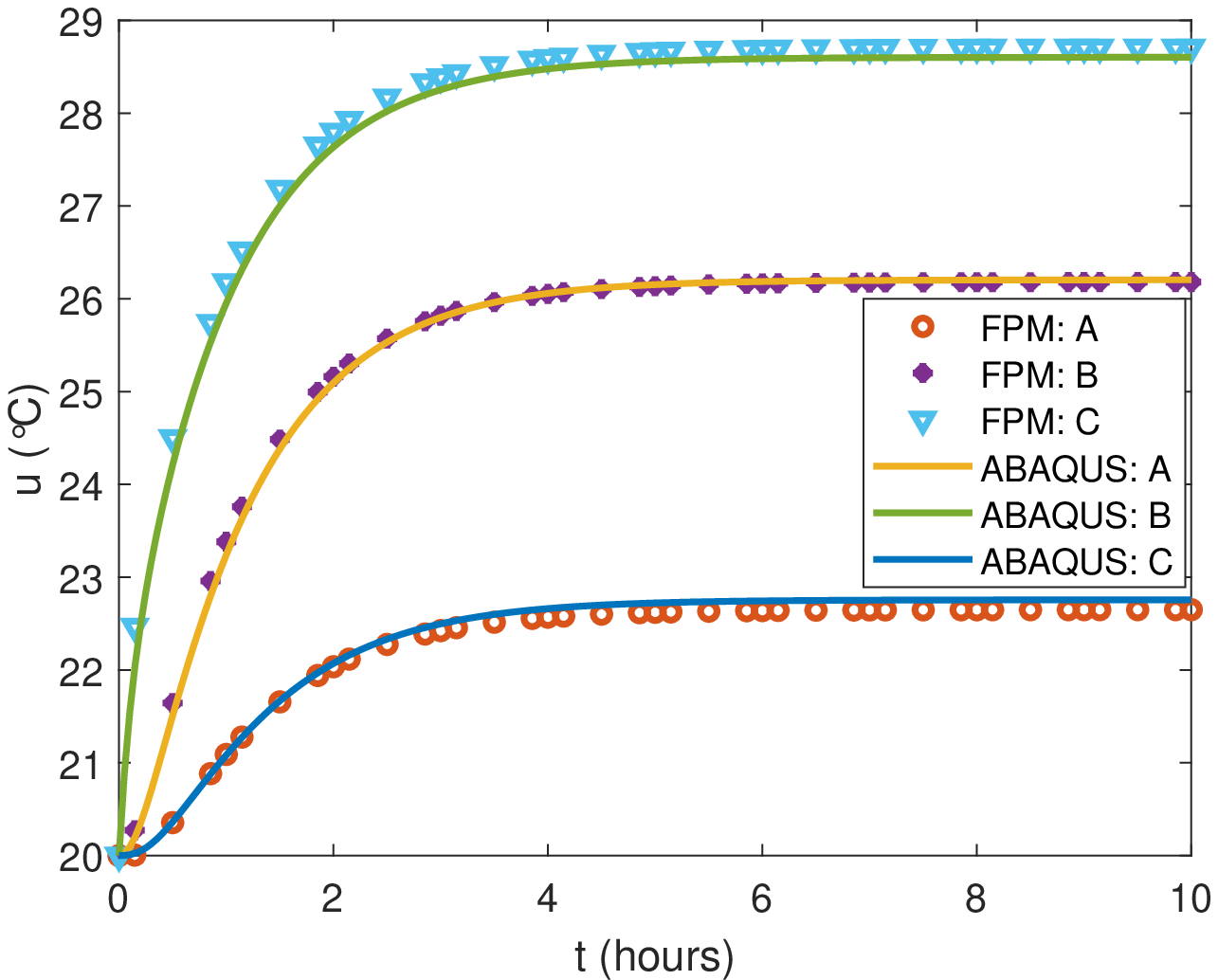}}  
    \subfigure[]{ 
    \label{fig:EX28-T1} 
    \includegraphics[width=0.48\textwidth]{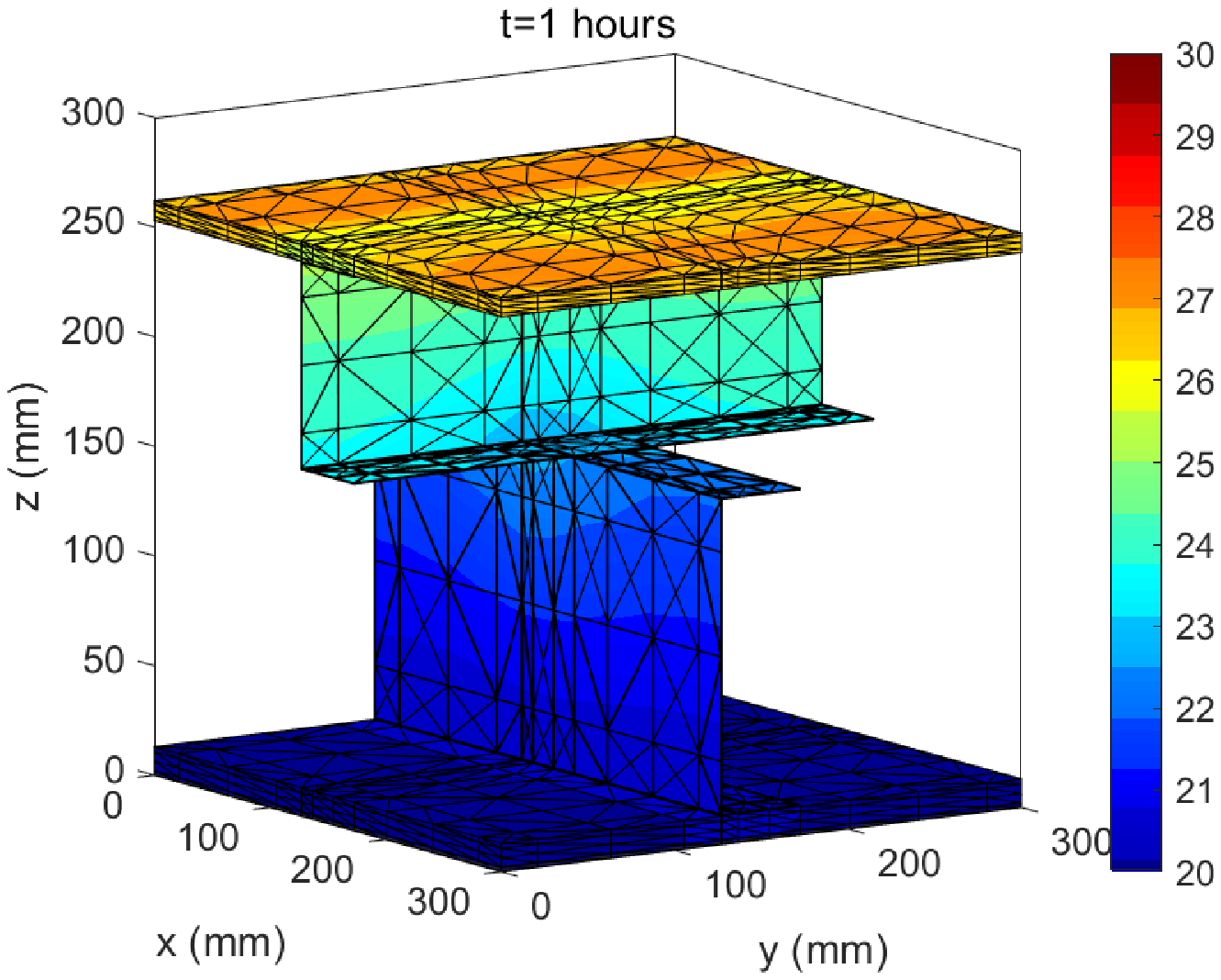}}  
    \subfigure[]{ 
    \label{fig:EX28-SS} 
    \includegraphics[width=0.48\textwidth]{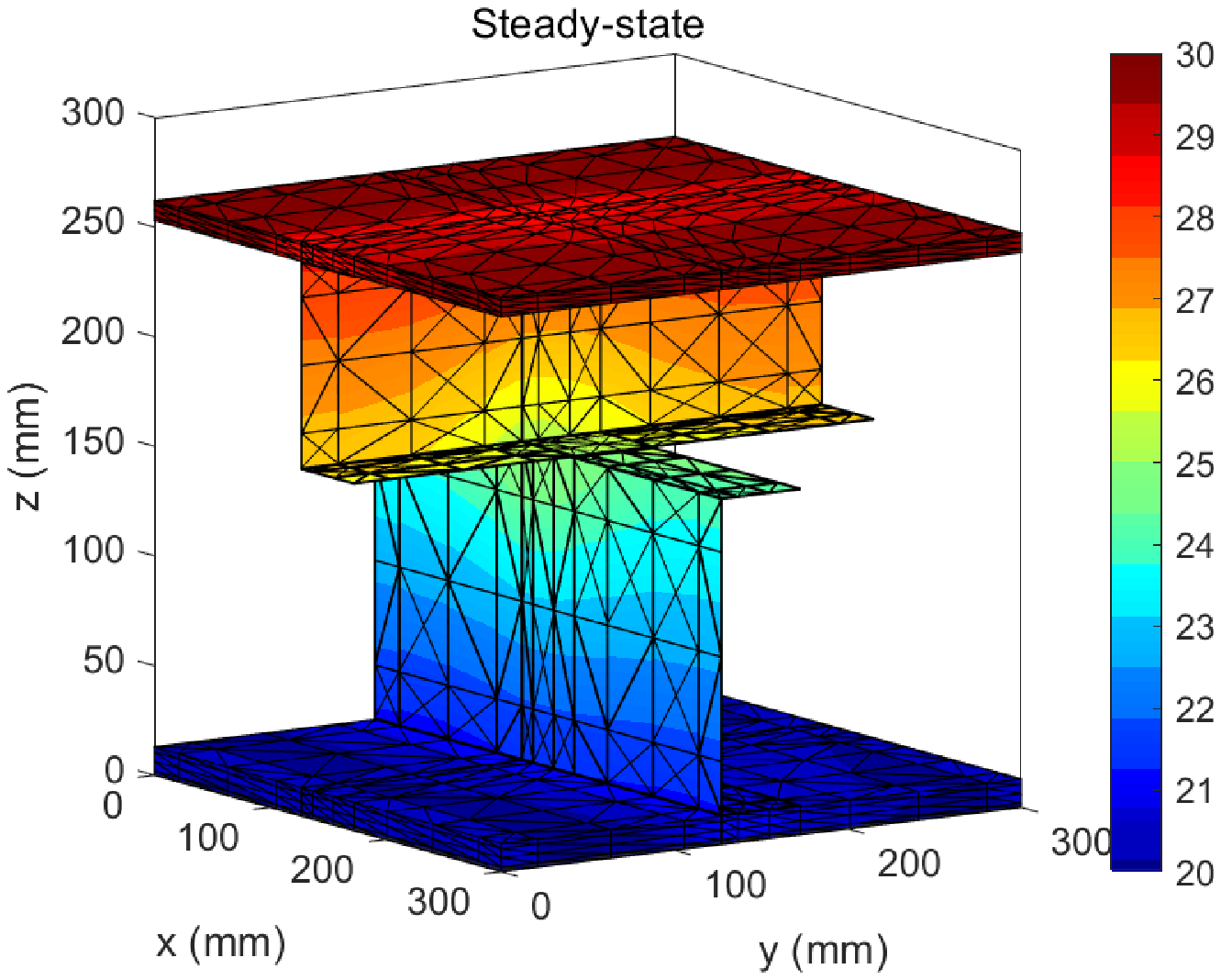}} 
  \caption{Ex. (2.8) – The problem and computed solutions. (a) the problem domain and boundary conditions. (b) transient temperature solution. (c) temperature distribution when $t=1$ hours. (d) steady-state result.} 
  \label{fig:EX28} 
\end{figure}

Finally, two practical examples with multiple materials and complicated geometries are considered. Ex.~(2.8) shows the heat conduction in a wall with crossed U-girders. The example is presented in \cite{Blomberg1996}. As shown in Fig.~\ref{fig:EX28-BC}, the wall is consisted of two gypsum wallboards, two steel crossed U-girders and insulation materials (the insulation material is not presented in the sketch). The U-girders are separated by $300~\mathrm{mm}$. Thus, we can only focus on a $300~\mathrm{mm} \times 300~\mathrm{mm} \times 262~\mathrm{mm}$ cell of the wall. The material properties are listed in Table~\ref{table:EX28-M}. The indoor ($z = 262~\mathrm{mm}$) and outdoor ($z = 0~\mathrm{mm}$) surfaces are under convection boundary conditions. The corresponding heat transfer coefficients $h$ and the temperatures outside the surfaces are shown in Table~\ref{table:EX28-BC}. All the other lateral surfaces are symmetric, i.e., $\widetilde{q}_N = 0$ in this case. The initial condition is $20\mathrm{^\circ C}$ in the whole domain.

A total of 2880 points are used in the FPM analysis. Notice that though the insulation material is not shown in the sketch, there are still points distributed in it. As a result of the uneven variation of material properties, the density of points used in the gypsum and steel are higher than the insulation. It should be pointed out that when the point distribution is extremely uneven, as in this example, it is highly recommended to define the boundary-dependent parameter $h_e$ in Eqn.~\ref{eqn:FPM} as the distance of the two points sharing the subdomain boundary. Fig.~\ref{fig:EX28-Trans} presents the time-variation of temperatures on three representative points A, B, and C (shown in Fig.~\ref{fig:EX28-BC}) in 10~hours. The FPM + LVIM solution shows an excellent consistency with the ABAQUS result obtained with 9702 DC3D8 elements (11132 nodes). The computed temperature distribution in the gypsum wallboards and U-girders when $t = 1~\mathrm{hours}$ and $t \geq 10~\mathrm{hours}$ (steady-state) are exhibited in Fig.~\ref{fig:EX28-T1} and \ref{fig:EX28-SS}. The results also agree well with ABAQUS. The computational parameters and times are shown in Table~\ref{table:EX28}. As can be seen, the LVIM approach helps to save approximately one half of the total computing time. Ex.~(2.8) demonstrates the accuracy and efficiency of the FPM + LVIM approach in solving complicated 3D transient heat conduction problems with multiple materials and highly uneven point distributions.

\begin{table}[htbp]
\begin{minipage}[t]{0.65\textwidth}
\caption{Material properties in Ex.~(2.8).}
\centering
{
\begin{tabular}{ c c c c }
\toprule[2pt]
Material & \tabincell{c}{$\rho$ \\ $(\mathrm(kg/m^3))$} &  \tabincell{c}{$c$ \\ $(\times 10^3~ \mathrm{J/(kg ^\circ C)})$} &  \tabincell{c}{$k$ \\ $(\mathrm{W/ (m^{2 \circ} C)})$} \\
\hline
gypsum & 2300 & 1.09 & 0.22 \\
steel & 7800 & 0.50 & 60 \\
insulation & 1.29 & 1.01 & 0.036 \\
\toprule[2pt]
\end{tabular}}
\label{table:EX28-M}
\end{minipage}
\begin{minipage}[t]{0.28\textwidth}
\caption{Boundary conditions in Ex.~(2.8).}
\centering
{
\begin{tabular}{ c c c }
\toprule[2pt]
bc & \tabincell{c}{$\widetilde{u}_R$\\ $(\mathrm{^\circ C})$} & \tabincell{c}{$h$ \\ $(\mathrm{W/(m^{2 \circ} C)})$} \\
\hline
outdoor & 20 & 25 \\
indoor & 30 & 7.7 \\
\toprule[2pt]
\end{tabular}}
\label{table:EX28-BC}
\end{minipage}
\end{table}

\begin{table}[htbp]
\caption{Computational time of FPM + LVIM / backward Euler approach (2880 points) in solving Ex.~(2.8).}
\centering
{
\begin{tabular}{ c c c c c }
\toprule[2pt]
Method & \tabincell{c}{Computational \\ parameters} & Time step & \ \tabincell{c}{Computational \\ time (s)} \\
\hline
FPM + LVIM & $\eta_1 = 11$, $M=3$, $tol = 10^{-8}$ & $\Delta t = 2~h $ & 31 \\
\hline
FPM + backward Euler & $\eta_1 = 11$  & $\Delta t = 0.1~h $ & 65 \\
\toprule[2pt]
\end{tabular}}
\label{table:EX28}
\end{table}

\begin{figure}[htbp] 
  \centering 
   \subfigure[]{ 
    \label{fig:EX29-BC} 
    \includegraphics[width=0.48\textwidth]{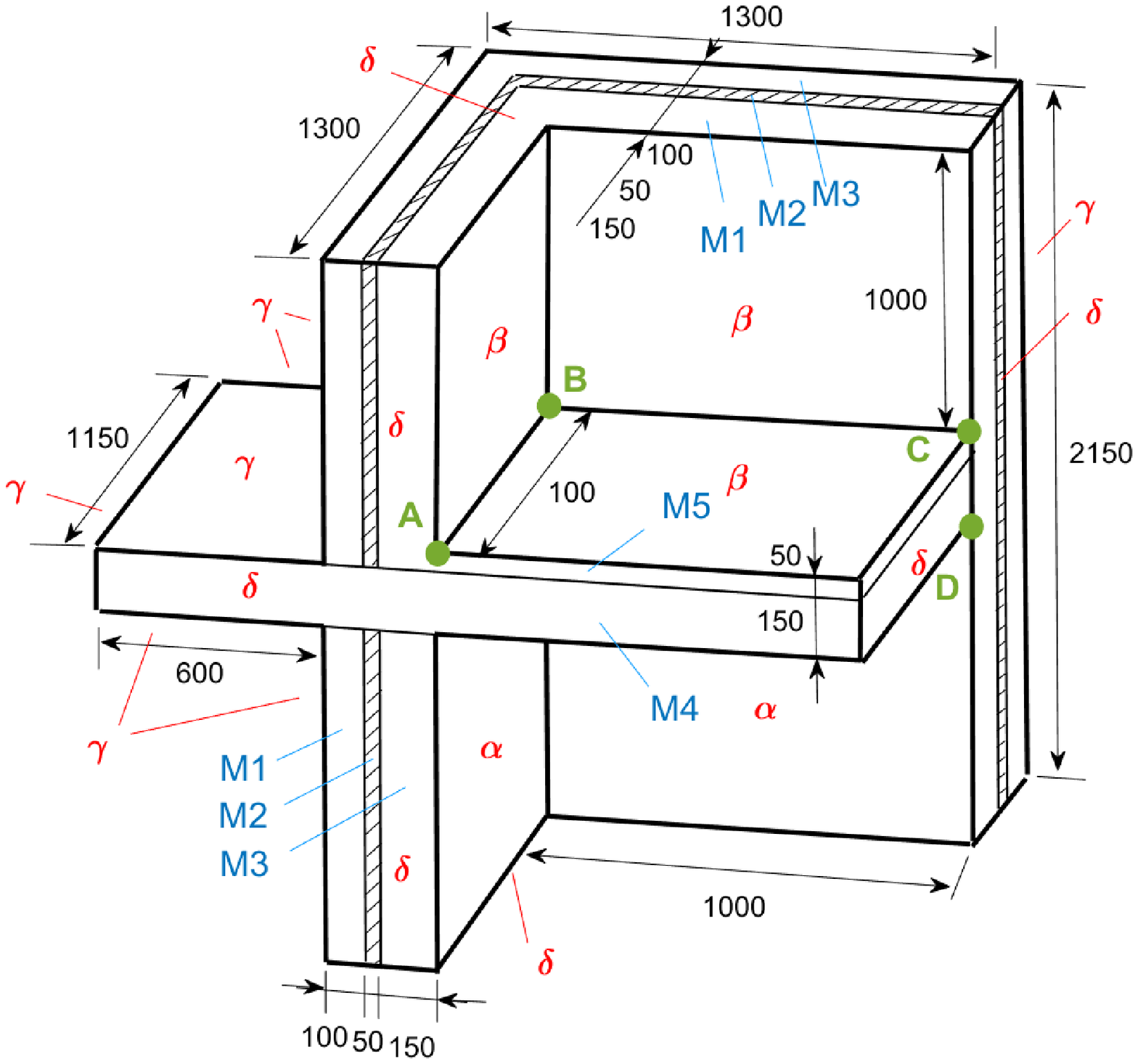}}  
   \subfigure[]{ 
    \label{fig:EX29-Trans} 
    \includegraphics[width=0.48\textwidth]{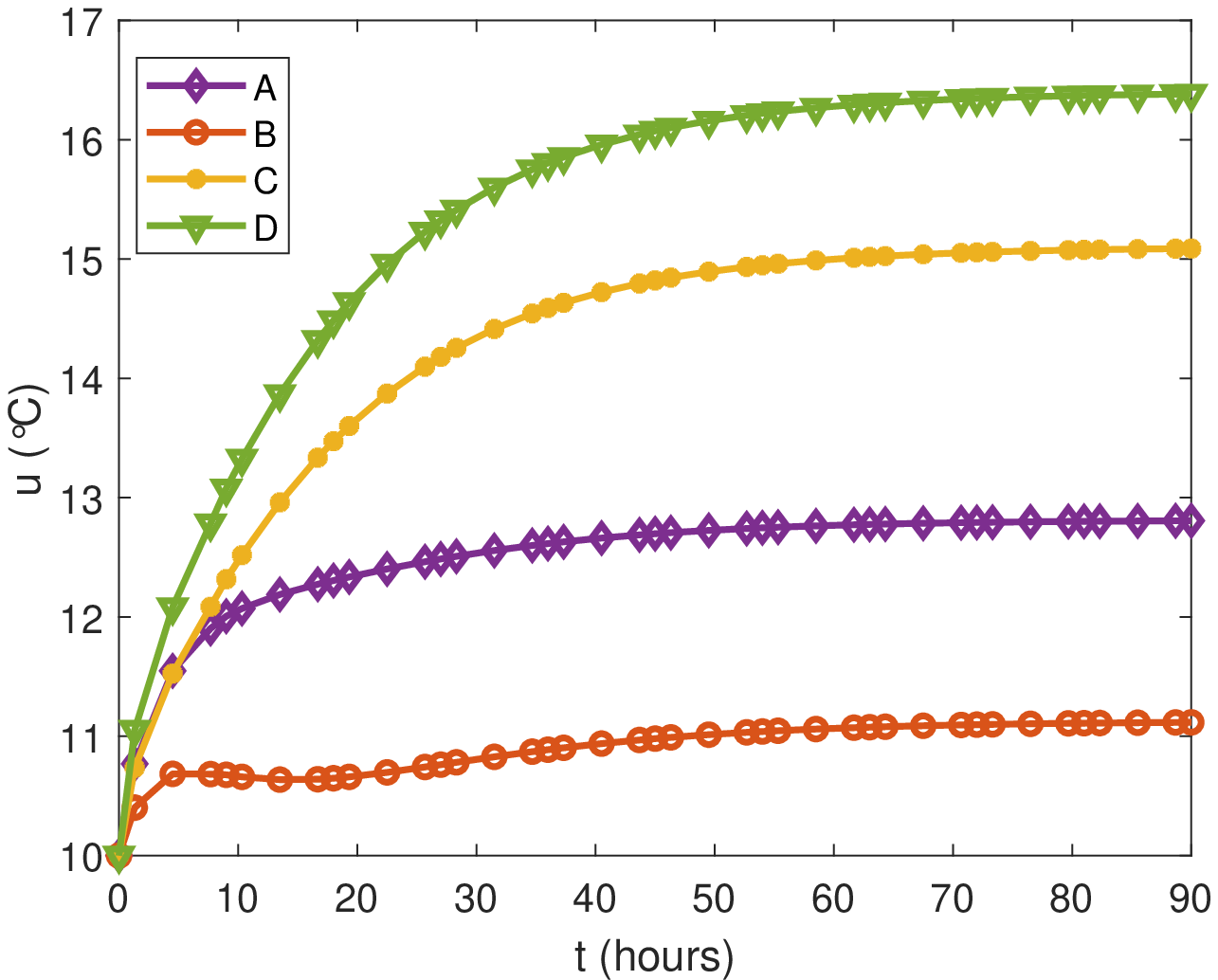}}  
    \subfigure[]{ 
    \label{fig:EX29-T6} 
    \includegraphics[width=0.48\textwidth]{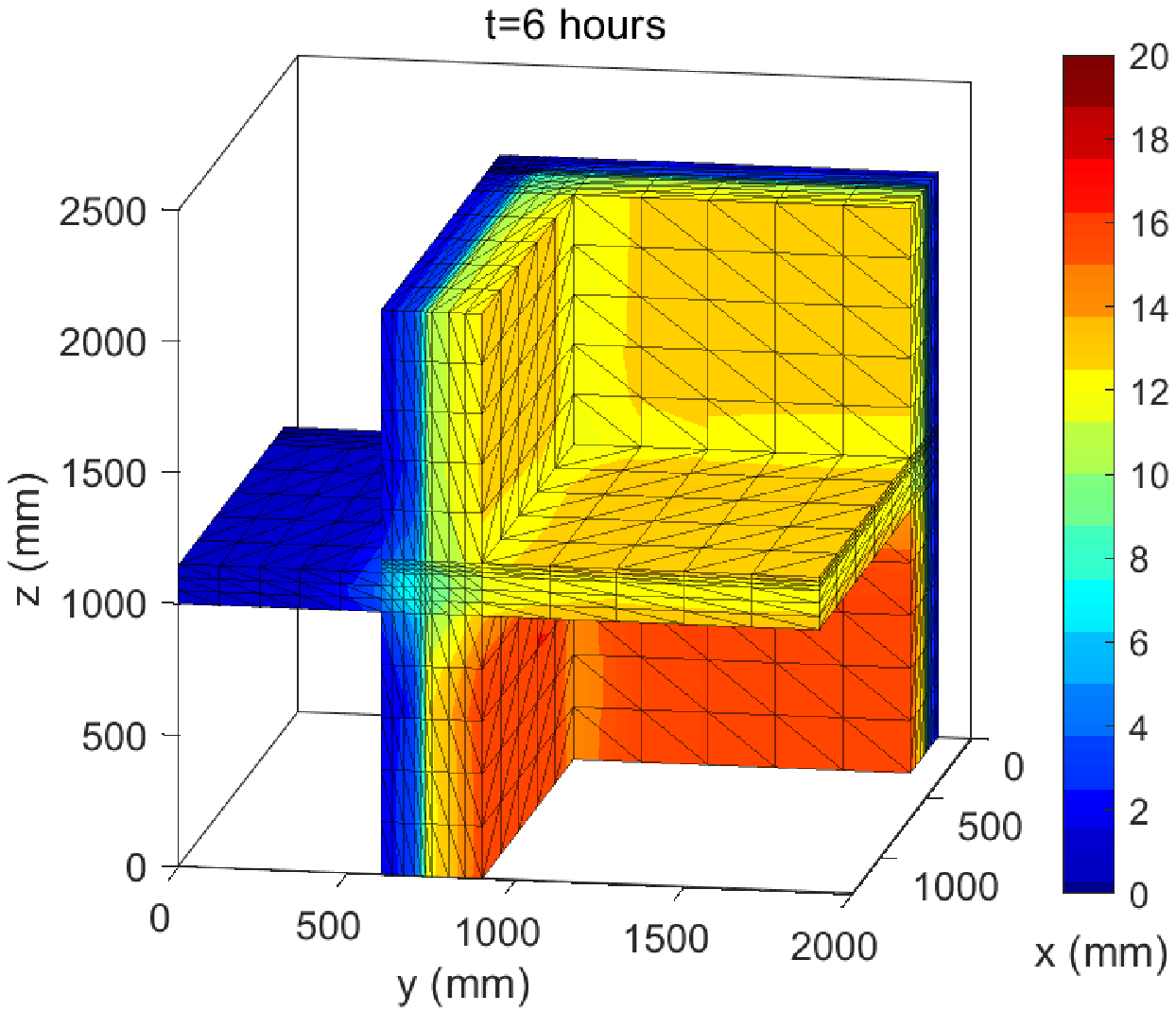}}  
    \subfigure[]{ 
    \label{fig:EX29-SS} 
    \includegraphics[width=0.48\textwidth]{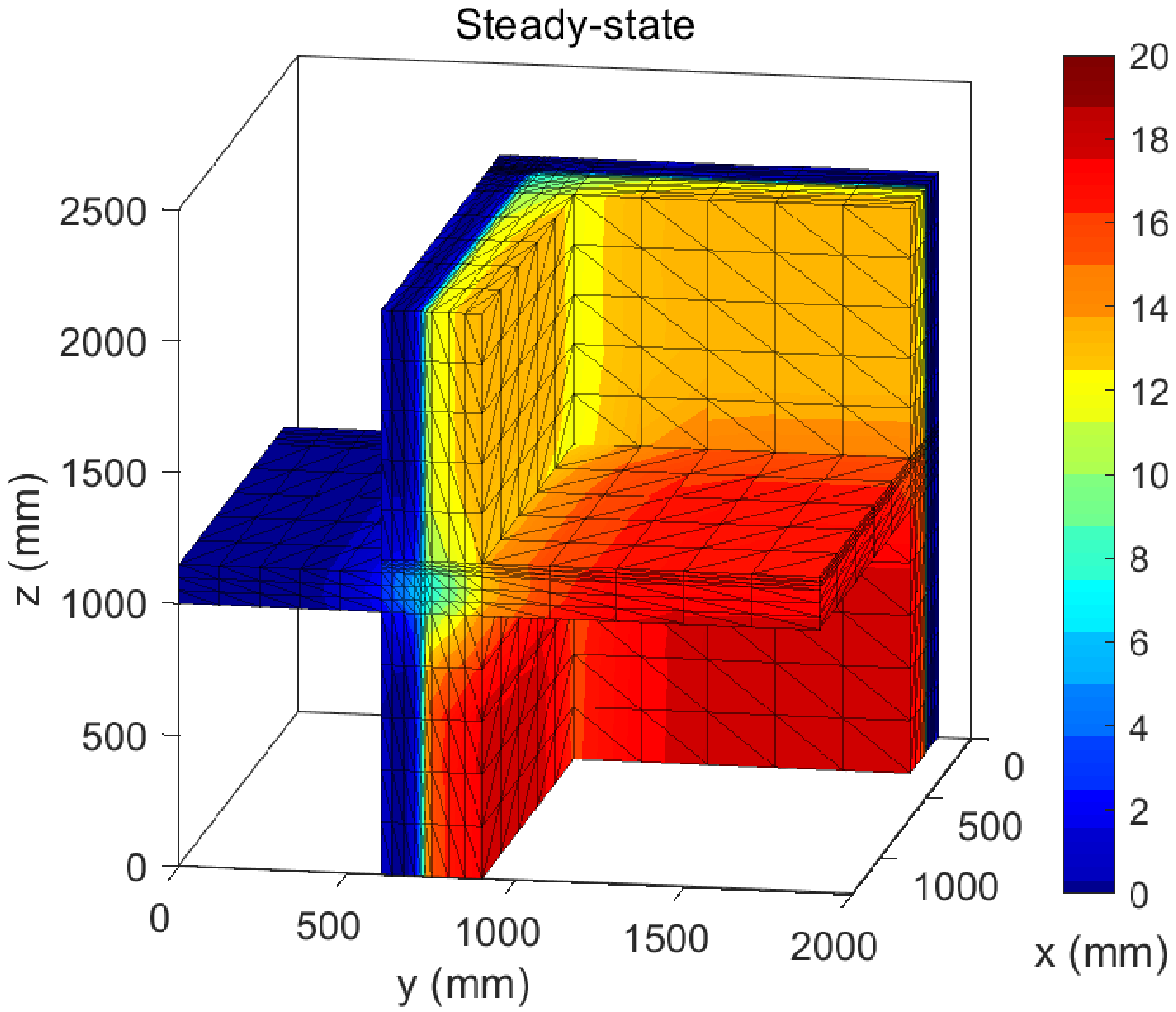}} 
  \caption{Ex. (2.9) – The problem and computed solutions. (a) the problem domain and boundary conditions. (b) transient temperature solution. (c) temperature distribution when $t=6 hours$ hours. (d) steady-state result.} 
  \label{fig:EX29} 
\end{figure}

Ex.~(2.9) is also given in \cite{Blomberg1996}. In this example, the heat transfer through a wall corner is studied. Fig.~\ref{fig:EX29-BC} shows the geometry and material distribution in the corner. Five kinds of materials are utilized. Their corresponding properties are listed in Table~\ref{table:EX29-M}. Four kinds of boundary conditions are presented in Fig.~\ref{fig:EX29-BC}, in which $\delta$ stands for adiabatic boundaries, while $\alpha$, $\beta$ and $\gamma$ are all convection boundaries. Table~\ref{table:EX29-BC} presents their heat transfer coefficients and surface temperatures. The initial condition is $10~\mathrm{^\circ C}$ in the whole domain.

First, we concentrate on the temperatures of four representative points (A, B, C, D) as shown in Fig.~\ref{fig:EX29-BC}. The time-variation of temperatures on these points is presented in Fig.~\ref{fig:EX29-Trans}. The result approaches steady state as time increases. Table~\ref{table:EX29-SS} illustrates how the number of points used in the FPM influences the steady-state solution. The results are compared with data in the European standards (CEN, 1995) \cite{Blomberg1996}. As can be seen, when the number of points rises, the solution approaches the reference solution gradually. With more than 6288 points, the result keeps stable and has no more than $0.1~\mathrm{^\circ C}$ error compared with the CEN solution. Fig.~\ref{fig:EX29-T6}  and Fig.~\ref{fig:EX29-SS} present the temperature distribution in the corner when $t = 6~\mathrm{hours}$ and after steady-state. Table~\ref{table:EX29-SS} shows the computational parameters and times comparing with the backward Euler scheme. Similar with the previous examples, the LVIM approach works well with large time intervals and has no stability problem. 

\begin{table}[htbp]
\begin{minipage}[t]{0.65\textwidth}
\caption{Material properties in Ex.~(2.9).}
\centering
{
\begin{tabular}{ c c c c }
\toprule[2pt]
Material & \tabincell{c}{$\rho$ \\ $(\mathrm(kg/m^3))$} &  \tabincell{c}{$c$ \\ $(\times 10^3~ \mathrm{J/(kg ^\circ C)})$} &  \tabincell{c}{$k$ \\ $(\mathrm{W/ (m^{2 \circ} C)})$} \\
\hline
M1 & 849 & 0.9 & 0.7 \\
M2 &  80 & 0.84 & 0.04 \\
M3 & 2000 &  0.8& 1.0 \\
M4 & 2711 & 0.88 & 2.5 \\
M5 & 2400 & 0.96 & 1.0 \\
\toprule[2pt]
\end{tabular}}
\label{table:EX29-M}
\end{minipage}
\begin{minipage}[t]{0.28\textwidth}
\caption{Boundary conditions in Ex.~(2.9).}
\centering
{
\begin{tabular}{ c c c }
\toprule[2pt]
bc & \tabincell{c}{$\widetilde{u}_R$\\ $(\mathrm{^\circ C})$} & \tabincell{c}{$h$ \\ $(\mathrm{W/(m^{2 \circ} C)})$} \\
\hline
$\alpha$ & 20 & 5 \\
$\beta$ & 15 & 5 \\
$\gamma$ & 0 & 20 \\
$\delta$ & -- & 0 \text{(adiabatic)} \\
\toprule[2pt]
\end{tabular}}
\label{table:EX29-BC}
\end{minipage}
\end{table}

\begin{table}[htbp]
\caption{The computed steady-state temperatures ($^\circ C$) obtained by the FPM with different numbers of points ($L$) - Ex.~(2.9).}
\centering
{
\begin{tabular}{ c c c c c c c }
\toprule[2pt]
Point & $L=1120$ & $L=3006$ & $L=6288$ & $L=11350$ & $L=28350$ & CEN \cite{Blomberg1996} \\
\hline
A & 12.7 & 12.8 & 12.7 & 12.6 & 12.6 & 12.6 \\
B & 10.9 & 11.1 & 11.1 & 11.0 & 11.0 & 11.1 \\
C & 12.7 & 14.6 & 15.1 & 15.2 & 15.2 & 15.3 \\
D & 15.9 & 16.4 & 16.5 & 16.4 & 16.4 & 16.4\\
\toprule[2pt]
\end{tabular}}
\label{table:EX29-SS}
\end{table}

\begin{table}[htbp]
\caption{Computational time of FPM + LVIM / backward Euler approach (3006 points) in solving Ex.~(2.9).}
\centering
{
\begin{tabular}{ c c c c c }
\toprule[2pt]
Method & \tabincell{c}{Computational \\ parameters} & Time step & \ \tabincell{c}{Computational \\ time (s)} \\
\hline
FPM + LVIM & $\eta_1 = 10$, $M=3$, $tol = 10^{-6}$ & $\Delta t = 30~h $ & 32 \\
\hline
FPM + backward Euler & $\eta_1 = 10$  & $\Delta t = 1.875~h $ & 59 \\
\toprule[2pt]
\end{tabular}}
\label{table:EX29}
\end{table}

\subsection{Discussion on computational parameters} \label{sec:PS}

\subsubsection{Penalty parameters}

As have been stated in the previous sections, the penalty parameters $\eta_1$ and $\eta_2$ have a significant influence on the accuracy and stability of the FPM. For example, if $\eta_1$ is too small, the method could be unstable and results in discontinuous solutions. If $\eta_2$ is too small, the Dirichlet boundary conditions may not be satisfied. On the contrary, if $\eta_1$ is very large, small jumps of temperature on the internal boundaries can be expected, but the accuracy of the solution is doubtable. In this section, parametric studies on $\eta_1$ and $\eta_2$ are carried out on both 2D and 3D examples.

\begin{figure}[htbp] 
  \centering 
   \subfigure{ 
    \label{fig:PS_EX3_00} 
    \includegraphics[width=0.48\textwidth]{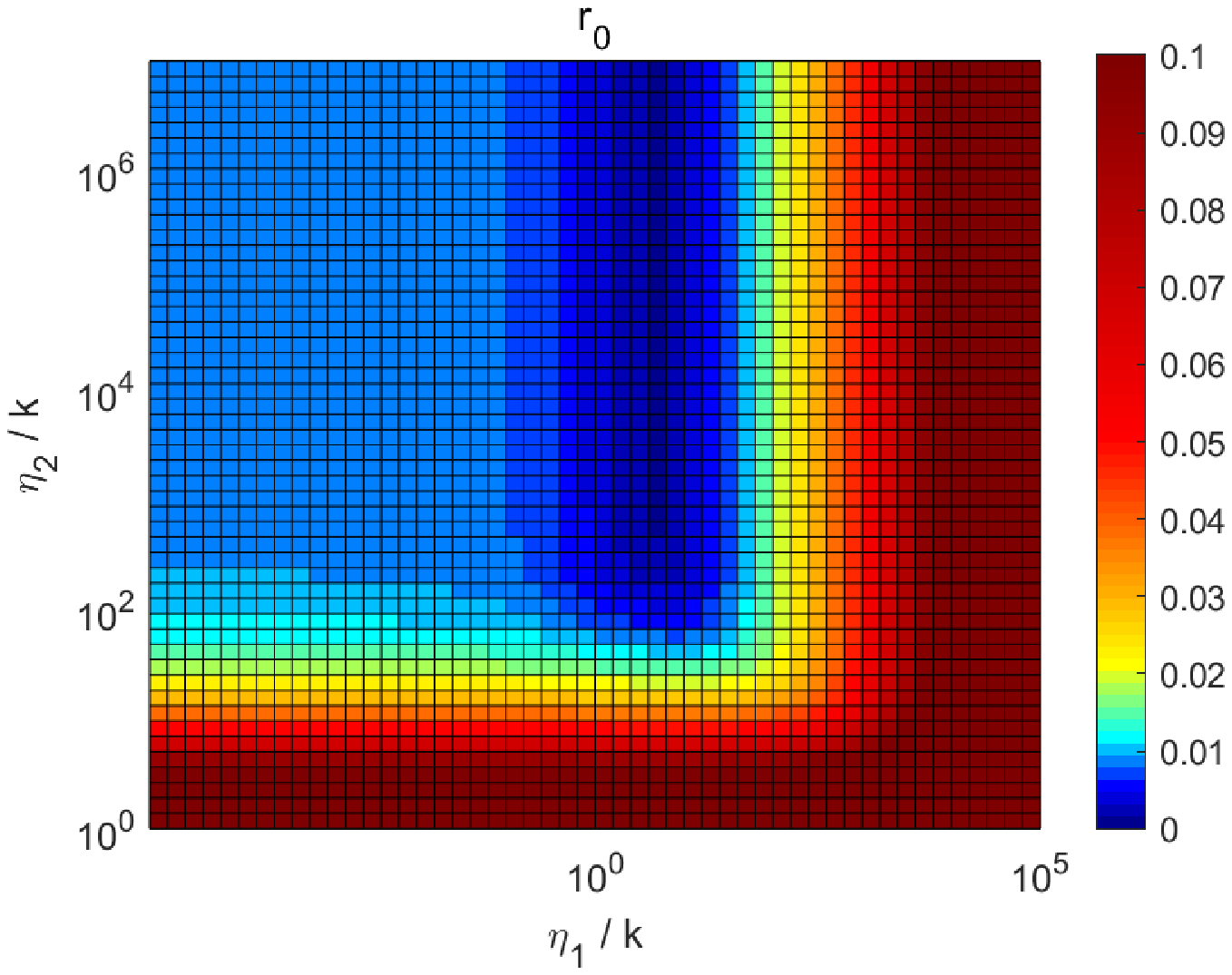}}  
   \subfigure{ 
    \label{fig:PS_EX21_00} 
    \includegraphics[width=0.48\textwidth]{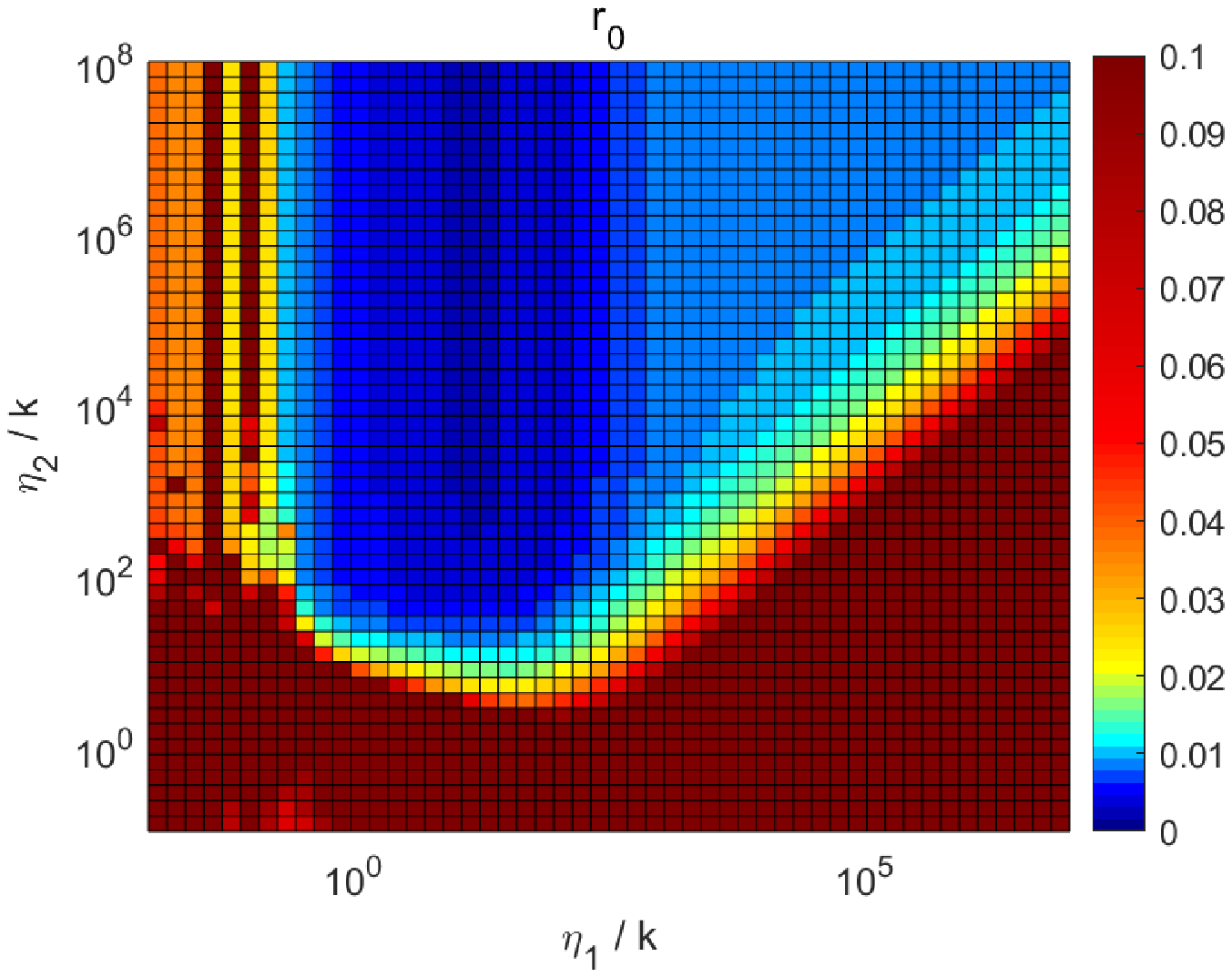}}  
   \subfigure{ 
    \label{fig:PS_EX3_01} 
    \includegraphics[width=0.48\textwidth]{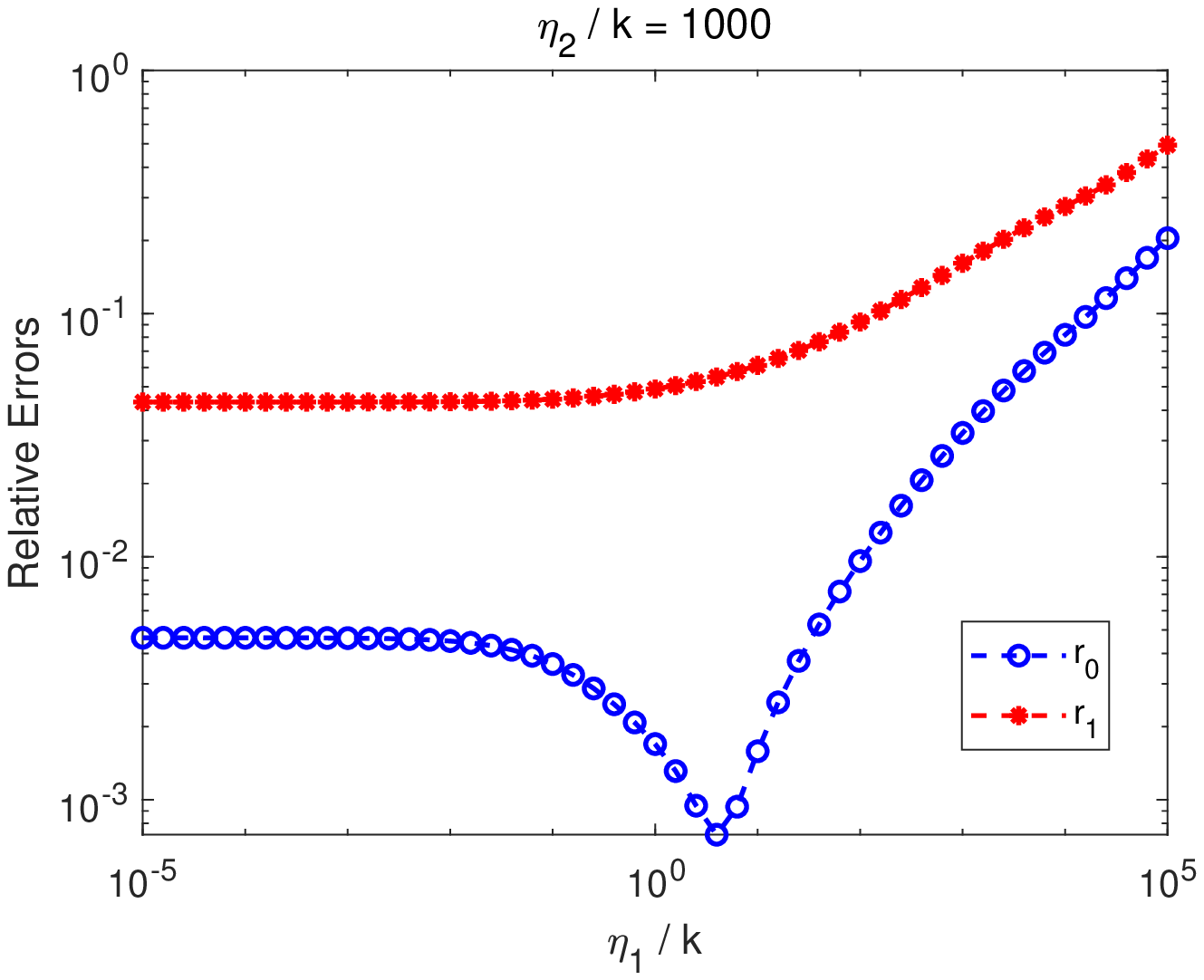}}  
   \subfigure{ 
    \label{fig:PS_EX21_01} 
    \includegraphics[width=0.48\textwidth]{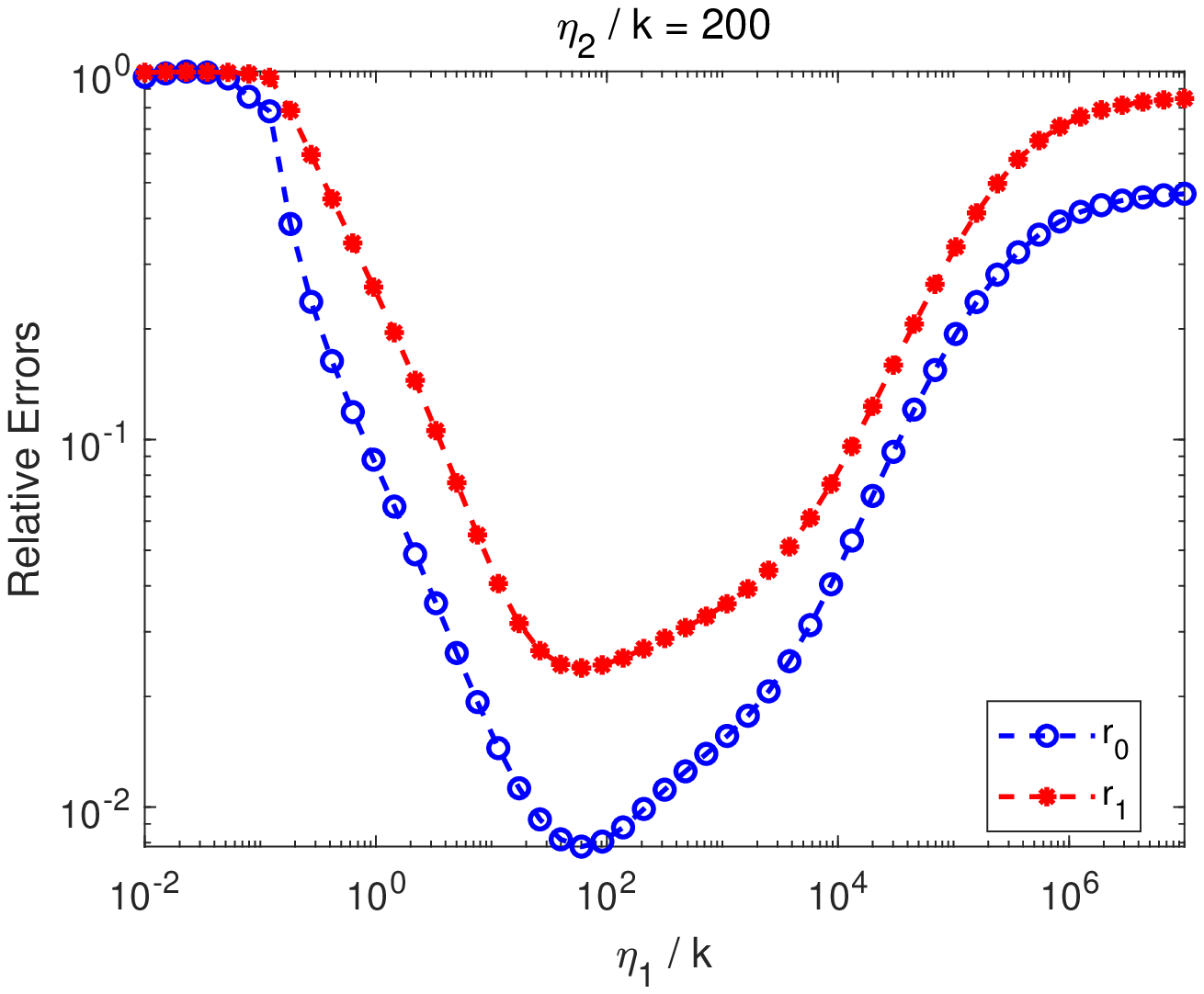}} 
    \addtocounter{subfigure}{-4}
    \subfigure[]{ 
    \label{fig:PS_EX3_02} 
    \includegraphics[width=0.48\textwidth]{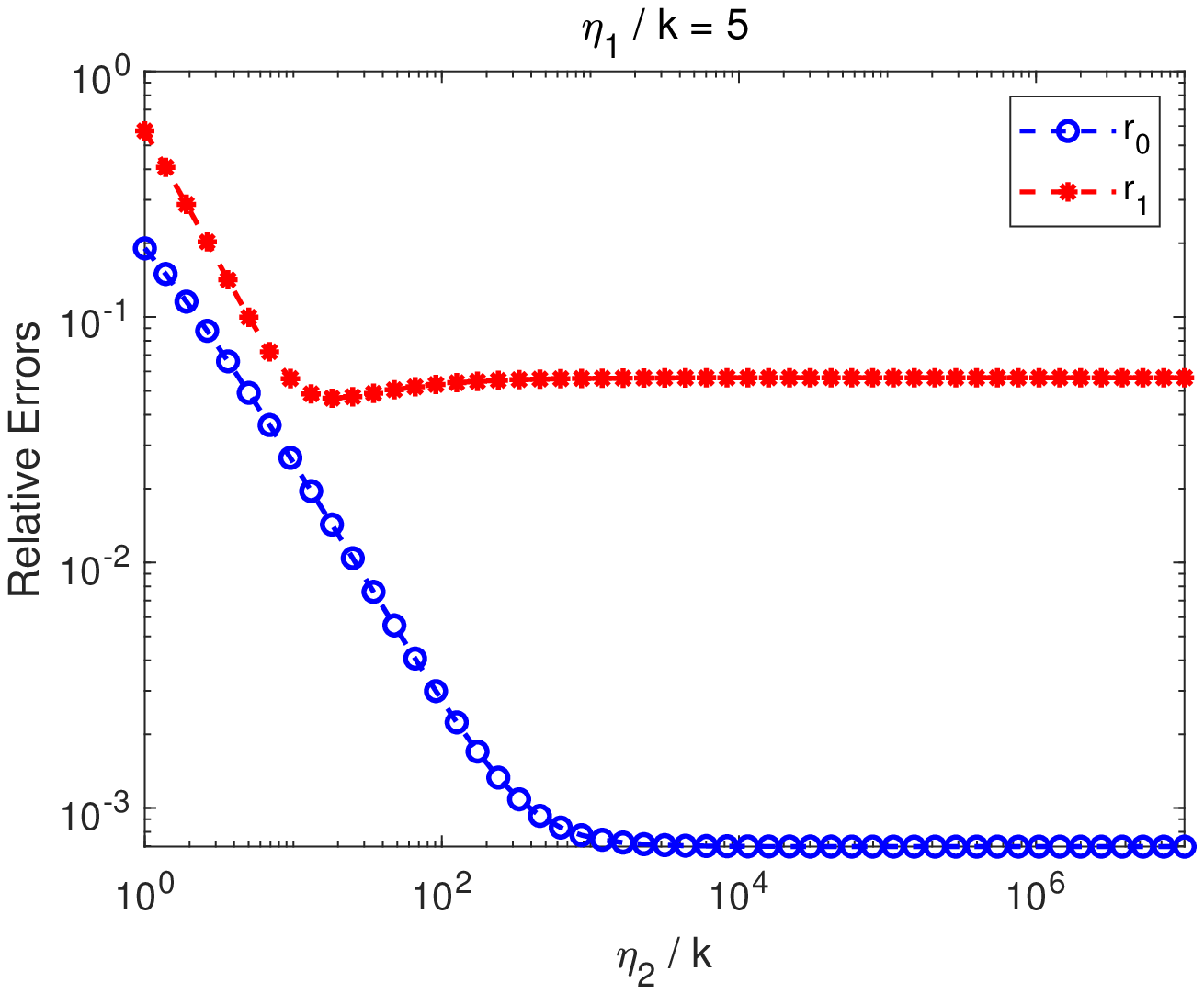}}   
    \subfigure[]{ 
    \label{fig:PS_EX21_02} 
    \includegraphics[width=0.48\textwidth]{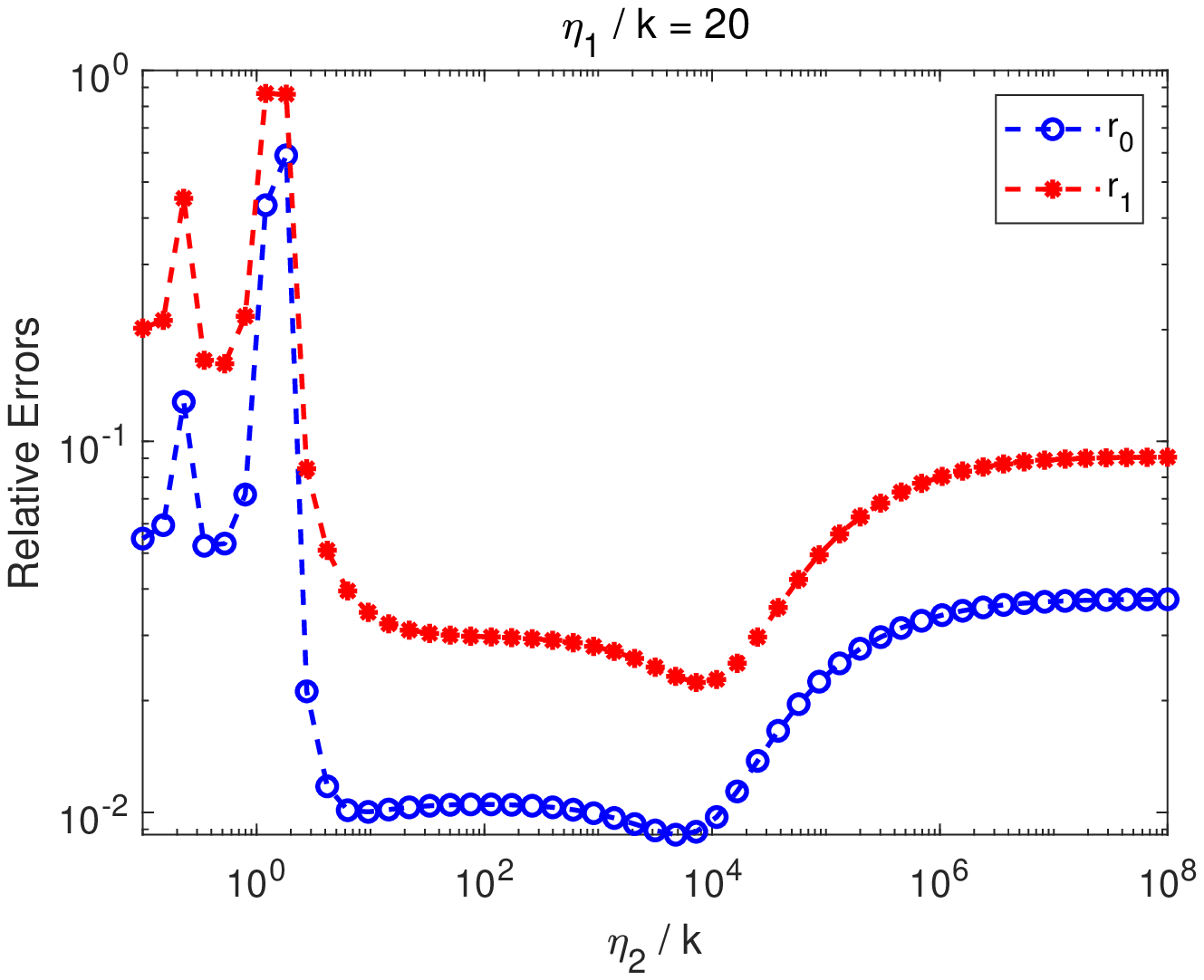}}  
  \caption{Parametric studies on $\eta_1$ and $\eta_2$. (a) 2D case: Ex.~(1.3). (b) 3D case: Ex.~(2.1).} 
  \label{fig:PS} 
\end{figure}

First, the steady-state solution of 2D example Ex.~(2.3) is considered. We concentrate on the nonhomogeneous anisotropic case, i.e., $\delta=3$, $\hat{k}_{11} = \hat{k}_{22} = 2, \hat{k}_{12} = \hat{k}_{21} = 1$. A total of 225 points are used in the FPM. Fig.~\ref{fig:PS_EX3_02} shows the influence of the penalty parameters on the relative errors $r_0$ and $r_1$. The penalty parameters are nondimensionalized by $k = \left( \hat{k}_{11} \hat{k}_{22} \right) ^{1/2} = 2$. As can be seen, in order to get a stable and accurate solution, it is recommended to define $\eta_1$ in the range of $0.1 k $ to $100 k$, and $\eta_2$ larger than $50 k$. The best choice in this example is $\eta_1 = 5 k$ and $\eta_2 > 500 k$. The accuracy decreases dramatically when $\eta_1$ is too large or $\eta_2$ is too small. Yet there is no upper limit of the recommended range of $\eta_2$. Notice that in homogenous or isotropic case, the effective range of $\eta_1$ and $\eta_2$ can be much larger.

In 3D case, the anisotropic steady-state example Ex.~(2.1) is considered. With 1000 points distributed uniformly in the domain, the relative errors $r_0$ and $r_1$ of the computed FPM solution under varying $\eta_1$ and $\eta_2$ are shown in Fig.~\ref{fig:PS_EX21_02}, in which the penalty parameters are nondimensionalized by $k = \left( {k}_{11} {k}_{22} {k}_{33}\right) ^{1/3} = 1 \times 10 ^{-4}$. To get a continuous and accurate computed solution, the penalty parameters should be defined in the range $0.5 k < \eta_1 < 1000 k$, and $5 k < \eta_2 < 20000 k$. In this example, the best choice is $\eta_1 = 15 k$ and $\eta_2 = 5000 k$. However, the best choice varies under different point distributions. As can be seen from the parametric studies, the relative errors shoot up when $\eta_1$ or $\eta_2$ is too small, as the continuity or essential boundary conditions may not be satisfied then. An excessively large $\eta_1$ should also be avoided. Whereas a large $\eta_2$ is still acceptable since it does not do much harm to the accuracy. 

In general, the recommended values of $\eta_1$ and $\eta_2$ are proportional to the thermal conductivity $\mathbf{k}$. The approximate effective ranges of $\eta_1$ and $\eta_2$ are $k < \eta_1 < 50 k$, and $50 k < \eta_2 < 1 \times 10^4 k$, where $k = \left( k_{11} k_{22} \right)^{1/2}$ in 2D case and $k = \left( k_{11} k_{22} k_{33} \right)^{1/3}$ in 3D case. Notice that the range may vary under different definitions of $h_e$ and different point distributions. Generally, the best choice of $\eta_2$ should be slightly larger than $\eta_1$ since a small discontinuity of temperature on the internal boundaries is acceptable, while the essential boundary conditions should be satisfied strictly. Homogenous and isotropic problem usually has less requirement on the effective penalty parameters.

\subsubsection{Number of collocation nodes in each time interval}

\begin{figure}[htbp] 
  \centering 
  \includegraphics[width=0.65\textwidth]{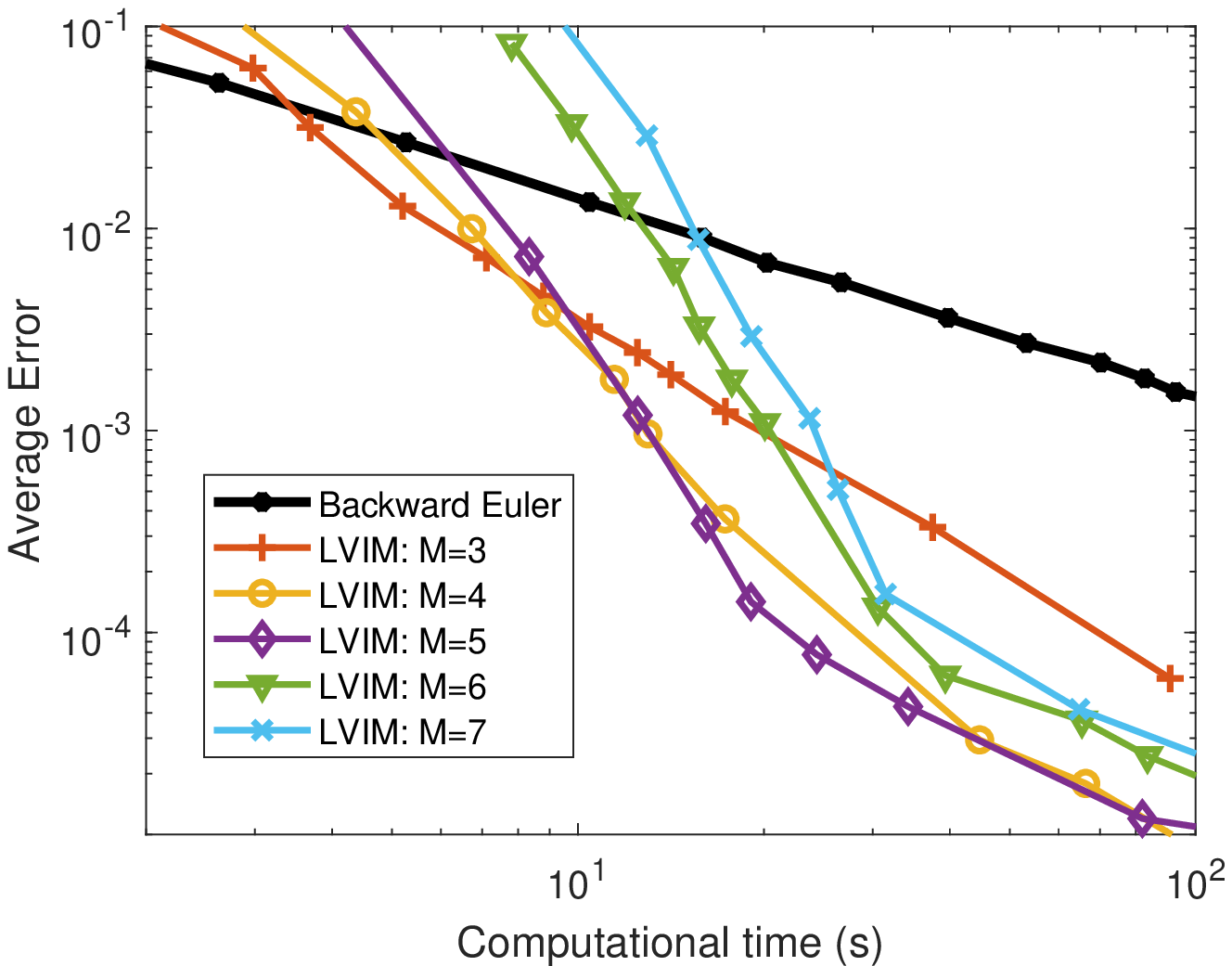}
  \caption{Parametric study on the number of collocation nodes in each time interval ($M$) - Ex.~(1.1).} 
  \label{fig:PS_M} 
\end{figure}

Next, the recommended value of the number of collocation nodes in each time interval ($M$) in the LVIM is discussed. Take Ex.~(1.1) as an example, Fig.~\ref{fig:PS_M} presents the relationship of computational time and average relative error $\overline{r}_0$ achieved by the LVIM approach with different $M$ and backward Euler scheme. The example is discretized with 601 uniform points in the FPM with $\eta_1 = 5$. And the error $\overline{r}_0$ is defined as the average value of relative error between the computed solution and the converged solution (obtained with an extremely small time step) in time scope $[0, 8]$. As can be seen, though the backward Euler scheme and LVIM approach with $M = 3$ have an advantage in computational time under low accuracy requirement, e.g., $\overline{r}_0 \leq 0.1$, their computational times increase rapidly when the required relative error decreases. As a result, large $M$ has a benefit in achieving relatively accurate solution, while small $M$ is more suitable for exploring a rough approximation. Notice that the backward Euler scheme is equivalent to the LVIM approach with $M = 2$, and follows the same tendency of accuracy and computational time for the LVIM.

Table~\ref{table:PS_M} shows the computational times required for the LVIM approach and backward Euler scheme ($M=2$) when obtaining the same relative errors. To get a solution with $\overline{r}_0 \approx 1 \times 10^{-2}$, the LVIM approach with $M = 3$ costs the least computational time, which is approximately one third of the computational time of the backward Euler scheme. On the other hand, in order to achieve higher accuracy, e.g. $\overline{r}_0 \approx 1 \times 10^{-3}$, the best choice would become $M = 5$. Comparing with the backward Euler scheme, the LVIM approach shows extraordinary efficiency under high accuracy requirement. Since the computational time rises rapidly with $M$, too many collocation nodes (e.g., $M > 5$) are not recommended. We usually apply $M$ in the range of 3 to 5. For problems with lower accuracy requirement and higher numbers of nodes, a small $M$ is recommended. Whereas for problems with higher accuracy requirement and less nodes, a larger $M$ could be more beneficial.

\begin{table}[htbp]
\caption{Computational time of FPM + LVIM / backward Euler approach under varying number of collocation nodes in each interval ($M$) in solving Ex.~(1.1).}
\centering
{
\begin{tabular}{ c c c c }
\toprule[2pt]
Method & $M$ & Time step $\Delta t$ & Computational time (s) \\
\toprule[2pt]
\multicolumn{4}{c}{Average relative error $\overline{r}_0 \approx 1 \times 10^{-2}$} \\
\hline
FPM + backward Euler & 2  & 0.013 & 16 \\
\hline
\multirow{5}*{FPM + LVIM} & 3 & 0.27 & 5 \\
~ & 4 & 0.53 & 7 \\
~ & 5 & 0.80 & 8 \\
~ & 6 & 1.33 & 12 \\
~ & 7 & 1.60 & 16 \\
\toprule[2pt]
\multicolumn{4}{c}{Average relative error $\overline{r}_0 \approx 1 \times 10^{-3}$} \\
\hline
FPM + backward Euler & 2  & 0.0016 & 144 \\
\hline
\multirow{5}*{FPM + LVIM} & 3 & 0.08 & 17 \\
~ & 4 & 0.27 & 13 \\
~ & 5 & 0.53 & 12 \\
~ & 6 & 0.8 & 20 \\
~ & 7 & 1.1 & 24 \\
\toprule[2pt]
\end{tabular}}
\label{table:PS_M}
\end{table}

\section{Conclusion}

A new computational approach is developed for analyzing 2D and 3D transient heat conduction problems in complex anisotropic nonhomogeneous media. The truly meshless Fragile Points Method (FPM) based on Galerkin weak-form formulation is employed for spatial discretization, while the Local Variational Iteration (LVI) scheme is used to achieve the solution in the time domain. The meshless FPM is a significant advancement over either the Element-Free Galerkin (EFG) Method or the Meshless Local Petrov-Galerkin (MLPG) Method. The EFG is based on Global Galerkin weak-form and requires back-ground cells to integrate the weak-form terms. The integration becomes tedious while using the meshless Moving Least Squares (MLS) approximations. Also when the mesh of back-ground cells is rotated, the EFG may not be an objective method. The MLPG is a truly meshless method, based on a local Petrov-Galerkin weak-form, and the integration of the weak-form is complicated when MLS approximations are used as trial functions and test functions are different from the trial functions. The FPM is also a truly meshless method based on a Galerkin weak-form, uses very simple polynomial discontinuous trial and test functions and the integration of the weak-form is simple. The imposition of essential boundary conditions in the FPM is similar to that in EFG and MLPG. The FPM leads to sparse symmetric matrices. The time integration scheme LVIM is considerablely superior to the finite difference methods. Thus, the FPM + LVIM method for transient heat conduction in anisotropic nonhomogeneous solids presented in this paper is a superior meshless method as compared to those in earlier literatures.

The FPM is generated by local, simple, polynomial, point-based (as opposed to element-based in the FEM) and piecewise-continuous trial and test functions. Numerical Flux Corrections are employed in terms of internal penalty functions. With large enough penalty parameters, the method presents its consistency and accuracy with both regularly and randomly distributed points. A simple domain partition is still required, but just for integral computation. Symmetric and sparse matrices can be achieved in most cases. In the time domain, the highly efficient LVIM is introduced. As a combination of the VIM and a collocation method in each time interval, the LVIM shows excellent accuracy and efficiency in solving nonlinear ODEs.

Plenty of numerical examples are presented both in 2D and 3D. Mixed boundary conditions are involved, including Dirichlet, Neumann, Robin, and purely symmetric boundary conditions. Both functionally graded materials and composite materials are considered. The computed solutions are compared with analytical results, equivalent 1D or 2D results, and FEM solutions obtained by a commercial software. The forward and backward Euler schemes are used together with the FPM as a comparison to the LVIM. The FPM + LVIM approach exhibits great accuracy and efficiency and has no stability problem under relatively large time intervals. The anisotropy and nonhomogeneity give rise to no difficulties in the current approach. The computing efficiency is extraordinary when the response varies dramatically, or a high accuracy is required. The approach is also capable of analyzing systems with preexisting cracks, even if the domain partition does not coincide on the crack geometry. This implies the further potential of the FPM + LVIM approach in solving crack propagation problems. At last, a recommended range of the computational parameters is given. We can conclude that, with suitable computational parameters, the FPM + LVIM approach shows excellent performance in analyzing transient heat conduction systems with anisotropy and nonhomogeneity.

\section*{References}
\bibliography{Paper}

\end{document}